\documentclass[usedcolumn,useAMS,usenatbib]{mn2e}
\usepackage{txfonts}
\usepackage{natbib}
\bibpunct{(}{)}{;}{a}{}{,}
\usepackage{graphicx}
\usepackage{epstopdf}
\usepackage{color}
\usepackage[usenames,dvipsnames]{xcolor}
\usepackage{multirow}
\usepackage{longtable,lscape}
\usepackage{xspace}
\usepackage{xargs}

\newcommand{\galapagos}{{\scshape galapagos}\xspace}
\newcommand{\sex}{{\scshape Sextractor}\xspace}
\newcommand{\galfit}{{\scshape galfit}\xspace}
\newcommand{\galfitm}{{\scshape galfitm}\xspace}
\newcommand{\sersic}{S\'ersic\xspace}
\newcommand{\red}{\textit{red}\xspace}
\newcommand{\green}{\textit{green}\xspace}
\newcommand{\blue}{\textit{blue}\xspace}
\newcommand{\Red}{\textit{Red}\xspace}
\newcommand{\Green}{\textit{Green}\xspace}
\newcommand{\Blue}{\textit{Blue}\xspace}
\newcommand{\lown}{\textit{low-$n$}\xspace}
\newcommand{\highn}{\textit{high-$n$}\xspace}
\newcommand{\Lown}{\textit{Low-$n$}\xspace}
\newcommand{\Highn}{\textit{High-$n$}\xspace}
\newcommand{\re}{R_{\rm e}}
\newcommandx{\N}[2][1= ,2= ]{$\mathcal{N}^{#1}_{#2}$\xspace}
\newcommandx{\R}[2][1= ,2= ]{$\mathcal{R}^{#1}_{#2}$\xspace}
\newcommandx{\Nang}[2][1= ,2= ]{$\langle\mathcal{N}^{#1}_{#2}\rangle$\xspace}
\newcommandx{\Rang}[2][1= ,2= ]{$\langle\mathcal{R}^{#1}_{#2}\rangle$\xspace}

\newcommand{\aj} {AJ}

\newcommand{\apjs} {ApJS}

\newcommand{\pasp} {PASP}

\title[Wavelength-dependent sizes and profiles of galaxies]{Galaxy And Mass Assembly (GAMA): the wavelength-dependent sizes and profiles of galaxies revealed by MegaMorph}

\author[Vulcani et al. ]{Benedetta Vulcani,$^{1,2}$\thanks{E-mail: benedetta.vulcani@ipmu.jp} Steven~P.~Bamford,$^{3}$ Boris H\"au\ss ler,$^{1,3,4,5}$ Marina Vika,$^{1}$\newauthor
Alex Rojas,$^{1}$ Nicola~K.~Agius,$^{6}$ Ivan Baldry,$^{7}$
Amanda~E.~Bauer,$^{8}$ Michael~J.~I.~Brown,$^{9}$\newauthor
Simon Driver,$^{10,11}$  Alister~W.~Graham,$^{12}$ Lee~S.~Kelvin,$^{13}$  Jochen Liske,$^{14}$ Jon Loveday,$^{15}$\newauthor
Cristina~C.~Popescu,$^{6}$ Aaron~S.~G.~Robotham$^{10,11}$ and Richard~J.~Tuffs$^{16}$
\\
\smallskip\\
$^{1}$Carnegie Mellon University Qatar, Education City, PO Box 24866, 
Doha, Qatar\\
$^{2}$Kavli Institute for the Physics and Mathematics of the Universe (WPI), Todai Institutes for Advanced Study, University of Tokyo, Kashiwa 277-8582, Japan\\
$^{3}$School of Physics \& Astronomy, The University of Nottingham, University Park, Nottingham, NG7 2RD, UK\\
$^{4}$Department of Physics, University of Oxford, Denys Wilkinson Building, Keble Road, Oxford, OX1 3RH, UK\\
$^{5}$University of Hertfordshire, Hatfield, Hertfordshire, AL10 9AB, UK\\
$^{6}$Jeremiah Horrocks Institute, University of Central Lancashire, Preston PR1 2HE, UK\\
$^{7}$Astrophysics Research Institute, Liverpool John Moores University, IC2, Liverpool Science Park, 146 Brownlow Hill, Liverpool, L3 5RF, UK\\
$^{8}$Australian Astronomical Observatory, PO Box 915, North Ryde, NSW 1670, Australia \\
$^{9}$School of Physics, Monash University, Clayton, Victoria 3800, Australia\\
$^{10}$SUPA - School of Physics and Astronomy, University of St Andrews, North Haugh, St Andrews KY16 9SS, UK\\
$^{11}$ICRAR - The University of Western Australia, 35 Stirling Highway, Crawley, WA 6009, Australia\\
$^{12}$Centre for Astrophysics and Supercomputing Swinburne University of Technology Hawthorn, Victoria 3122, Australia\\
$^{13}$Institut f\"{u}r Astro- und Teilchenphysik, Universit\"{a}t Innsbruck, Technikerstra{\ss}e 25, 6020 Innsbruck, Austria \\
$^{14}$European Southern Observatory, Karl-Schwarzschild-Str. 2, 85748 Garching, Germany \\
$^{15}$Astronomy Centre, University of Sussex, Falmer, Brighton BN1 9QH, UK \\
$^{16}$Max Planck Institut fuer Kernphysik, Saupfercheckweg 1, 69117 Heidelberg, Germany}

\begin{document}

\date{Accepted 2014 March 31.  Received 2014 February 27; in original form 2013 November 26}
\pagerange{\pageref{firstpage}--\pageref{lastpage}} \pubyear{2014}

\maketitle

\label{firstpage}

\begin{abstract}
We investigate the relationship between colour and structure within galaxies using a large, volume-limited sample of bright, low-redshift galaxies with optical--near-infrared imaging from the GAMA survey.  We fit single-component, wavelength-dependent, elliptical \sersic models to all passbands simultaneously, using software developed by the MegaMorph project.  
Dividing our sample by $n$ and colour, 
the recovered wavelength variations in effective radius ($\re$) and \sersic index ($n$)
reveal the internal structure, and hence formation history, of different types of galaxies. All these trends depend on $n$; some have an additional dependence on galaxy colour.  Late-type galaxies ($n_r < 2.5$) show a dramatic increase in \sersic index with wavelength.  This might be  a result of their two-component (bulge-disk) nature, though stellar population gradients within each component and dust attenuation are likely to play a role.  All galaxies show a substantial decrease in $\re$ with wavelength.  This is strongest for early-types ($n_r > 2.5$), even though they maintain constant $n$ with wavelength, revealing that ellipticals are a superimposition of different stellar populations associated with multiple collapse and merging events.  Processes leading to structures with larger $\re$ must be associated with lower metallicity or younger stellar populations.  This appears to rule out the formation of young cores through dissipative gas accretion as an important mechanism in the recent lives of luminous elliptical galaxies.
\end{abstract}


\vspace*{3\baselineskip}  

\section{Introduction}

It is common to describe galaxies by a few key properties, such as luminosity, colour, size and ellipticity, which are (at least in the local universe and  at optical and near-infrared wavelengths) indicative of the total mass, average age, and spatial extent of their stellar contents. 
At a slightly more detailed level, one may characterise a galaxy in terms of its internal structure, or morphology: the presence and relative strength of features such as disks, bulges, bars, rings and spiral arms. Each of these features represents a distinctive distribution of stars, in terms of both their positions and velocities. Unfortunately, resolved stellar velocities are difficult to obtain, particularly for large samples of distant galaxies.  The largest studies of the internal structure of galaxies must therefore inevitably rely on imaging data.

Galaxies principally come in two morphological types: ellipticals and disk galaxies. Ellipticals have a simple, smooth appearance,
show old stellar populations and live preferably in denser environments (see, e.g., \citealt{dressler97, kauffmann03, brinchmann04}). Spirals primarily comprise a thin disk, containing a spiral arm pattern; possess younger stellar populations, and avoid dense regions 
(e.g. \citealt{freeman70, blanton09} and references therein). However, spirals often also contain a spheroidal `bulge', a structure similar in colour and shape to a small elliptical galaxy, at their centre \citep{andredakis95}; though bulges appear to be denser than present day elliptical galaxies, (e.g., \citealt{graham08,graham13}). The relative size and importance of the bulge and the appearance of the spiral features vary substantially, resulting in a range of spiral morphologies \citep{dejong96}.  Lenticular galaxies sit at the intersection of ellipticals and spirals.  They display similarities to both, often being dominated by a bulge, but also by definition, containing a spiral-free, disk-like component.  Ellipticals and lenticulars can be hard to distinguish from their appearance.  Indeed, over the past decade many galaxies that were thought to be ellipticals have been discovered to contain rotating disks, and thus are actually  lenticular galaxies.

From the latest models of galaxy formation and evolution, it is becoming clear that there is a more fundamental distinction in the galaxy population than that between elliptical, lenticular,  and spiral galaxies (e.g., \citealt{ellis01}). To truly understand galaxy evolution, one must separate the components of galaxies: spheroids and disks. Elliptical galaxies are dominated by a single spheroidal component, while most spiral and lenticular galaxies contain both a spheroid (the bulge) and a disk. The striking difference between galaxy types may primarily be a result of variation in the relative prominence of the spheroid and disk components.

A common way of describing the structure of galaxies is through the \sersic index $n$, which describes the radial concentration of the projected light distribution.
The \citet{sersic68} function is given by:
$$
 I(r)=I_{\rm e} \exp \left \{   -b_n \left [ \left ( \frac{r}{\re} \right )  ^{\frac{1}{n}}  -1 \right ] \right \}
$$ 
where $\re$ is the effective radius (that is, the radius containing half of the model light),  $I_{\rm e}$ is the intensity at the effective radius, $n$ is the \sersic index, and $b_n$ 
is a function of \sersic index and is such that $\Gamma(2n) = 2\gamma(2n, b_n)$, where $\Gamma$ and $\gamma$ represent the complete and incomplete gamma functions, respectively \citep{ciotti91}. 
{\citet{andrae11} showed that this is the simplest function one could consider to model a galaxy profile, being the first-order Taylor expansion of any real light profile.}
When $n$ is equal to the values $0.5$, $1$ and $4$ (and $b_n$ assumes tha values of 0.676, 1.676 and 7.676, respectively), the \sersic profile is equivalent to a Gaussian, exponential and de Vaucouleurs profile, respectively.  A \sersic projected profile implies a 3D luminosity density profile involving the Fox $H$ function \citep{baes11}.

Elliptical galaxies are typically regarded as single-component structures, hence $n$ is a reasonably fundamental description of the profile (but see \citealt{cappellari11} and \citealt{huang13a}).
However, when fitting a single-\sersic profile to a disk galaxy the resulting $n$ reflects the contribution of both the disk and the bulge to the light distribution.

{We note that some degeneracy exists, in that appropriate small variations in Re and n can produce similar surface brightness profiles, which are indistinguishable at the resolution and depth of the data we consider.  The two quantities are also found to be correlated in the galaxy population.  However, e.g., \cite{trujillo01} showed that the correlation between $n$ and $\re$ is definitely not explained by parameter coupling in the fitting process; this trend between galaxy structure and size exists also when one uses model-independent values. Indeed, it is a natural consequence from the  relations existing between the model-independent parameters: total luminosity, effective radius and effective surface brightness.}

All galaxies show variations in colour to some degree. 
These often appear as radial colour gradients, with the centres of galaxies generally redder than their outer regions (at least in the more luminous galaxy population; \citealt{jansen00}).
Even the colour of elliptical galaxies varies significantly with radius (e.g., \citealt{barbera10b}).

Colour gradients provide important information for understanding how galaxies form and evolve.  For example, steep gradients are expected when stars form during strongly-dissipative (monolithic) collapse in galaxy cores.  In such a scenario, gas is retained by the deep potential well, with consequently extended star formation activity and greater chemical enrichment of the inner regions (e.g., \citealt{kobayashi04}).

In early-type galaxies the main driver of colour gradients is probably metallicity (see,  e.g., \citealt{saglia00, tamura00, barbera03,spolaor09, rawle10, brok11} and references therein), even though the contribution of the age  (see, e.g., \citealt{barbera02, barbera09, barbera12}) and even dust (see, e.g., \citealt{guo11, Pastrav13}) cannot be ignored.
Structurally, ellipticals may appear to be single-component systems, but that component certainly does not possess a homogeneous stellar population.

Colour gradients are typically stronger in bulge-disk systems.
Bulges tend to contain older, redder stars of higher metallicity than those found in the disks of galaxies (e.g., \citealt{johnston12} and references therein). Shorter optical wavelengths are more sensitive to the young, blue populations typically found in disks,
whereas longer (optical and near-infrared) wavelengths increasingly trace the old stellar population of both components.  Most late-type galaxies therefore display a substantial colour gradient from red centres to blue outskirts, \citep{dejong96b,gadotti01,mac04}, though peculiar galaxies can show opposing trends \citep{taylor05}. 

In addition to stellar population gradients, the dust content of a galaxy can produce colour gradients and influence the measured structural parameters in a wavelength dependent manner \citep{peletier96,moll06,graham08, kelvin12, Pastrav13,rodriguez13}. However, it remains unclear whether dust attenuation or stellar population gradients are the dominant factor in determining how observed galaxy structure varies with wavelength.

\citet{park05} have shown how colour gradients may be used as a morphological classifier, while \citet{lee08} found that steeper colour gradients appear within star-forming galaxies, in both late and early-types.
 Atypical gradients can be used to identify interesting populations.  For example, \citet{suh10}, studying  early-type galaxies drawn from the SDSS DR6, found a tight correlation between the existence of steep colour gradients and centrally-concentrated residual star formation. They suggest a relation such that elliptical galaxies with bluer cores present globally bluer colours than average. \citet{ferreras09} found the same relation for spheroidal galaxies observed by the HST at $0.4 < z < 1.5$.

Most of the studies mentioned so far have focused on measuring radial colour gradients, rather than considering how their profile shapes depend on wavelength.
However, variations in colour with radius directly imply that the light profile must change with wavelength, in terms of varying $n$ and $\re$.\footnote{Of course, it may also depart from a \sersic profile, but if we assume \sersic profiles at all wavelengths -- as is reasonably well-justified {\citep{andrae11}} -- then most variations will be captured by changes in $n$ and $\re$.}
In general, therefore, the values measured for $n$ and $\re$ will depend on the observed wavelength.

Some works have started investigating the wavelength dependence of galaxy structure, fitting \sersic models independently to images in different wavebands. 
\citet{barbera10b} presented for the first time a careful analysis of the structural parameters of low-redshift early-type galaxies as a function of wavelength.
They showed that $\re$ decreases by 35\% from the optical to the near-infrared (NIR), reflecting the internal colour gradients in these systems. Their \sersic indices span a domain from $\sim 2$--$10$, with a median of $6$ for all wavebands. 
\citet{kelvin12} presented similar two-dimensional, single-\sersic model fits to a large sample of low-redshift galaxies. 
They found that the mean \sersic index of early-types shows a smooth variation with wavelength, increasing by 30\% from $g$ through $K$, while the effective radius decreases by 38\% across the same range. In contrast, late-types exhibit a more extreme change in \sersic index, increasing by 52\%, but a more gentle variation in effective radius, decreasing by 25\%.
These trends are interpreted as due to the effects of dust attenuation and stellar population/metallicity gradients within galaxy populations.

When fitting images at different wavelengths completely independently, statistical and systematic variations in the recovered centre, ellipticity, position angle, etc. will translate into noise in the recovered trends for $n$ and $\re$.  This is particularly an issue if one wants estimates of the wavelength dependence of structure on a galaxy-by-galaxy basis.  To address this, some parameters may be fixed to fiducial values, while the parameters of interest are allowed to vary.  However, it is not clear how to fairly determine such fiducial values.

In this paper we build upon the work of \citet{barbera10b} and \citet{kelvin12}, by fitting a single, wavelength-dependent model to all the data simultaneously.  This is made possible using an extended version of \galfit \citep{peng02,peng10} developed by the MegaMorph project (\citealt{bamford12}, \citealt[][hereafter H13]{Haussler13}, \citealt{vika13} and Bamford et al. in prep.).
This approach maximises the signal-to-noise available to constrain the model and ensures that the parameters are optimal for the whole dataset.
The greatest advantages of this approach are expected when decomposing galaxies with multiple components (e.g., \citealt{cameron09}).
Making colour information available to the fit increases the robustness with which the components can be separated, and hence improves the accuracy and physical meaningfulness of their recovered parameters.  However, to begin we have explored the performance of our technique in fitting single-component models.
H13 have tested the new method on large datasets, automating both the preparation of the data and the fitting process itself by extending the existing \galapagos \citep{barden12} code. 
\citet{vika13}, on the other hand, studied nearby galaxies and assessed our method by fitting artificially-redshifted versions of these galaxies.

Having at our disposal structural parameters obtained via a consistent multi-wavelength approach from MegaMorph,  we can now
study the wavelength dependence of  \sersic index and effective radius for both individual galaxies and sample populations. In this paper we quantify these  trends using the ratio of $n$ and $\re$ at two wavelengths. These ratios provide a simple but powerful parametric way of considering galaxy colour gradients, which reveal information about the  prevalence of galaxies with different internal structures, and how this varies with other galaxy properties.
At a more practical level, quantifying how $n$ and $\re$ vary with wavelength is crucial for removing biases when comparing measurements made using different bandpasses or at different redshifts.

The paper is organised as follows: in \S~\ref{data} we present the dataset used for the analysis and describe the properties of the sample we analysed.
In \S~\ref{results} we show our results for the wavelength dependence of \sersic indices and  effective radii, first by considering trends for subsamples of the galaxy population and then by quantifying the behaviour of individual galaxies.
In \S~\ref{stacks} we discuss the interpretation of our measurements and demonstrate how they relate to the visual appearance of galaxies.
In \S~\ref{discussion} we discuss our results in the context of previous work. Finally, in \S 6 we summarise and present our conclusions.

The analysis has been carried out using a cosmology with ($\Omega_m$, $\Omega_\Lambda$, $h$) = (0.3, 0.7, 0.7) and AB magnitudes.

\section{Data}\label{data}
\subsection{\sersic models and parent sample selection}

The sample of galaxies considered in this paper has been previously presented in H13. A detailed description of the selection criteria, { robustness of the fits} and properties of our sample can be found in that paper; here we give a brief overview.  Our sample is drawn from the Galaxy And Mass Assembly survey II (GAMA; \citealt{driver09, driver11}), the largest homogeneous multi-wavelength dataset currently available, in terms of both spatial volume and wavelength coverage. GAMA is primarily a redshift survey, but it is supplemented by a highly consistent and complete set of multi-wavelength data, spanning from the far-UV to radio.

The GAMA imaging data includes five-band optical ($ugriz$) imaging from SDSS \citep{york00} and four-band near-infrared ($YJHK$) imaging from the Large Area Survey (LAS) component of the UKIRT Infrared Deep Sky Survey (UKIDSS; Lawrence et al. 2007).
All of these bands have a depth and resolution amenable to \sersic-profile fitting (as demonstrated by \citealt{kelvin12}). Conveniently, the images for all nine bands have been `micro-registered' onto the same pixel grid (using SWarp, \citealt{bertin02,bertin10}, as described in \citealt{hill11}), and rescaled to a common zero point of $30$~mag, which are essential for our purposes.

Our sample is limited to galaxies in one region of GAMA, the equatorial field at 9h R.A. (known as {\it G09}). The multi-wavelength imaging data provided by GAMA is analysed using a modified version of \galapagos \citep{barden12}. This version (\galapagos-2) has been adapted for use on multi-wavelength, ground-based data, as described in detail in H13. This software enables the automated measurement of wavelength-dependent \sersic profile parameters for very large samples of galaxies in a homogeneous and consistent manner. To fit each galaxy, \galapagos-2 utilizes a recently-developed multi-wavelength version of \galfit (which we refer to as \galfitm). This extended version of \galfit was introduced in H13 and is described in detail in Bamford et al. (in prep). We refer the interested reader to this paper and H13 for further details and technical background reading.

\galfitm fits a single wavelength-dependent model to all the provided images simultaneously.  Rather than fit the parameter values at the wavelength of each band, \galfitm fits the coefficients of a smooth function describing the wavelength dependence of each parameter.  Specifically, \galfitm employs Chebyshev polynomials, with a user-specifiable order for each parameter, which controls the desired smoothness of their wavelength dependence.  For example, the $x$ and $y$ coordinates of the galaxy centre may be allowed to vary, but remain a constant function of wavelength.  Magnitudes could be allowed to vary completely freely, by setting a sufficiently high polynomial order.  Other parameters may be allowed to vary at a lower order, providing flexibility to obtain a good fit, while avoiding unjustified (physically or by the data quality) variability and reducing the dimensionality of the model.

In the fitting process for this paper, we allow full freedom in magnitudes, while \sersic index and $\re$ are allowed to vary with wavelength as second order polynomials. All other parameters ($x$ and $y$ position, position angle and axis ratio) are selected to be constant with wavelength.

\begin{figure}
\centering
\includegraphics[scale=0.6]{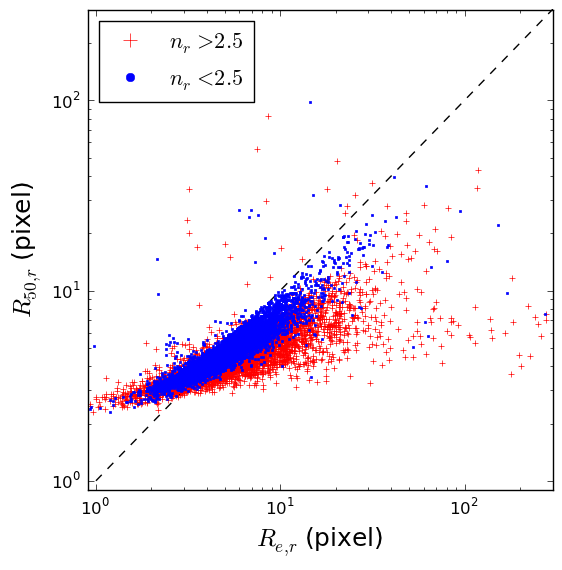}
\caption{Comparison between the effective radius in $r$-band obtained by \galfitm  and the radius containing half the flux within the Kron aperture in $r$-band, for the galaxy sample described in \S\ref{gal_sel}. Red crosses: galaxies with $n_r>2.5$. Blue dots: galaxies with $n_r<2.5$. Dashed line represents the 1:1.
\label{cfr_rsize}}
\end{figure}

{Amongst other data, \galapagos-2 returns magnitudes ($m$), \sersic indices ($n$) and effective radii ($\re$), calculated at the wavelength of each input image. To investigate the reliability of our size determinations, we have compared our $\re$ measurements to an alternative, less model-dependent, estimate of galaxy size. For convenience we use the standard half-light radius, $R_{50}$, provided by \sex during the initial stages of \galapagos. This is the radius containing half of the total flux within the Kron aperture \citep{kron80,bertin96}, obtained independently for each bandpass. Figure \ref{cfr_rsize} shows the comparison of these sizes in the $r$-band, for galaxies with $n_r<2.5$ and $n_r>2.5$.  We see very similar relations in all other bands. There is a clear correlation between the two size estimates, but also obvious deviations from a one-to-one line.  $R_{50}$ tends to $\sim 2$ pixels for small objects,  due to the effect of the PSF and pixelisation. Both of these are accounted for by the \sersic fit, and hence $\re$ suffers less bias at small sizes.  For larger galaxies, $R_{50}$ is consistently smaller than $\re$.  This is a result of Kron apertures missing flux in the outer regions of galaxies. The half-light radii are thus systematically underestimated.  The fraction of missed flux depends on the profile shape \citep{graham05}, such that the size is underestimated more, and with greater variation, for higher-$n$ objects.
\sersic sizes are therefore often regarded as being more reliable than aperture-based measurements \citep{graham05b}.
}

\galapagos-2 also returns the coefficients that describe the polynomial wavelength-dependence of each parameter. As described below, this function can be used to `interpolate' between the bands, e.g.\ in order to derive rest-frame values.
On high signal-to-noise data, one could in principle fit each band independently and then fit a polynomial to the individual-band results in order to derive a continuous function of wavelength. On high S/N data, for single-component fits, this would give more or less equivalent results to our approach.  However, on lower signal-to-noise images our approach is substantially more robust (H13, \citealt{vika13}, Bamford et al. 2013).  Nevertheless, there are cases where directly using the fit polynomial function to interpolate between bands is inadvisable.

If a high degree of freedom is given to a certain parameter (e.g.\ magnitude in our case), the polynomial may oscillate between the wavelengths at which it is constrained by data, and especially in regions close to the edge of the wavelength range of the input data. This is a general issue encountered when fitting polynomials, known as `Runge's phenomenon'.\footnote{This effect is discussed further in Bamford et al. (in prep.).} As a consequence, we do not use the polynomial derived for the magnitudes to directly calculate rest-frame magnitudes for all objects. Instead, we recommend using the magnitudes derived for each input band to accurately estimate rest-frame magnitudes and colours using a stellar population fitting code. 

\subsection{Our galaxy selection}\label{gal_sel}
We match our sample to the GAMA redshift catalogue (SpecObjv21). For the galaxies in the catalogue, we compute absolute magnitudes by applying {\scshape INTERREST} \citep{taylor09} to our measured photometry.  This tool determines rest-frame colours by interpolating the observed photometry, using a number of template spectra as guides (for details, see \citealt{rudnick03}).

For \sersic index and effective radii, the polynomial order is sufficiently low that Runge's phenomenon is not an issue.  We can therefore conveniently calculate rest-frame values directly using the respective Chebyshev functions returned by \galfitm, provided the required wavelength is in the range covered by the input data.  For the remainder of this paper, all mentions of wavelength, or `band', refer to rest-frame quantities, unless explicitly stated otherwise.

As in H13, the catalogue resulting from \galapagos-2 has been cleaned in order to select only the objects that have been successfully fit by \galfitm. In particular, we wish to identify and discard fits with one or more parameters lying on (or very close to) a fitting constraint. Such a fit is unlikely to have found a true minimum in $\chi^2$ space and is indicative of a serious mismatch between the model profile and the object in question. The selection also serves to remove stars from the catalogue. Following H13, we extract those galaxies that satisfy the following criteria:
\begin{itemize}
\item[$\bullet$] $0<m<40$ at all wavelengths, where $m$ is the total apparent output  magnitude in each band.
\item[$\bullet$] $m_{\rm input}-5<m<m_{\rm input}+5$, where $m_{\rm input}$ is the starting value of the magnitude in each band.  These are derived by scaling an average galaxy SED (obtained from previous fits) by the ${\rm mag\_best}$ value measured by \sex during the object detection.  As it is known \citep[e.g.][]{Haeussler07} that \galfit is very stable against changing the starting values, such an approach is valid and was easy to implement into \galapagos-2 (see H13 for details). During the fit, we allow a generous $\pm 5$ magnitudes variation from these starting values.
\item[$\bullet$] $0.201 < n < 7.99$, since fits with values outside these ranges are rarely meaningful (\citealt{vika13}; though note that some elliptical galaxies are indeed measured to have $n > 8$, e.g., as shown in \citealt{caon93}.)
\item[$\bullet$] $0.301 < \re < 399.0$ pixels, which maintains values in a physically meaningful range and prevents the code from fitting very small sizes, where, due to oversampling issues, the fitting iterations become very slow.
\item[$\bullet$] $0.001<q\leq1.0$, where $q$ is the axis ratio,  to ensure the fit value is physically meaningful. In practice, this constraint is barely ever encountered.
\end{itemize}
This catalogue cleaning is done on all bands simultaneously, i.e.\ if any value of the fit fails to meet the above criteria for any band, the entire fit is considered unsuccessful.\footnote{Please note that the last criterion used in H13 ($95 - 5 \times {\rm mag\_best} < {\rm fwhm\_image}$ and ${\rm fwhm\_image} < 1$~pixel) is not  used in this paper. The purpose of this cut was to separate (mainly saturated) stars from galaxies in order to derive a cleaner galaxy sample for code testing and comparing single- and multi-band fitting techniques. In this paper, this step is not necessary as galaxies are selected by a redshift criterion.  Avoiding it prevents us from accidentally removing bright, compact galaxies from our sample.}

We additionally exclude objects with: 
\begin{itemize}
\item[$\bullet$] $132.6 \leq \alpha \leq 142.0 $ and $ -1.55 \leq \delta  \leq -0.50$; 
\item[$\bullet$] $135.2 \leq \alpha \leq 135.7 $ and $ -0.35 \leq \delta  \leq +0.20$; 
\item[$\bullet$] $130.0 \leq \alpha \leq 131.2 $ and $ -2.50 \leq \delta  \leq -1.20$; 
\end{itemize}
which are objects close to areas for which at least one of the infrared bands ($YJHK$) does not provide good data quality. The first one of these areas has no data coverage in the J-band image, the other two are close to very bright stars which caused scattered light in the image, essentially making them unusable in some of the band in these areas. Rather than fitting different galaxies using different combinations of bands, we prefer to maintain a homogeneous dataset and excluded galaxies in these areas.

\subsection{Galaxy samples}
To make quantitative, representative statements we choose to volume limit our sample.  We also wish to demonstrate our ability to obtain meaningful multi-wavelength structural measurements with low-resolution data, so we elect to push our sample selection to relatively high redshifts. We therefore focus on the intrinsically  bright galaxy population in this paper.
\label{samples}
\begin{figure}
\centering
\includegraphics[scale=0.45]{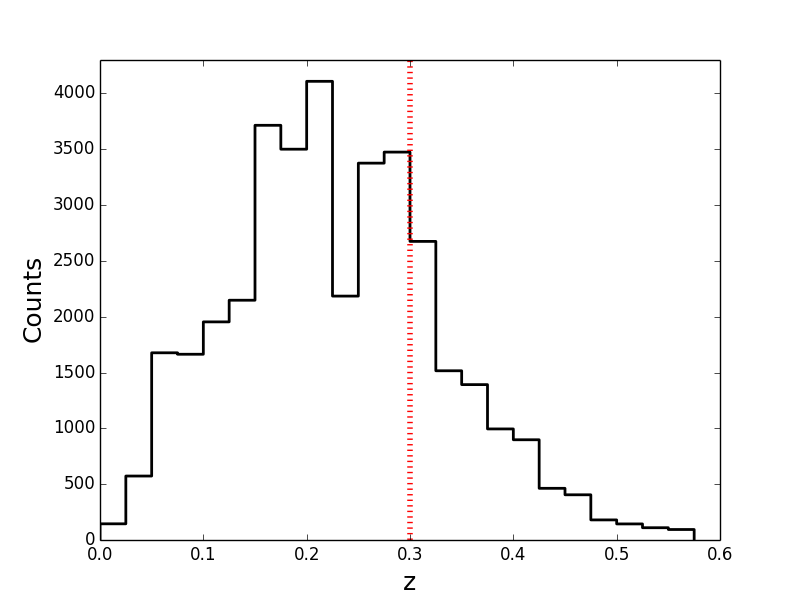}
\caption{Redshift histogram of our parent sample. Only galaxies with 0$<$$z$$<$0.3 are included in the subsequent analysis. \label{z_histo}}
\end{figure}

\begin{figure}
\centering
\includegraphics[scale=0.45]{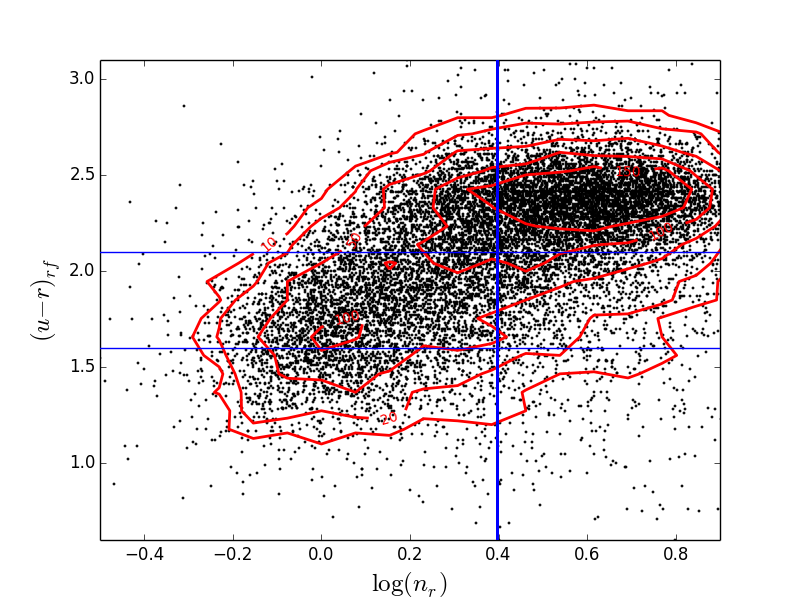}
\caption{$(u-r)_{rf}$ versus $n_r$ for the galaxies in our sample. Lines illustrate the cuts we apply to divide the galaxies by colour and \sersic index.
\label{col}}
\end{figure}

Figure~\ref{z_histo} shows the redshift histogram of all galaxies in the H13 sample with a reliable spectroscopic redshift estimate. We limit our analysis to galaxies at $0.01<z<0.3$.
For the selected redshift range, the rest-frame equivalents of all bands up to and including $H$ are within the observed ($u$--$K$) wavelength range, and thus their values can be interpolated.  The rest-frame equivalent for our reddest observed band, $K$, is obviously always extrapolated, hence unreliable, and we therefore exclude it from our analysis.

The nominal apparent magnitude limit of the GAMA II redshift survey is $r < 19.8 \, {\rm mag}$, which corresponds to a total absolute magnitude $M_r < -21.2 \, {\rm mag}$ at $z<0.3$.  Imposing this absolute magnitude limit gives us a complete, volume-limited sample of 14,274 galaxies.  Note that this selection means our results are limited to the relatively high-luminosity galaxy population.

The aim of this paper is to carefully analyse how galaxy structural parameters change as a function of  wavelength.  We expect different types of galaxies to behave in contrasting ways.  We therefore separate galaxies according to their colour and structure.  Throughout the paper, we 
use the rest-frame $(u-r)_{rf}$ colour,
and the $r$-band \sersic index, $n_r$,
as our reference for separating populations. The choice of colour is not critical.  However, bracketing the $4000\,$\AA break, $(u-r)_{rf}$ performs well for separating red and blue populations, despite the noisiness in the SDSS $u$-band.  After some experimentation, $n_r$ turned out to be the best choice for  separating high- and low-concentration populations, which roughly correspond to early- and late-type morphologies in our luminous galaxy sample.  Choosing $(u-r)_{rf}$ and $n_r$ also facilitates comparison with previous SDSS studies.

Figure~\ref{col} plots colour versus \sersic index for our sample. These are clearly correlated, and the well-known bimodality (e.g., \citealt{kauffmann03, driver06}) is evident.  Galaxies may be optimally divided into two types using a sloping cut in this plane.  However, we find it is informative to study the behaviour with respect to colour and \sersic index in more detail, and thus apply selections to these two quantities independently.
We empirically divide galaxies into red and blue using the cut $(u - r) = 2.1$.
Traditionally the division by colour also includes a magnitude dependence; however, this is a small effect, particularly at the bright magnitudes considered in this paper.  To avoid complicating our analysis we prefer this simple cut.  In addition to separating red and blue galaxies, we are also interested in the bluest population, which may host starbursts.  We therefore further subdivide the $(u - r) < 2.1$ population into `green' and `blue' galaxies, using a cut at $(u - r) = 1.6$. We note that our `green' galaxies do not correspond to what is commonly understood as the `green valley', between the blue and red modes.  We simply adopt this name to describe objects with blue, but not extremely blue, colours.  Henceforth we use italics to explicitly indicate when we are referring to the \red, \green or \blue samples.

We further divide galaxies by \sersic index, with the aim of better separating disk galaxies from ellipticals.
Following previous works (e.g., \citealt{barden05}), we adopt a cut of $n_r = 2.5$. From now on, we will refer to `\lown' galaxies with $n_r < 2.5$ and `\highn' galaxies with $n_r > 2.5$.

With these combined cuts we can compare the two main galaxy populations: \green\ \lown  systems, which  correspond to star-forming, disk-dominated galaxies, and \red\ \highn systems, which are typically passive spheroid-dominated galaxies. We can also examine the properties of galaxies which do not respect the majority behaviour.

We caution the reader that these definitions do not extend down to the fainter dwarf galaxies.  Our sample selection is limited to luminous galaxies, so trends discussed in this paper do not represent the faint population.

\begin{table}
\centering
\begin{tabular}{l|ll|ll}
Colour & \multicolumn{2}{c|}{$n_r<2.5$}&  \multicolumn{2}{c}{$n_r>2.5$}\\ 
\hline
\blue  & $1391$  & $9.7 \pm 0.2$ \% & $309$  & $2.2 \pm 0.1$ \% \\
\green  & $3183$ & $22.3 \pm 0.4$ \% & $1263$ & $8.8 \pm 0.2$ \%\\
\red  & $2321$ & $16.3 \pm 0.3$ \% & $5807$ & $40.7 \pm 0.4$ \%\\
\end{tabular}
\caption{Number count and fraction (of the total sample) for galaxies with different combinations of colour and \sersic index. Errors are binomial. The choice of the band adopted for the \sersic index separation does not considerably alter the fractions. \label{tab_frac}}
\end{table}

Table \ref{tab_frac} summarises the fraction of galaxies corresponding to our different colour and \sersic index selections.  As expected, most \red galaxies ($\sim 70$\%) are characterised by large values of \sersic index, while the vast majority of \blue  ($\sim 82$\%) and \green galaxies ($\sim 73$\%) display small values of \sersic index.

In the next section we study how the structural properties of galaxies, characterised by different colours and surface brightness profiles, vary with wavelength.

\section{Results}
\label{results}

Trends in galaxy properties with wavelength may be studied from two perspectives.  One can investigate correlations for an entire galaxy population, or measure variations in individual galaxies.  The first approach is useful when the variations in individual galaxies are noisy.  Combining large samples allows one to identify trends which would be obscure for individual objects.  Such population trends do not guarantee that galaxies individually follow the average trend, but they can place constraints on the range of variation for individual galaxies.  Averaged trends also do not allow one to easily select subsamples with specific properties. The second approach is possible when one has sufficiently reliable measurements. In this case it is  preferable to study the distributions of those measurements directly.  In this way, one can examine how consistent the trends are across the galaxy population and isolate subsamples with different behaviour.

We first consider trends in \sersic index and effective radius versus wavelength, averaged across various subsets of the galaxy population.  We then consider the distributions of these trends for individual galaxies. These individual measurements allow us to study correlations of the trends with other quantities, and consider their use in classifying galaxies. We show that the population trends, like those presented by \citet{barbera10b} and \citet{kelvin12}, do successfully capture the typical wavelength variation of \sersic index and effective radius in individual galaxies.  Finally, we investigate the correlations between wavelength variations in \sersic index and effective radius, finding highly contrasting behaviour for early- and late- type galaxies.

This section will focus on presenting the observed variation of \sersic index and effective radius and their respective ratios at different wavelengths.  First we consider \sersic index and effective radius individually, in \S\ref{res:n} and \S\ref{res:re}, respectively. Then, in \S\ref{res:n_re} we examine correlations between their behaviour.  In the subsequent section, \S\ref{stacks} we will confirm the meaning of our results, and discuss their interpretation, with the aid of stacked images.

\subsection{Wavelength dependence of the \sersic index}
\label{res:n}
\subsubsection{Population trends}

\begin{figure}
\centering
\includegraphics[scale=0.7]{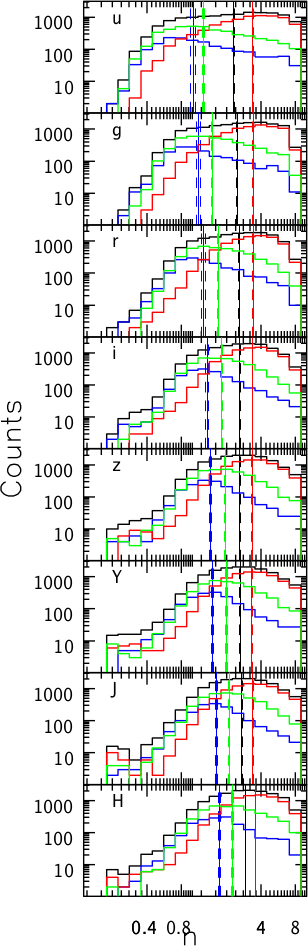}
\caption{\sersic index distribution for our full sample (black line) and for galaxies in each of our  colour subsamples (\red, \green, \blue). Each panel shows the distribution in a different band: $ugrizYJH$.
The median of each distribution, and its uncertainty, are indicated by vertical solid and dashed lines, respectively.  The median \sersic index for \red galaxies does not depend on wavelength, while for our \green and \blue subsamples the \sersic index increases significantly toward redder wavelengths.
\label{n_distr}}
\end{figure}

To begin, we examine the distribution of \sersic index, and its variation with wavelength, for a variety of samples.  We first consider the overall galaxy population divided using the colour cuts described in \S\ref{samples}.  These distributions are plotted for each passband in Fig.~\ref{n_distr}. The wavelength dependence of the \sersic index distribution for the full magnitude-limited sample of GAMA galaxies has previously been presented by \citet{kelvin12}, based on independent fits to each band.  \citeauthor{kelvin12} also divide into spheroid and disk dominated systems using a joint cut on $(u-r)$ and $n_K$.  Here we show in Fig.~\ref{n_distr}  a volume-limited sample with measurements obtained using our multi-band method, meaning that we have measurements in every band for the whole sample. Furthermore, unlike \citet{kelvin12}, we divide populations using colour alone, which results in a less clean division of galaxy type, but avoids the ambiguity of plotting distributions for samples that have been selected using the plotted quantity.  Despite these differences, the behaviour we see is very similar to that found by \citet{kelvin12}.

As it is well known, and also as seen in Fig.~\ref{col}, galaxies of different colour are characterised by very different \sersic index distributions: the median \sersic index across all wavelengths is $3.39\pm0.07$, $1.8\pm0.3$ and $1.4\pm0.2$ for \red, \green and \blue galaxies respectively. The quoted uncertainties on the median are estimated as $1.253 \sigma/\sqrt{N}$, where $\sigma$ is the standard deviation about the median and $N$ is the number of galaxies in the sample under consideration \citep{rider60}.

In addition to their different medians, Fig.~\ref{n_distr} shows that the \blue and \red samples are skewed in opposite directions. The \green sample follows the same distribution as the \blue sample at \lown, but has a higher proportion of \highn objects. Galaxies of different colour thus dominate at different \sersic indices. Colour is therefore a reasonably good predictor of galaxy structure, and vice versa.  However, remember that often we wish to study colour (e.g., as an indicator of star-formation history) versus galaxy structure, and so frequently need to identify galaxies with contrasting structure without reference to their colour.  Also, note that there is a significant region of overlap at intermediate \sersic index, where the majority of galaxies are found.

The overall distribution becomes more strongly peaked and moves to higher values of $n$ with increasing wavelength. A similar trend is observed for each of the \red, \green and \blue subsamples. However, whereas the whole distribution shifts for \blue and \green galaxies, the only change for \red galaxies is a slight decline in the proportion of \lown galaxies.  The median $n$ for red galaxies is therefore almost perfectly constant with wavelength, while the medians for the bluer samples vary significantly.  Many blue galaxies must significantly change appearance from the $u$- to the $H$-band.
\begin{figure}
\centering
\includegraphics[scale=0.3,angle=-90]{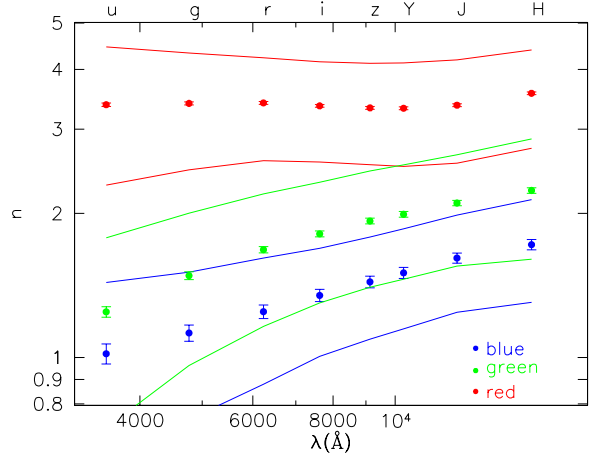}
\caption{Median \sersic index as a function of wavelength for our \red, \green and \blue galaxy subsamples. Error bars represent the uncertainty on the median. Solid lines indicate the 16th- and 84th-percentiles of the distribution. The \red subsample clearly behaves differently to our \green and \blue subsamples: the \sersic index of red galaxies is very stable across all optical--NIR wavelengths, while  \sersic index shows a steady increase with wavelength for bluer galaxies.
 \label{n_med}}
\end{figure}

Having examined the full distributions in Fig.~\ref{n_distr}, we present the variation in median \sersic index as a function of wavelength more compactly in Fig.~\ref{n_med}. The lack of variation in \sersic index with wavelength for \red galaxies indicates that they
 principally comprise one-component objects, i.e.\ elliptical galaxies, or two-component galaxies in which the components possess very similar colours, i.e.\ lenticulars.  In contrast, the systematically lower values and more substantial trends exhibited by \green and \blue galaxies are consistent with them being two-component systems, comprising both a blue, low-$n$ disk and red, high-$n$ bulge, with the disk being more dominant for bluer galaxies. 
Of course, various other types of galaxies may be present in the galaxy population, e.g. spirals with blue, low-$n$ bulges; but the simple picture above might be enough to explain much of the behaviour displayed by the samples we consider.

However, at least some of the variation in \sersic index with wavelength could also be attributed to dust attenuation.  The low \sersic index and blue colours of these systems may indicate that they contain star-forming disks. An increase in \sersic index with wavelength is in qualitative agreement with predictions of the effect of dust from radiative transfer models of galaxy disks \citep{Pastrav13}.

Finally, the observed trends may include variations in the disk stellar population (i.e.\ age and metallicity) with radius (as found by \citealt{dejong96b, beckman96, pompei97, waller03, mac04}).  However, it is not clear whether such gradients would manifest themselves as variations in \sersic index with wavelength.

We remind the reader that in the above analysis we have not applied any cut in $n$: a cut in the colour-magnitude plane results in a natural separation of galaxies with small and large \sersic index.  However, the correlation is not perfect, as is clear from Fig.~\ref{col}.  As previously mentioned, \citet{kelvin12} address this by defining a joint cut in $(u-r)$ versus $n_K$ to divide all galaxies into one of two classes.  We build on their analysis by taking a complementary approach.  Objects with combinations of colour and \sersic index contrary to the majority, e.g., red spirals and blue ellipticals, potentially correspond to interesting stages of galaxy evolution.  Studying these may illuminate the physical processes responsible for the structural and star-formation histories of galaxies.  We therefore proceed by considering subsamples defined by independent cuts in both colour and \sersic index, as described in \S\ref{samples}.
\begin{figure}
\centering
\includegraphics[scale=0.3,angle=-90]{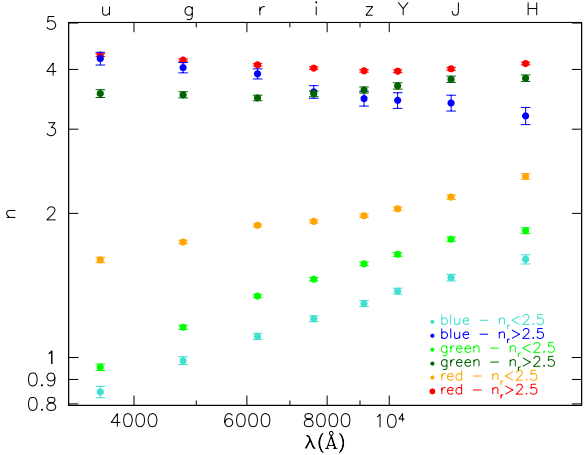}
\caption{Median \sersic index as a function of wavelength for galaxies of different colour and \sersic index. Red, dark green, and blue symbols represent \highn galaxies; orange, light green, and turquoise symbols \lown
galaxies. Error bars represent the uncertainty on the median. Differences between \lown and \highn galaxies are striking: the \sersic index for \highn objects only slightly depends on wavelength and it is systematically higher than the \sersic index of \lown galaxies, which strongly depends on wavelength.  \label{n_med_n}}
\end{figure}

Figure~\ref{n_med_n} shows the dependence of median \sersic index on wavelength for these six different populations. The different behaviour of galaxies with low and high \sersic index is striking.  \Highn galaxies consistently display much less \sersic index variation with wavelength than \lown galaxies.  \Red, \highn galaxies show no significant dependence on wavelength (in agreement with the results of \citealt{barbera10b} for early-type galaxies). The median value across all wavelengths is $4.10\pm0.09$.  Therefore, on average, they possess a classic de Vaucouleur profile at all wavelengths. \Green and \blue galaxies with \highn show a mild dependence on wavelength.  Both have a median \sersic index of $\sim 3.5$, and are equal in the $i$-band.  However, the median $n$ for \blue (high-$n$) galaxies decreases toward redder wavelengths (the Pearson product-moment correlation coefficient is $r = -0.98$, indicating a high significance), while \green (high-$n$)  galaxies show the opposite trend, increasing toward redder wavelengths ($r=0.8$, significant at the $99$\% level). This interesting contrast, which is also seen for individual galaxies in the next section, is our main motivation for considering \green and \blue galaxies as separate samples.

For \lown galaxies, the median \sersic index shows a substantial trend with wavelength, increasing from the $u$- to $H$-band. The median \sersic index and the strength of its variation with wavelength depend on the colour of the population considered. The median value systematically increases from \blue to \green to \red galaxies, while the strength of the trend with wavelength is similar for \blue and \green samples, but shallower for \red galaxies.  Comparing to Figure~\ref{n_med}, it is apparent that adopting a cut only in colour mixes populations and thus loses useful information.  \Blue and \green galaxies with \highn actually show similar trends to the overall \red population, although they number too few to influence the trends when no cut in \sersic index is applied.  Similarly, the trend for \red galaxies with \lown resembles those of \blue and \green galaxies.  \Lown galaxies make up a significant proportion of the red population, and thus they substantially reduce the median value of red galaxies when \sersic index is not taken into account.  The degree of variation of \sersic index with wavelength is more closely related to the overal value of \sersic index than to colour.

The adopted \sersic index cut at $n_r = 2.5$ does not appear to significantly alter the trends we recover for the \lown and \highn populations.  The $n$ distributions for each colour sample are sufficiently distinct and narrow that only a small fraction of each distribution is affected by the \sersic index cut.  This is apparent from the observations that the standard deviations of the \red and \green distributions only just cross in Fig.~\ref{n_med}, and the average of the \red, \lown sample in Fig.~\ref{n_med_n} extends to almost $n = 2.5$ in the $H$-band, without any sign that the trend is being suppressed.
Given that our subsample definitions are based on
\sersic indices in only one band ($r$), it is also reassuring to note that the spheroidal population retains high \sersic index values across all wavelengths. This indicates that performing the \sersic index separation in different bands would not substantially affect the samples, nor the above results.  However, Fig.~\ref{n_med_n} does suggest that separating in bluer bands, where the populations are better separated, is advisable.

\subsubsection{\sersic index versus wavelength for individual galaxies}
\begin{figure*}
\centering
\includegraphics[scale=0.39,clip=true]{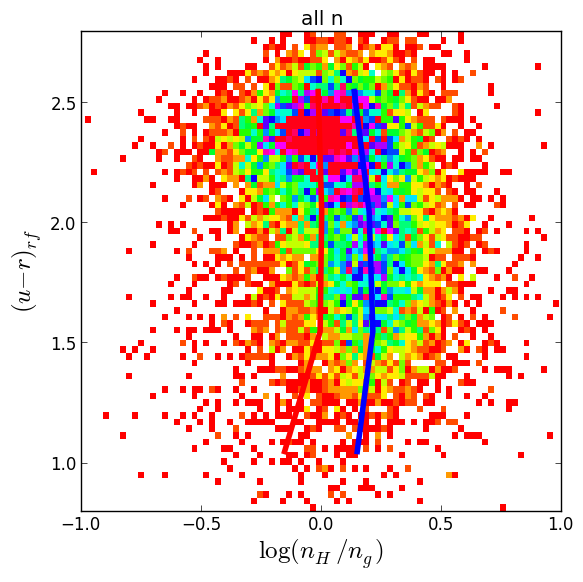}
\includegraphics[scale=0.39,clip=true]{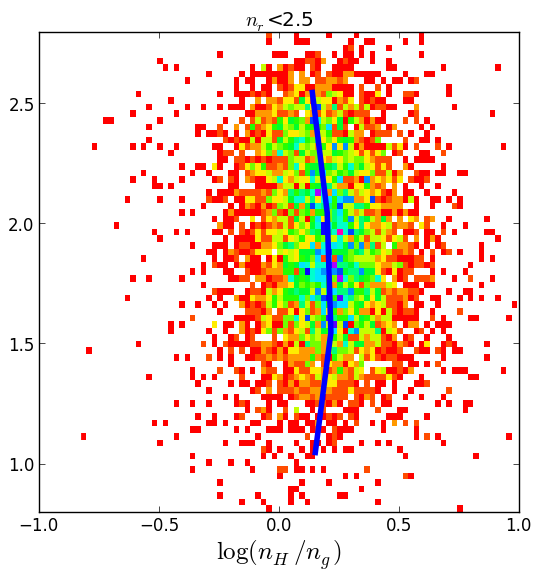}
\includegraphics[scale=0.39,clip=true]{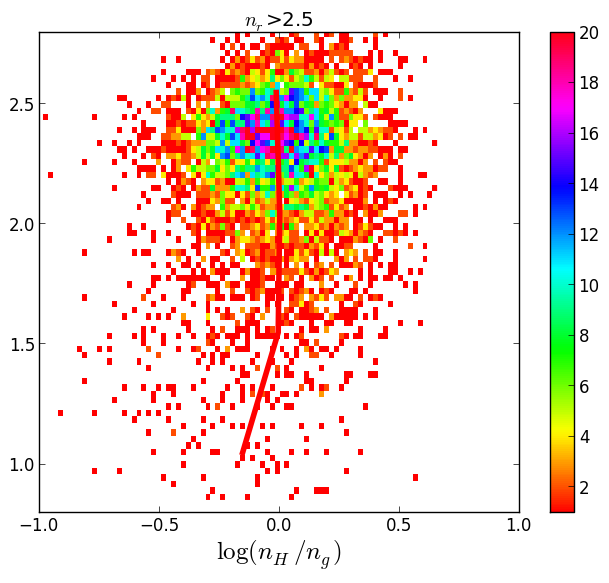}
\caption{Density map of $(u-r)$ colour  versus the ratio of the $H$-band \sersic index and $g$-band \sersic index, defined here as \N[H][g]=$n_H/n_g$. The left panel displays all galaxies together, while centre and right panels show only galaxies with low and high \sersic indices, respectively. The median \N[H][g] as a function of colour is also plotted, for both the \lown (blue line) and \highn (red line) samples. These two populations show different distributions, indicating that their constituent galaxies possess different internal structures.
\Highn galaxies have \N$\sim 1$, indicating that the central concentration of their profile tends not to change with wavelength.  On the other hand, \lown galaxies generally display peakier profiles in the red, particularly those with intermediate colour.
 \label{ur_nn}}
\end{figure*}

The analysis we have presented so far considered average trends across a sample of galaxies.  If all galaxies in a given sample behave in the same way, then the trends in the average do reflect the trends in the individual galaxies.  However, there is the possibility that individual galaxies have behaviours which cannot be easily identified from the distributions of \sersic index in different bands.
For example, if the average $n$ for a sample does not change between two bands, it could be that all the constituent galaxies have constant $n$, but it is also plausible that half the galaxies increase their $n$ while the other half decrease, or that  most remain constant while a minority display a significant trend, etc.  We have already seen such behaviour when dividing by colour: different trends have been revealed for particular subsamples.  In this case we were able to discover this by dividing by \sersic index, but ideally we would like to look for such trends directly in the full sample.

In this section we aim to explore the wavelength dependence of \sersic index for individual galaxies.  This is difficult when fitting galaxies independently in each band.  However, as shown by H13, our multi-band method dramatically reduces the noise on such measurements and avoids cases where galaxies are successfully fit in some bands, but not in others.\footnote{We have confirmed this by making versions of the figures in this paper based on single-band fits.  These all contain dramatically fewer objects which pass our quality control cuts, and the distributions show much larger scatter.}

We thus have measurements of \sersic index in every band for all objects in our sample.  With this information it is possible to devise various ways of quantifying the wavelength dependence of \sersic index.  In this paper we take the conceptually and computationally simple approach of comparing values at particular pairs of wavelengths, by taking their ratio.
We will investigate what these ratios can tell us about the correlations between galaxy stellar populations and spatial structure.

For compactness we adopt the notation \N[r][b]= $n(r)/n(b)$, where $n(r)$ and $n(b)$ are the \sersic indices in some (rest-frame) red and blue band, respectively (we thus generally maintain the same convention as colour).  
Using a ratio, rather than a difference, is appropriate, as the physical distinction between profiles separated by constant $\Delta n$ decreases for larger $n$; i.e.\ there is more `difference' between $n=1$ and $n=1.5$ profiles, than between $n=4$ and  $n=4.5$ profiles.  This is the same reason why $n$ is usually plotted with a logarithmic scale.  However, the use of a ratio  results in asymmetric behaviour when plotted linearly: low ratios are confined to between $0$ and $1$, while high ratios may extend to infinity.  To avoid this we usually plot \N on a logarithmic scale.  These choices result in distributions that look much more symmetrical and Gaussian than otherwise, justifying our approach.

We begin by inspecting how galaxy colour varies with \N, and specifically focus on
the dependence of $(u-r)$ on \N[H][g] $= n_H/n_g$. This combination of bands ($H$ and $g$) is favoured here, and throughout the paper, as it is the pair with the longest robust wavelength baseline ($u$ is noisy and $K$ is extrapolated).

The left panel of Fig.~\ref{ur_nn} shows that the colour bimodality is accompanied by a shift of the blue population to higher values of \N[H][g] than the red population.  The remaining two panels divide the sample by \sersic index, and demonstrate that \lown and \highn galaxies display different distributions in this diagram. For \lown galaxies, which span a wide range of $(u-r)$, the median \N[H][g] at any colour is greater than one.
This agrees with the \lown sample being disk galaxies, comprising a blue $n=1$ disk and red $n>1$ bulge, such that the disk is more prominent in $g$ than in $H$.  Although subtle, a curve is apparent in the median \N[H][g] versus $(u-r)$.  \Lown galaxies with intermediate colours have the greatest variations of \sersic index with wavelength, consistent with the bulge and disk being of similar prominence, but differing colour, in these systems.

In contrast, the \highn galaxies, which are mostly red, display a distribution that is centred on \N[H][g]$\sim 1$, indicating that the profile of these galaxies is similar at all wavelengths.  There is a tail of \highn galaxies to blue colours, which tend to have \N[H][g]$< 1$.  This suggests that \blue, \highn galaxies have peakier profiles at blue wavelengths, possibly indicating the presence of a central starburst.

The segregation of galaxy populations seen in Fig.~\ref{ur_nn}
suggests that \N could be used as a proxy to separate galaxy types or identify specific populations.  For example, `red spirals', disk galaxies with suppressed star-formation, should possess a disk with similar colour to their bulge.  They may therefore be isolated from early-type spirals, with comparable colour and \sersic index, by their lower \N.   

However, we reiterate that our definitions of \blue versus \red and \lown versus \highn depend on our sample selection and adopted magnitude cut.  Our present results are therefore limited to the intrinsically bright galaxy population.

\begin{figure*}
\centering
\includegraphics[scale=0.35]{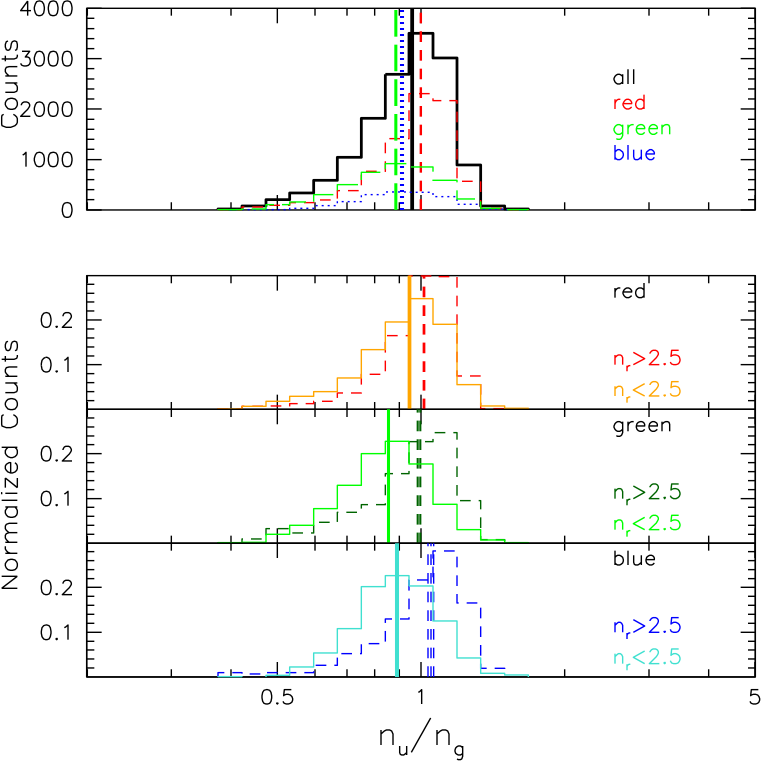}
\hspace{1cm}
\includegraphics[scale=0.35]{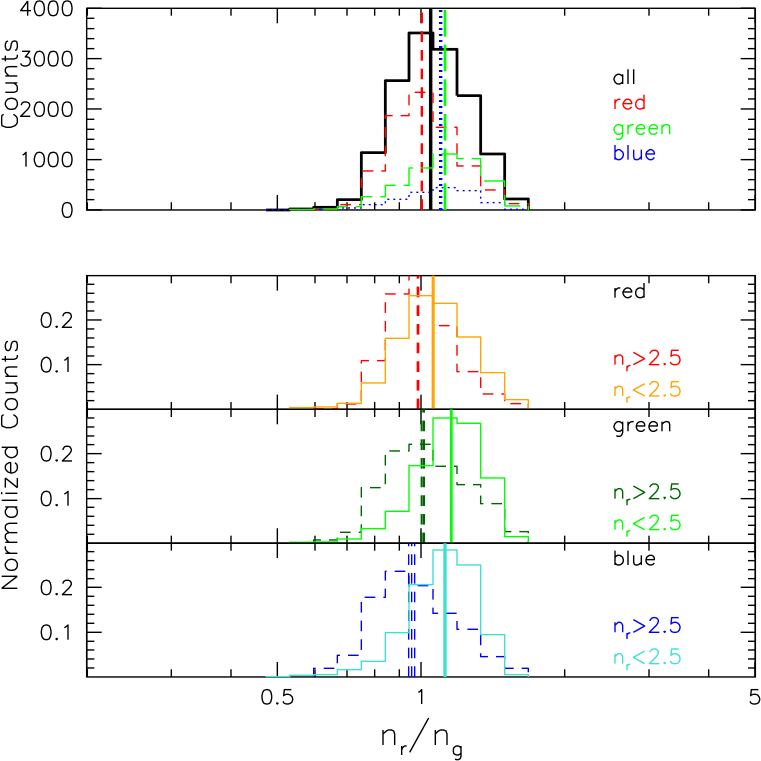}\\
\includegraphics[scale=0.35]{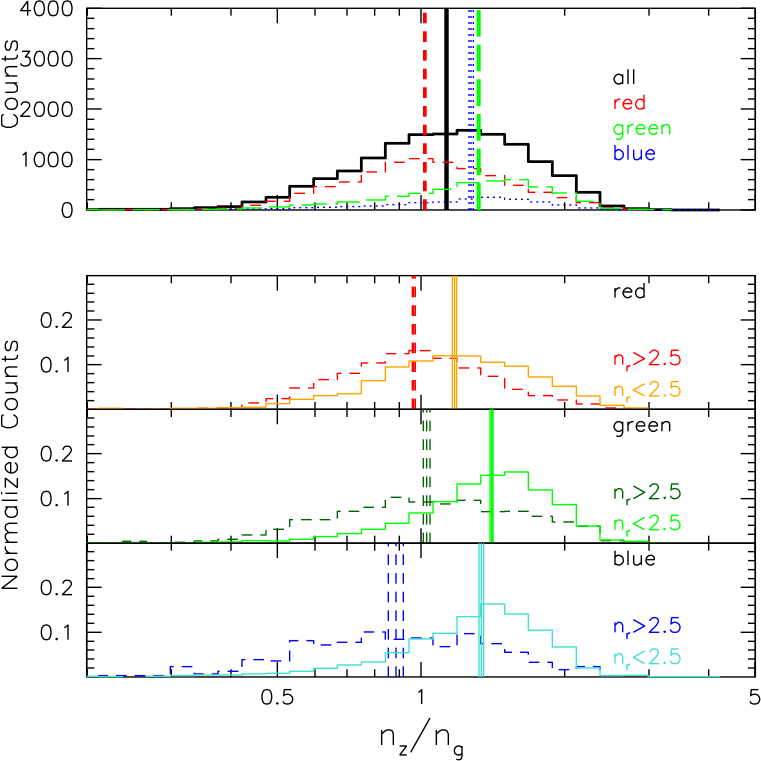}
\hspace{1cm}
\includegraphics[scale=0.35]{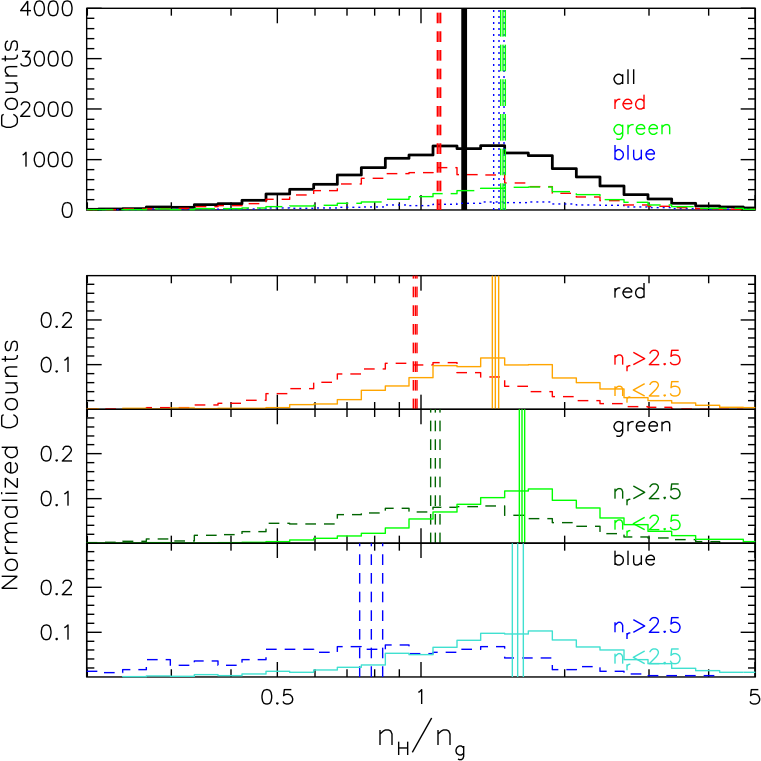}
\caption{Distributions of \N for a representative selection of bands.  The four groups of panels show (\textit{1}) \N[u][g], (\textit{2}) \N[r][g], (\textit{3}) \N[z][g] and (\textit{4}) \N[H][g]. Within each group, the upper panel (\textit{a}) shows the distribution of \N for the sample overall and divided by colour, without taking into account the \sersic index. The lower windows show the normalised histograms of \N for \lown and \highn galaxies, for \red (\textit{b}), \green (\textit{c}) and \blue (\textit{d}) samples separately.
\label{gra}}
\end{figure*}

So far we have focused on \N[H][g], but now we consider other combinations of bands, keeping $g$ as our reference, to be sure that our conclusions are not driven by a particular choice of the bands.
Figure~\ref{gra} shows the distributions of \N[u][g], \N[r][g], \N[z][g], \N[H][g]. We chose to show $u$, $z$ and $H$ as examples because they are
the bluest, an intermediate and the reddest bands, respectively, and $r$ because it is the band we use to define our \lown and \highn samples.  
As well as showing the overall distributions we also divide by colour and \sersic index, as described in \S\ref{samples}.

The distributions are unimodal, with widths that strongly depend on the passbands involved: the redder the wavelength (relative to $g$), the broader the distributions.  This is somewhat inevitable given that observations at neighbouring wavelengths will be dominated by the same stellar population, while longer wavelength baselines permit greater variability. In addition, the \sersic index has been constrained to vary as a smooth polynomial, which will act to suppress large variations between neighbouring bands.
Focusing only on colours (upper subpanels), \N for \red galaxies always remains close to one, as does \N for \blue and \green galaxies for short wavelength baselines.  Increasing the wavelength range spanned ($g$ versus $z$- and $H$-bands), \blue and \green galaxies have systematically higher \N than \red galaxies, and are thus easier to differentiate, despite the increasing width of the the \N distributions.

Further separating galaxies for $n$ (lower subpanels), we see that the \lown and \highn normalised distributions differentiate, at any fixed colour. 
As for colour, the difference between \lown and \highn is more evident with longer wavelength baselines. For \red galaxies the width of the \N distributions are very similar for \lown and \highn galaxies, at any wavelength. However, for \green and \blue objects the \highn distribution is significantly broader than for \lown, particularly for long wavelength baselines.  Objects with high \sersic index and blue colours appear to be a varied population, although the bluest examples are rare, and hence their distribution is uncertain.

We note that at any wavelength, \green\ \highn galaxy trends resemble those of \red\ \highn galaxies, while \green\ \lown galaxy trends resemble those of \blue\ \lown galaxies, further indicating that our \green sample mixes objects with different structural properties.

To assess the differences between the different distributions, we
perform a Kolmogorov-Smirnov (K-S) test on each pair.
The K-S tests indicate that all the subsamples are drawn from different parent \N distributions, except for \blue and \green galaxies with \lown, suggesting that these galaxies share common properties.

\begin{figure}
\centering
\includegraphics[scale=0.3, angle=-90]{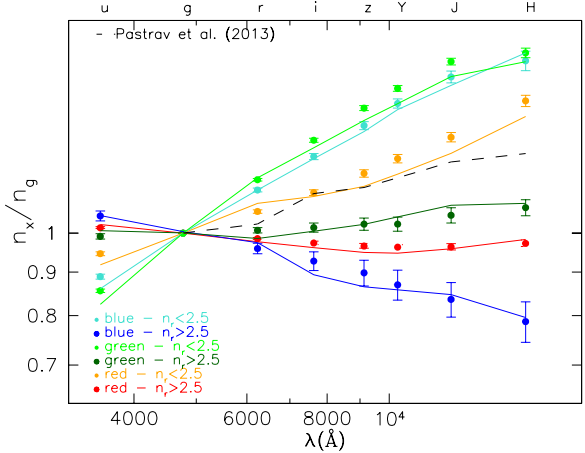}
\caption{Median values of \N[x][g] as a function of wavelength, where $x$ denotes the band corresponding to wavelength $\lambda$. As indicated by the legend, red, dark green, and blue symbols and lines represent \highn galaxies; orange, light green, and turquoise symbols and lines represent \lown galaxies. 
Median \N[x][g]$= n_x/n_g$ are plotted by points.  Error bars give the uncertainty on the median. Lines plot ${\rm median}[n_x]/{\rm median}[n_g]$, i.e.\ the points from Fig.~\ref{n_med_n} normalised to the $g$-band. The black dashed line represent the wavelength dependence of \N for a disk population due to the effects of dust, as predicted by \citet{Pastrav13}. We do not plot the prediction for the $u$-band since it has not been calculated in the model, but it is a linear interpolation from the $B$-band to the near-UV. The trends depend on both colour and \sersic index, indicating that the typical internal structure differs for galaxies in each population. 
\label{nn_med}}
\end{figure}

As was done for \sersic index in Fig.~\ref{n_med_n}, we show the variation in median \N as a function of wavelength, $\lambda$, more compactly in Fig.~\ref{nn_med}. All the ratios are normalised to the $g$-band.
Points represent the median of the \N[x][g] values for the galaxies in each subsample, where $x$ is the band corresponding to wavelength $\lambda$.  For comparison, the lines in this figure show the ratio of the median $n(g)$ to the median $n(x)$ (as individually plotted in Fig.~ \ref{n_med_n}).  One can see that, as expected statistically, the median \N reveals the same behaviour as considering the ratio of the median trends in $n$.
Fig.~\ref{nn_med} is therefore an alternative way of presenting Fig.~\ref{n_med_n}.  The unimodality of the distributions in Fig.~\ref{gra}, and the consistency shown in Fig.~ \ref{n_med_n} support the use of population trends to study the variation of galaxy structure with wavelength.
However, remember that the \N quantifies the dependence of \sersic index on wavelength for \emph{individual} galaxies in a robust manner. 

Galaxies with different \sersic indices and colours follow different trends.
For \green and \blue galaxies with \lown, \N correlates strongly with wavelength, such that \sersic index is higher at redder wavelengths.
\Red, \lown galaxies display a weaker correlation with wavelength, while for \green, \highn objects it is weaker still, and entirely absent for \red, \highn galaxies.  Finally, for \highn, \blue galaxies \N anti-correlates with wavelength.  This behaviour is consistent with a progression from galaxies containing significant red bulge and blue disk components, to bulge-disk systems that are increasingly dominated by a red bulge, or contain components with less contrasting colours, to galaxies possessing a single component, or homogeneous colour, and finally to systems with a bulge bluer than their disk.

As mentioned earlier, the relative prominence and colour of bulges and disks are unlikely to completely explain the trends we see.  Even in pure exponential disk systems, dust attenuation can result in measured \sersic indices that vary with wavelength.
In Fig.~\ref{nn_med} we overplot the wavelength dependence of \N expected for a disk population due to the effects of dust, as computed by \citet{Pastrav13} from the radiative transfer model simulations of \cite{popescu11}. Their model is for the disk in a typical spiral galaxy, with inclination $i=60$ degrees and central B-band face-on optical depth $\tau^{\rm f}_{B} = 4$ (see their Figure 25). Note that since \citet{Pastrav13} did not calculate the model for the  $u$-band, but simply linearly interpolated from the $B$-band to the near-UV, we do not consider the value for the $u$-band. Their prediction lies below the majority of our disk-dominated (\lown) galaxies, but is consistent with the trend for our \red, \lown sample.
This suggests that dust may be responsible for part, but not all, of the trends for \lown galaxies.

\begin{figure}
\centering
\includegraphics[scale=0.3,clip=true,trim=25 15 15 25,angle=-90]{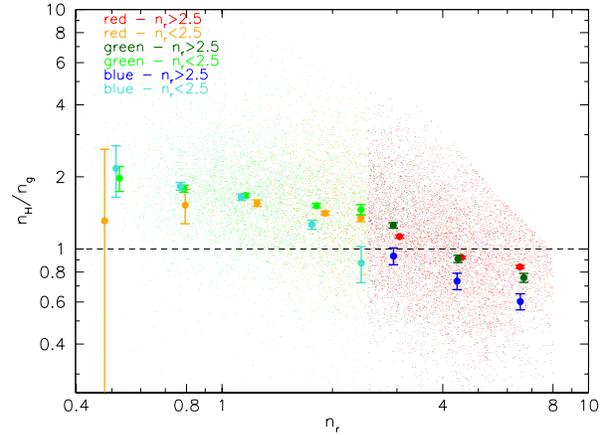}
\caption{Median \N[H][g] as a function of $n_r$ for galaxies with each colour and \sersic index subsample, coloured as indicated in the legend.  Small points show individual galaxies, while larger points indicate the median in bins of $n_r$.  Error bars represent the uncertainty on the median. The dashed line shows the \N[H][g]=1.
\label{nnn}}
\end{figure}

In the above analysis we have seen that galaxies with high \sersic indices typically have smaller variations in \sersic index with wavelength.  We demonstrate this explicitly in Fig.~\ref{nnn} by plotting \N[H][g] versus $n_r$.  This plot shows that galaxies with $n_r \sim 4$ have \N[H][g]$\sim 1$ on average, though with a significant scatter. While departures to lower $n_r$ are consistently associated with a trend to higher \N[H][g], the correlation also continues to $n_r > 4$ and lower \N[H][g].
This would appear to indicate that \emph{all} truly homogeneous, one-component, systems have $n \sim 4$ at all optical and NIR wavelengths.  However, our sample is limited to intrinsically luminous galaxies, and such a conclusion would be at odds with the established magnitude-\sersic index relation for elliptical galaxies (e.g., \citealt{GrahamGuzman03}, and references therein).  It will be interesting to explore this result over a wider luminosity range.
Finally, we note again that blue \highn objects tend to have \N[H][g]$< 1$, and hence become more concentrated at bluer wavelengths, indicative of central star-formation. 

Many of the results in this subsection echo those we found earlier by considering the population trends, which are less reliant on multi-band fits.  However, by reliably measuring the wavelength dependence of \sersic index, \N, for individual galaxies we are able to study the distribution of \N, rather than simple averages, and even select objects with contrasting \N values.

Next we consider the wavelength dependence of the effective radius in a similar manner to above, before considering the joint behaviour of $n$ and $\re$.

\subsection{Wavelength dependence of the effective radius}
\label{res:re}
\subsubsection{Population trends}

The effective, or half-light, radius is simply a measure of the size, and thus not normally considered to represent galaxy structure directly.
However, the amount by which the size of a galaxy varies can be due to its inclination, and hence it depends on the opacity, or to its internal structure.
In particular, depending upon the wavelength considered, it reveals radial variations of colour, and hence stellar population or attenuation, within a galaxy.  These may be related to the presence of multiple structural components with different colours, e.g., a red bulge and blue disk, or gradients within a single structural element.

In this section, following the same approach as above for \sersic index, we 
investigate how $\re$ varies with wavelength for galaxies divided by colour and \sersic index, as described in \S~\ref{samples}.

\begin{figure}
\centering
\includegraphics[scale=0.7]{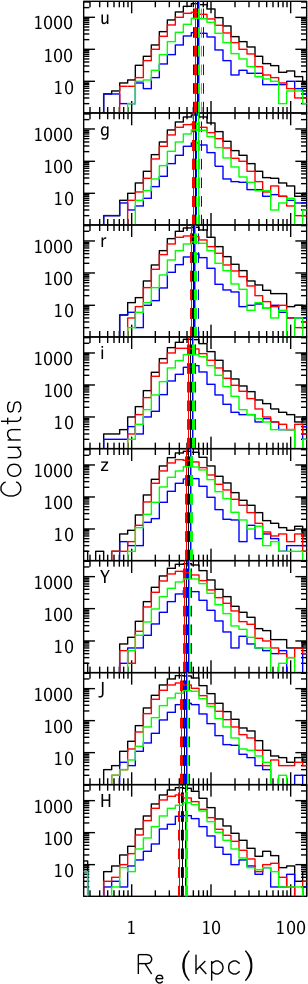}
\caption{Effective radius distribution for our full sample (black line) and for galaxies in each of our colour subsamples (\red, \green, \blue). Each panel shows the distribution in a different band: $ugrizYJH$.
The median of each distribution, and its uncertainty, are indicated by vertical solid and dashed lines, respectively.
The $\re$ distributions for each colour sample are very similar, and the medians vary only weakly with wavelength. 
\label{r_distr}}
\end{figure}

Figure~\ref{r_distr} shows the distribution of $\re$ in each waveband, for the full galaxy sample and for our three colour subsamples. In contrast to $n$ (Fig.~\ref{n_distr}) the distribution of $\re$ does not depend strongly on galaxy colour. The median and shape of
the $\re$ distribution are very similar for the three samples in all wavebands.

\begin{figure}
\centering
\includegraphics[scale=0.3,angle=-90]{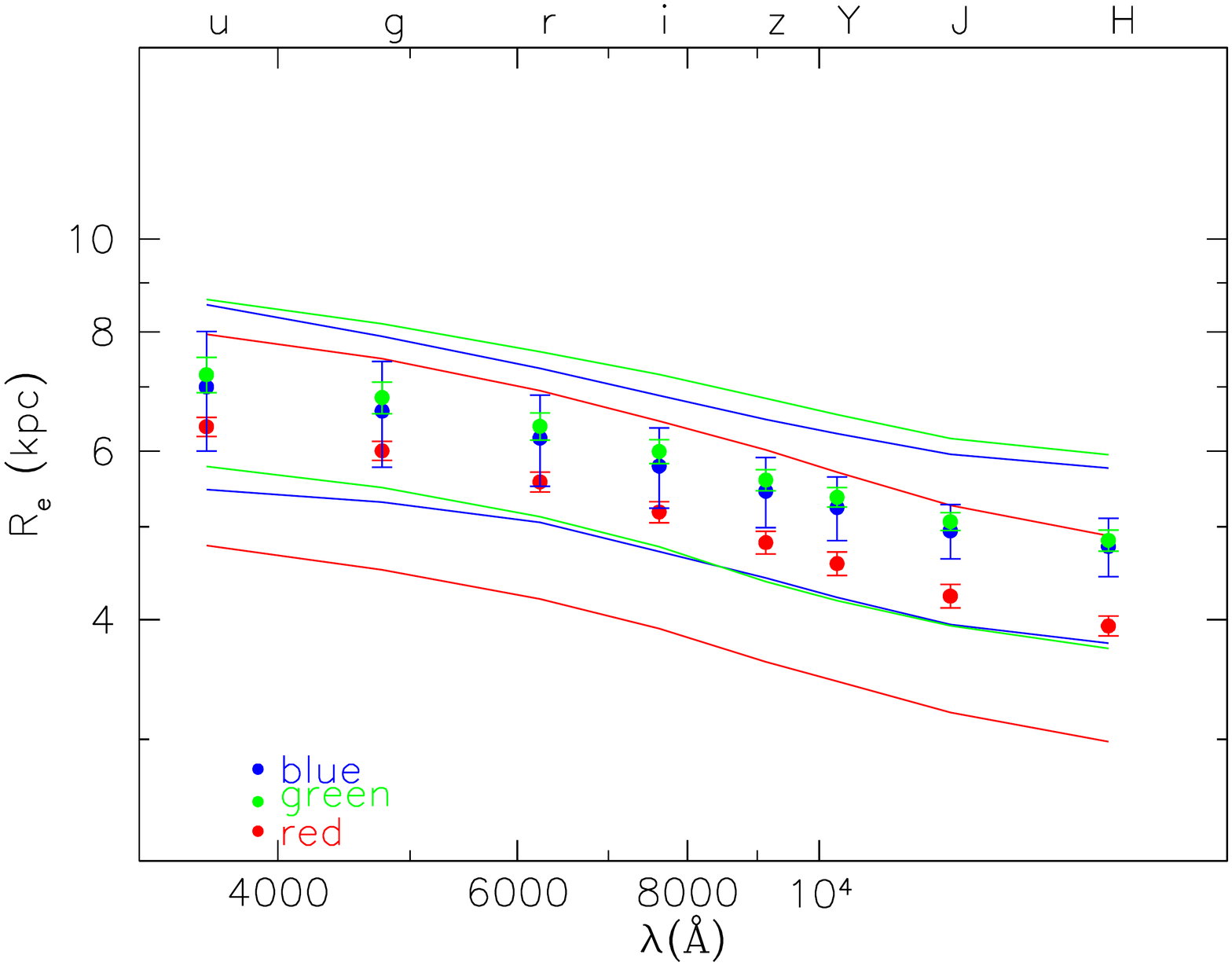}
\caption{Median effective radius as a function of wavelength for our \red, \green and \blue galaxy samples. Error bars represent the uncertainty on the median. Solid lines indicate the 16th- and 84th-percentiles of the distribution. The effective radius clearly depends on wavelength. \Blue and \green galaxies show very similar behaviour, while \red galaxies have systematically smaller effective radii and a slightly steeper dependence on wavelength. \label{r_med}}
\end{figure}

\begin{figure}
\centering
\includegraphics[scale=0.3,angle=-90]{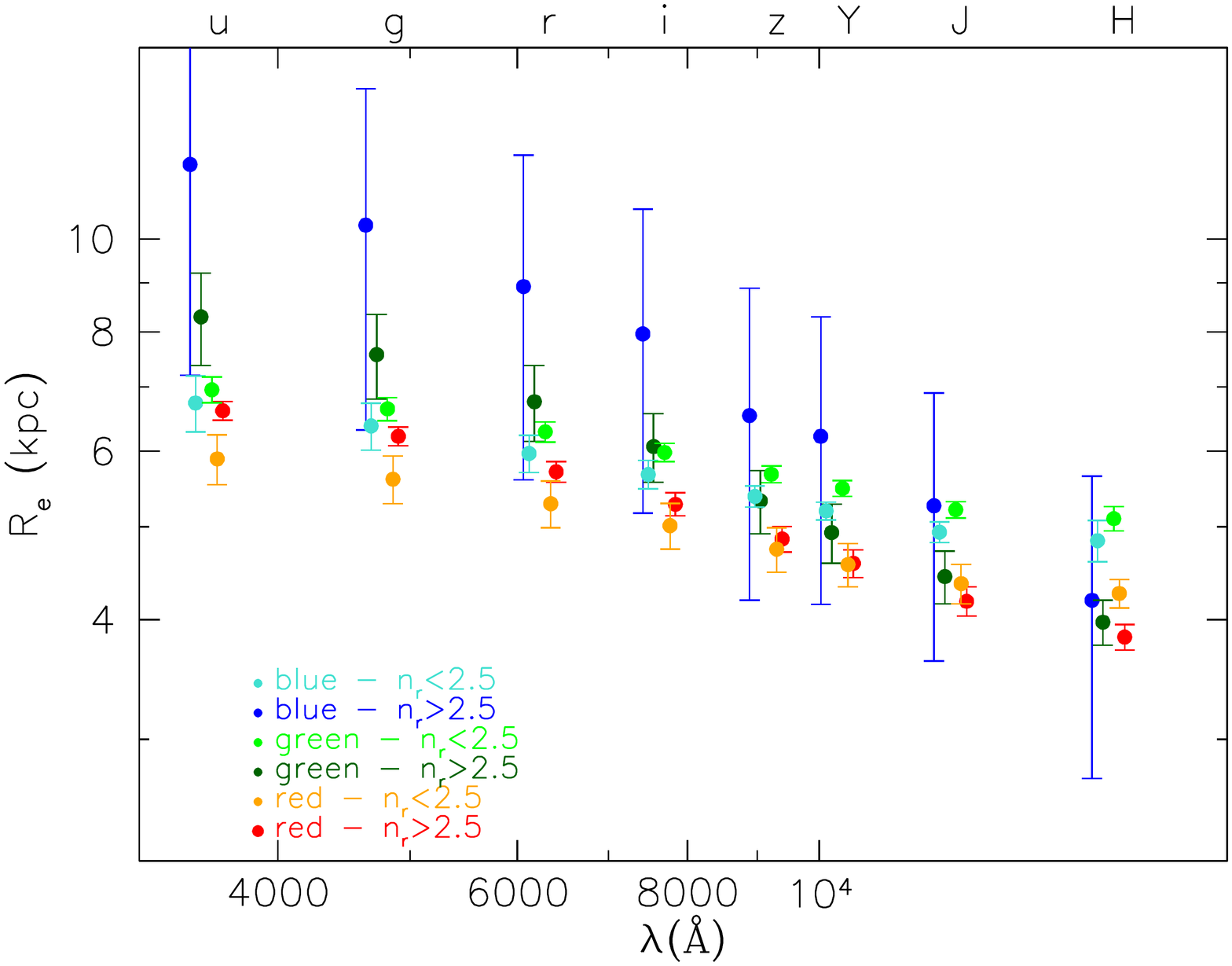}
\caption{Median effective radius as a function of wavelength for galaxies of different colour and \sersic index. Red, dark green, and blue symbols represent \highn galaxies; orange, light green, and turquoise symbols \lown
galaxies. Error bars represent the uncertainty on the median. The median effective radius depends on wavelength. Small offsets in wavelength have been applied to all points for clarity.
\Lown samples display similar behaviour, while the \highn samples are show stronger, and more varied, trends.
\label{r_med_n}}
\end{figure}

The median $\re$ decreases toward redder wavelengths. This behaviour is slightly more pronounced for the \red sample.
This is more apparent in Fig.~\ref{r_med}, where the dependence of the median effective radius on wavelength is presented more compactly. The slope of the relations is similar for galaxies of different colour, just slightly steeper for \red galaxies. 
\Red galaxies also have a smaller median $\re$ at all wavelengths, while \blue and \green galaxies show very similar values.  
The median $\re$ in $u$-band are $6.4 \pm 0.1$, $7.2 \pm 0.3$ and $7 \pm 1$ kpc for \red, \green and \blue subsamples, respectively. 
In $H$-band the corresponding figures are $3.94 \pm 0.09$, $4.9 \pm 0.1$ and $4.8 \pm 0.3$ kpc. 

In contrast to colour and \sersic index, in our sample there is no obvious separation of populations in effective radius.  We therefore do not attempt to consider samples divided by $\re$, but continue to use $n$ to isolate bulge- and disk-dominated populations.

Figure~\ref{r_med_n} shows the dependance of the median effective radius on wavelength for galaxies characterised by different colour and \sersic index. 
Each population shows a decrease in effective radius toward redder wavelengths.
Galaxies with \highn are characterised by a steeper trend (a $\sim 45$\% reduction in $\re$ from $u$ to $H$) than those with \lown ($\sim 25$\% reduction).
The three \lown samples behave very similarly; the variation of median $\re$ with wavelength does not depend on colour for \lown galaxies.  The \highn samples display more variety, with bluer galaxies showing a stronger wavelength dependence.

\subsubsection{Effective radius versus wavelength for individual galaxies}

\begin{figure*}
\centering
\includegraphics[scale=0.39,clip=true]{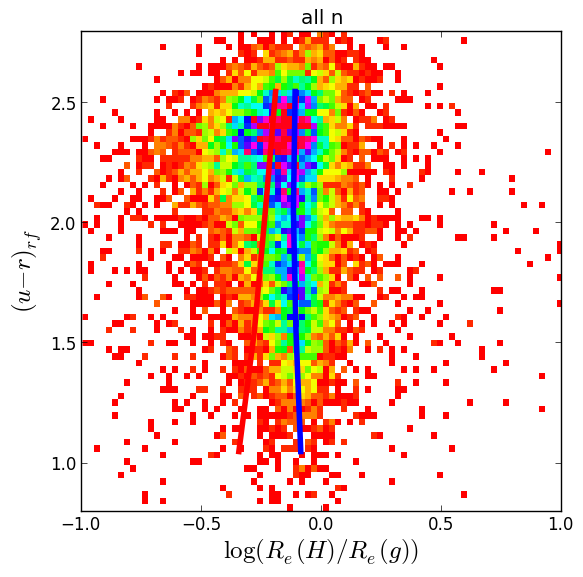}
\includegraphics[scale=0.39,clip=true]{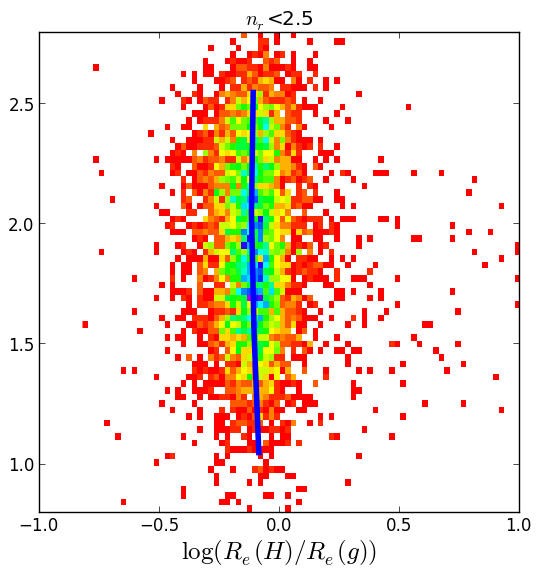}
\includegraphics[scale=0.39,clip=true]{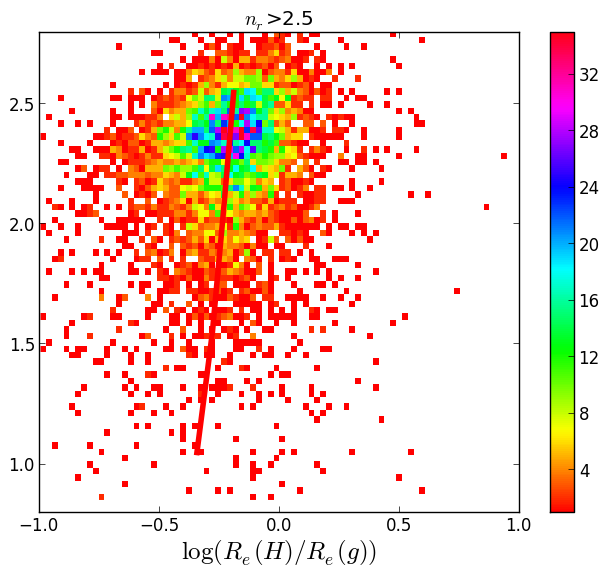}
\caption{Density map of $(u-r)$ colour  versus the ratio of the $H$-band effective radius and $g$-band effective radius, defined here as
 \R[H][g]$=\re(H)/\re(g)$. The left panel considers all galaxies together, while centre and right panels show only galaxies with low and high \sersic index respectively. The median \R[H][g] as a function of colour is also plotted, for both the \lown (blue line) and \highn (red line) samples. The two populations show different distributions, indicating that they are characterised by different internal structures.  Most galaxies have \R$< 1$, indicating that they appear smaller at redder wavelengths, with \highn galaxies showing the greatest effect.   
\label{ur_rr}}
\end{figure*}

Following the same reasoning and approach for \sersic index, we now move to quantifying the dependence of effective radius on wavelength for individual galaxies.
We adopt the similar notation \R[r][b]$= \re(r)/\re(b)$, where $\re(r)$ and $\re(b)$ are the effective radii in some (rest-frame) red and blue band, respectively (again maintaining the standard colour convention where sensible).
Using a ratio is appropriate, as we are more interested in fractional, rather than absolute, variations in $\re$.  As with \N, to avoid an asymmetry when plotting \R, we use a logarithmic scale.  The resulting distributions appear symmetrical and approximately Gaussian.

To start, we examine the relationship between galaxy colour $(u-r)$ and \R[H][g]$=\re(H)/\re(g)$ in Fig.~\ref{ur_rr}.  This displays similar behaviour to the distribution of $(u-r)$ versus \N[H][g] (Fig.~\ref{ur_nn}).  The distribution of \R[H][g] is different for galaxies on either side of the usual colour bimodality, and the two overlapping modes can be well separated by dividing by \sersic index. 
\Lown galaxies occupy a narrow vertical locus, indicating that they can assume a wide range of colours, but are limited to a narrow range around \R[H][g]$\sim 0.8$. In contrast, \highn galaxies, which have preferentially red colour, display a broader distribution offset to lower \R[H][g]$\sim 0.6$.  Bluer \highn galaxies are offset to even lower \R[H][g].
As with \N, this figure indicates that galaxies of different \sersic index have $\re$ with different wavelength dependence, suggesting that \R could be useful for classifying galaxies.

\begin{figure*}
\centering
\includegraphics[scale=0.35]{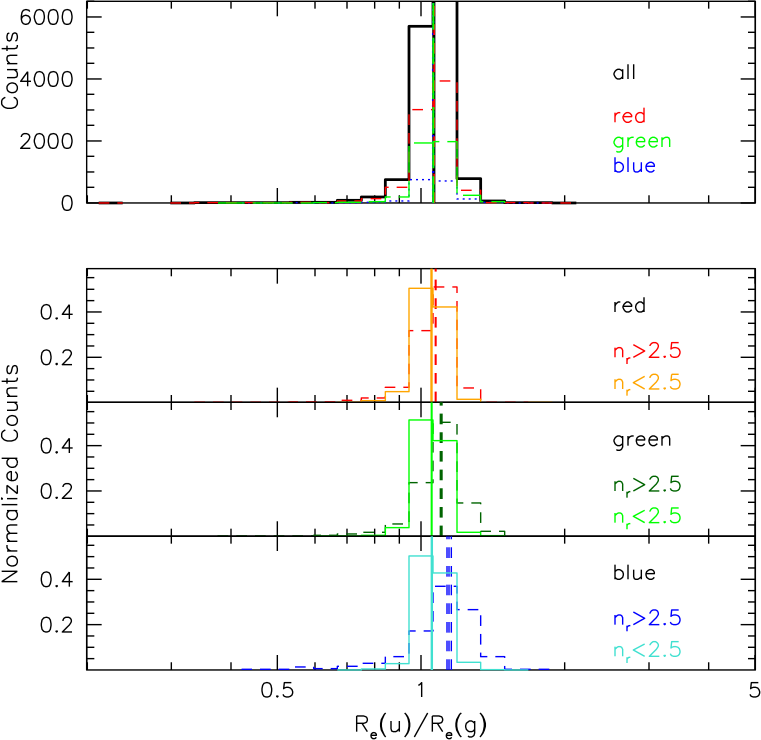}
\hspace{1cm}
\includegraphics[scale=0.35]{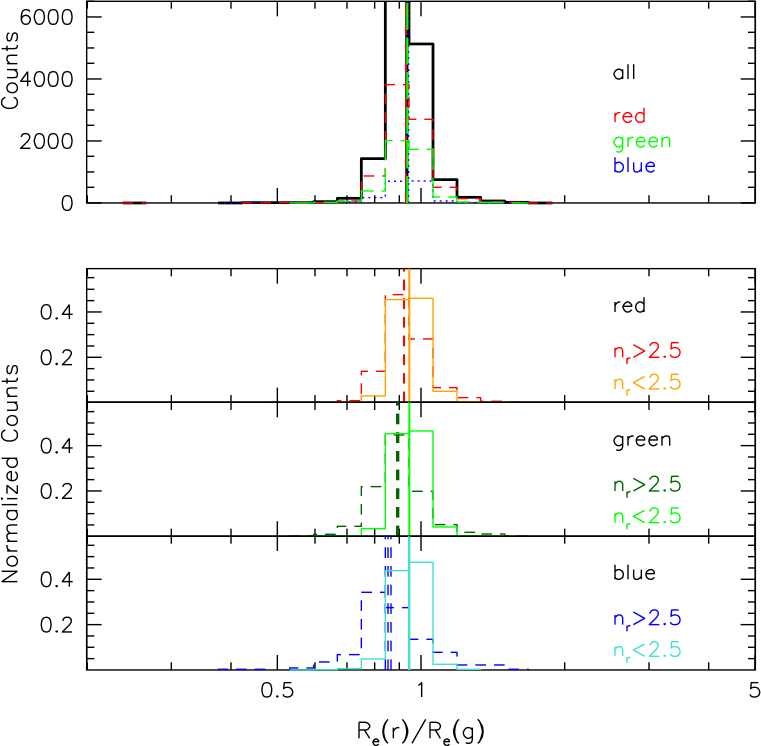}\\
\includegraphics[scale=0.35]{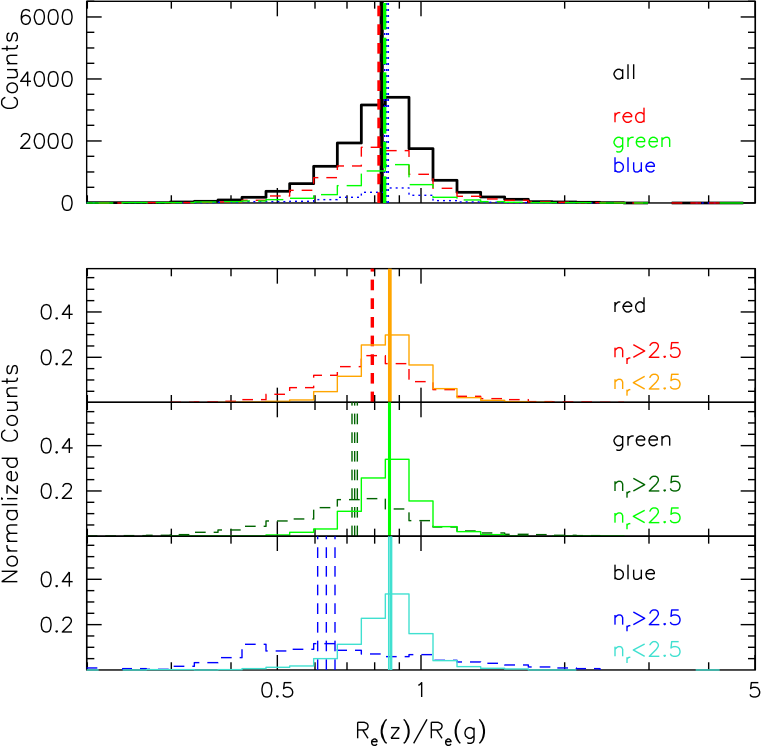}
\hspace{1cm}
\includegraphics[scale=0.35]{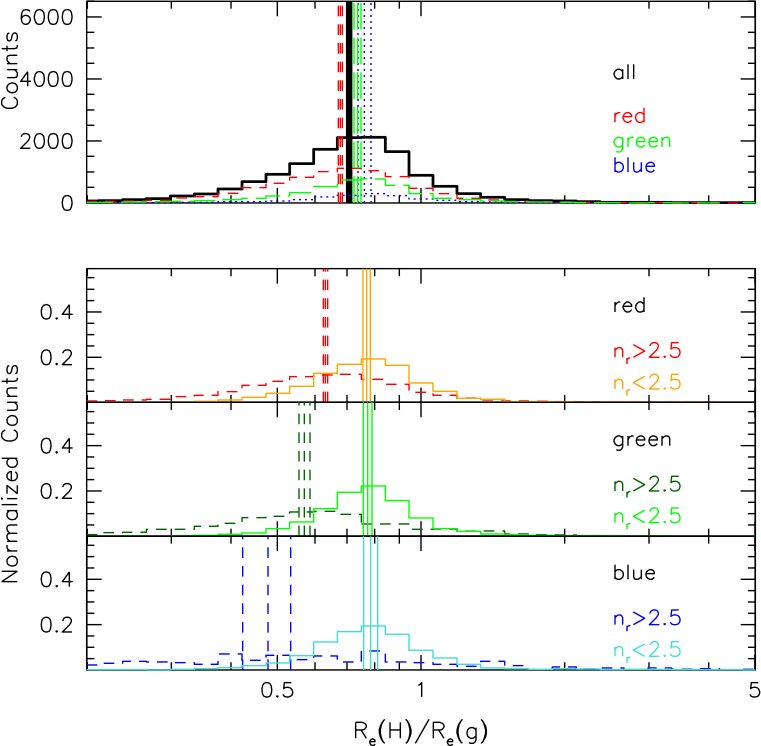}
\caption{
Distributions of \R for a representative selection of bands.  The four groups of panels show (\textit{1}) \R[u][g], (\textit{2}) \R[r][g], (\textit{3}) \R[z][g] and (\textit{4}) \R[H][g]. Within each group, the upper panel (\textit{a}) shows the distribution of \R for the sample overall and divided by colour, without taking into account the \sersic index. The lower windows show the normalised histograms of \R for \lown and \highn galaxies, for \red (\textit{b}), \green (\textit{c}) and \blue (\textit{d}) samples separately. 
\label{grar2}}
\end{figure*}

We now compare the \R distributions for various combinations of bands, keeping $g$ as our reference, for our full sample of galaxies and subsamples divided by colour and \sersic index, as described in \S\ref{samples}.
Figure~\ref{grar2} shows that galaxies of different colours show very similar \R distributions, for any wavelength baseline.  As for \N, the distributions are unimodal, with widths that depend on the passbands involved: the longer the wavelength baseline, the broader the distributions.
The medians of these distributions become lower for longer wavelength baselines. 
Focusing on galaxies with \lown and \highn, we find that \lown galaxies are characterised by an \R distribution which is completely independent of colour and varies only slightly with wavelength baseline. In contrast, the median \R for \highn galaxies depends on colour, becoming lower for bluer galaxies.  Similar to \N, the \R distribution for \highn galaxies is typically broader than for \lown galaxies.

\begin{figure}
\centering
\includegraphics[scale=0.3,angle=-90]{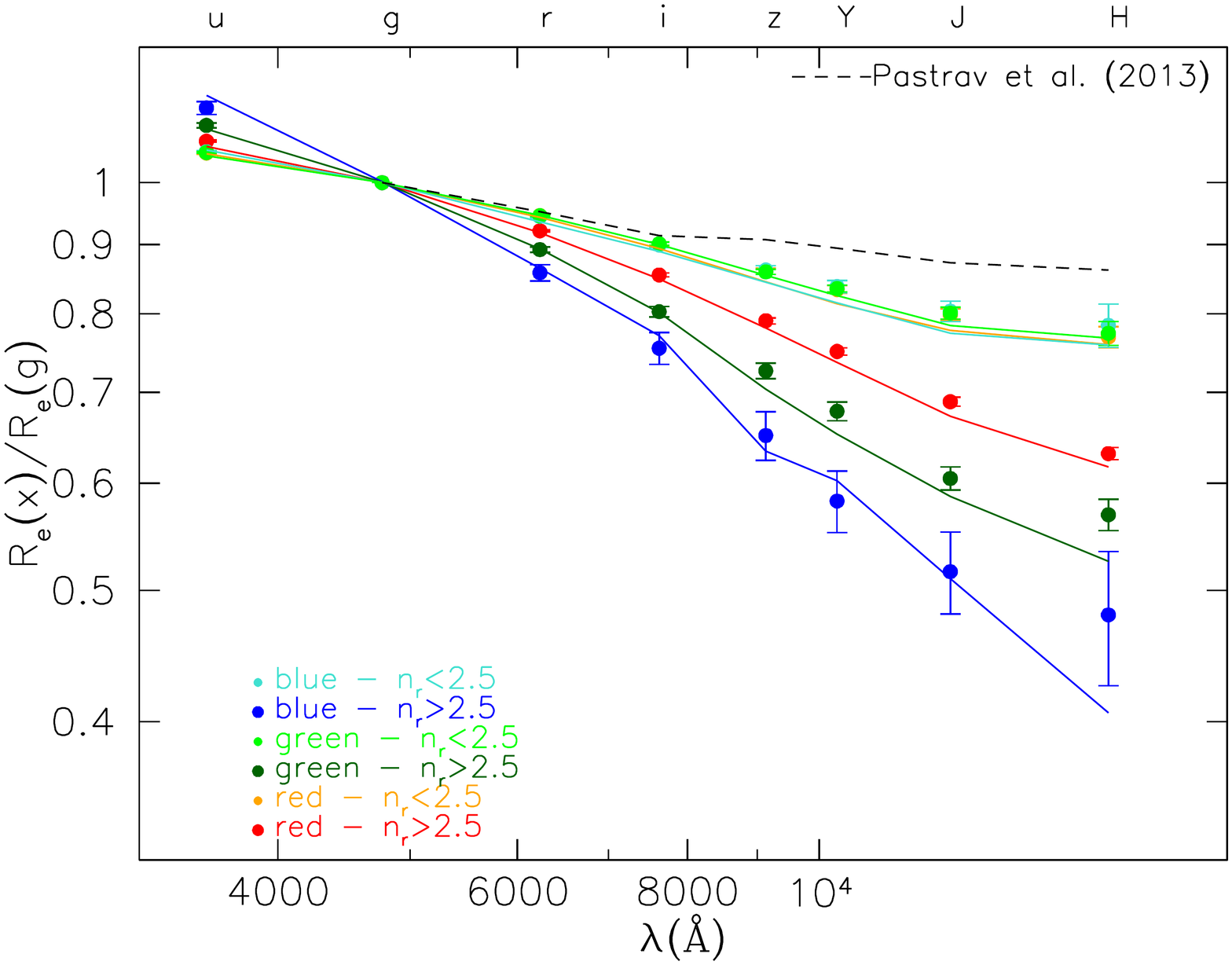}
\caption{Median values of \R[x][g] as a function of wavelength, where $x$ denotes the band corresponding to wavelength $\lambda$. As indicated by the legend, red, dark green, and blue symbols and lines represent \highn galaxies; orange, light green, and turquoise symbols and lines represent \lown galaxies. 
Median \R[x][g]$= \re(x)/\re(g)$ are plotted by points.  Error bars give the uncertainty on the median. Lines plot ${\rm median}[\re(x)]/{\rm median}[\re(g)]$, i.e.\ the points from Fig.~\ref{r_med_n} normalised to the $g$-band.  Small offsets in wavelength have been applied to all points for clarity. The black dashed line represent the wavelength dependence of \R for a disk population due to the effects of dust, as predicted by \citet{Pastrav13}. We do not plot the prediction for the $u$-band since it has not been calculated in the model, but it is a linear interpolation from the $B$-band to the near-UV.  The trends for the \lown samples are indistinguishable, irrespective of galaxy colour. The \highn samples show steeper trends which depend on colour.
\label{rr_med}}
\end{figure}

We summarise the median trends of \R versus wavelength in Fig.~\ref{rr_med}. Points represent the median of the \R[x][g] values for the galaxies in each subsample, where $x$ is the band corresponding to wavelength $\lambda$.  For comparison, the lines in this figure show the ratio of the median $\re(g)$ to the median $\re(x)$ (as individually plotted in Fig.~ \ref{r_med_n}). As before, thanks to the simplicity of the distributions in Fig.~\ref{grar2}, these two approaches give equivalent results.

Galaxies with \lown and \highn clearly display different trends: the former show a relatively mild decrease in \R with wavelength, which is indistinguishable for the three colour samples.  \Lown galaxies are characterised by very similar size distributions at any wavelength, regardless of their colour.  
In contrast, trends for \highn galaxies are more pronounced and become steeper for bluer samples.  Blue, \highn galaxies decrease in effective radius by over a factor of two from $u$- to $H$-band.

Figure~\ref{rr_med}  also plots the wavelength dependence of \R predicted for a disk population, due solely to the effect of dust attenuation, as computed in \citet{Pastrav13} from the radiative transfer model simulations of \cite{popescu11}.
Their prediction is significantly flatter than our disk-like (\lown) points, suggesting that only part of the slope we measure might be ascribable to the presence of dust.  Note that, although the model of \cite{popescu11} contain colour gradients, here we only plot the wavelength dependence due to dust alone.
While the trends for disk galaxies may be due to their two-component nature, the strength of our \highn trends indicate that this subsample comprises galaxies with strong colour gradients.  These may be due to multiple components with different SEDs, or substantial stellar population or dust gradients within a single structural component \citep{Pastrav13b}.

\subsection{The joint wavelength dependence of \sersic index and effective radius}
\label{res:n_re}
\begin{figure}
\centering
\includegraphics[scale=0.42]{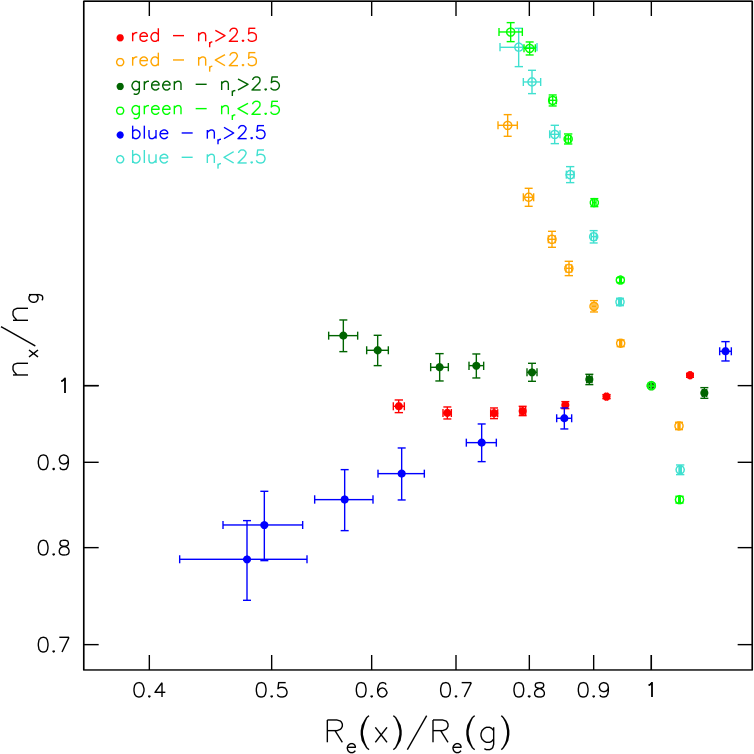}
\caption{Median \N[x][g] vs. median \R[x][g] for galaxies in each of our different subsamples divided by colour and \sersic index. Symbols are as usual.  Multiple points of the same symbol correspond to different wavebands, $x$.  These consistently follow the wavelength sequence from $H$ on the left to $u$ on the right.  All the $g$-band points lie at (1, 1) by definition.
\label{nn_rr}
}
\end{figure}

So far, we have considered the wavelength dependence of \sersic index, \N, and effective radius, \R, separately. Summarising the results obtained from Figs.~\ref{ur_nn} and \ref{ur_rr}, we find that \lown galaxies, which can assume a wide range of colour, mostly have \N above one (\Nang[H][g]~$\sim 1.5$; $0.8 \la$~\N[H][g]~$\la 3$) and occupy a narrow \R range (\Rang[H][g]~$\sim 0.8$; $0.5 \la$~\R[H][g]~$\la 1.1$).
At redder wavelengths they therefore typically display smaller and peakier (higher $n$) profiles.  This is consistent with \lown galaxies typically being bulge-disk systems, comprising a small, $n \ga 2$, red bulge and larger, exponential, blue disk.  However, dust may also play a role in these trends.

In contrast, \highn galaxies, which span the full range of colour but are preferentially red, have \N centred on unity (\Nang[H][g]~$\sim 1.0$; $0.5 \la$~$<$\N[H][g]$>$~$\la 2.0$), and \R offset to slightly lower values (\Rang[H][g]~$\sim 0.6$; $0.3 \la$~\R[H][g]~$\la 1.1$).
At the reddest wavelengths their profiles typically maintain the same shape ($n$) but become substantially smaller.  This is initially surprising, as a stellar population gradient would, in general, be expected to result in both the \sersic index and effective radius varying with wavelength.
Our \highn galaxies are consistent with comprising either a single spheroidal structural component with a strong stellar population gradient, or multiple spheroidal components with different $\re$ and stellar populations, such that the larger components have bluer colours. The former may be identified with a single monolithic collapse, for which the colour gradient would be driven by metallicity.  The latter may correspond to distinct phases of collapse or merging.

We can generalize our findings for \N[H][g] and \R[H][g] by considering \N and \R for multiple wavebands.  
Figure~\ref{nn_rr} shows how the median of the \N-\R distribution varies with wavelength for galaxies belonging to each of our six subsamples, and neatly presents many of the trends we have explored above.
\Lown and \highn galaxies show different behaviour with respect to wavelength: \lown galaxies vary substantially in \N, but over only a narrow range in \R.  On the other hand, \highn galaxies vary widely in \R, but only slightly in \N.  Colour has a more subtle influence on these trends.  For \lown galaxies, redder galaxies vary less in \N, but equally in \R.  \Highn galaxies show a particularly interesting behaviour with colour.  \Red galaxies, which comprise the majority of the \highn population, show almost no variation in \N, but vary significantly in \R.  \Green\ \highn galaxies show a slight increase in \N with wavelength, potentially indicating the presence of a blue disk, but show an even greater variation in \R, in contrast to the \lown objects.  \Blue\ \highn galaxies continue the trend of a strengthening wavelength dependence for \R, but \emph{reverse} their behaviour in terms of \N, appearing peakier at shorter wavelengths.
The trends with wavelength are relatively linear in \N versus \R.  They are therefore well-described by any one pair of wavebands, justifying our approach of focussing on $H$ and $g$ throughout this work.

\subsection{\N and \R as a classifier}

Above we find, somewhat unsurprisingly, that the wavelength-dependence of \lown galaxies is consistent with them containing both a spheroid and disk, while \highn galaxies comprise one or more spheroidal components.  This allows the possibility of using \N and \R to directly identify bulge-disk systems, irrespective of their overall colour or \sersic index.  Furthermore, we can combine information from \N and \R with colour and \sersic index to isolate galaxies with specific structural properties.  We now briefly investigate the use of \N[H][g] and \R[H][g] alone to separate galaxy populations with distinct internal structure.

\begin{figure*}
\centering
\includegraphics[scale=0.37,clip=true]{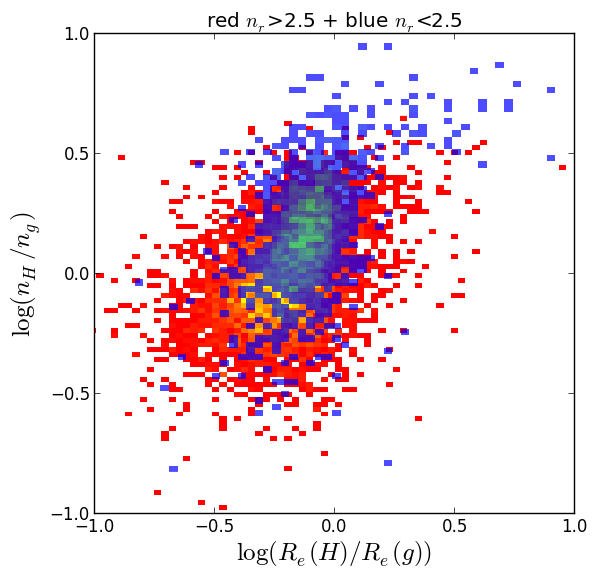}
\hfill
\includegraphics[scale=0.37, clip=true]{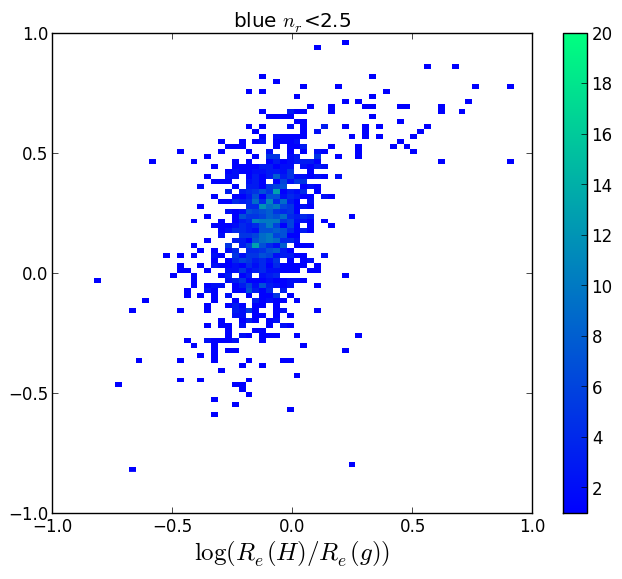}
\includegraphics[scale=0.37,clip=true]{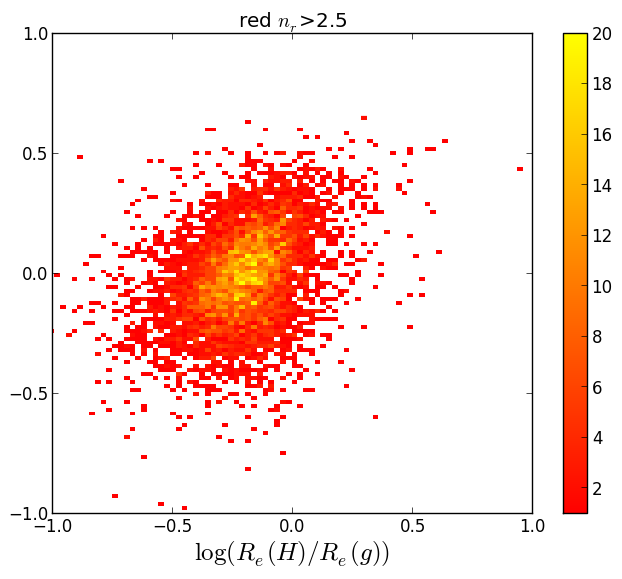}
\caption{\N[H][g] versus \R[H][g] for \blue\ \lown galaxies (blue points) and \red\ \highn galaxies (red points). In the left panel the two samples are plotted together, while in the central and right panels separately plot the \blue\ \lown and \red\ \highn samples, respectively.  These are the main contrasting populations in terms of colour and \sersic index, and it can be seen that they also occupy different, though overlapping, loci in this plane.
\label{nn_rr_col}}
\end{figure*}
In Fig.~\ref{nn_rr_col} we consider the main contrasting populations: \blue\ \lown galaxies and \red\ \highn galaxies. 
Although the averages of the two populations are significantly offset (see Fig.~\ref{nn_rr}), their distributions are sufficiently broad that they overlap considerably.  \Blue\ \lown galaxies inhabit a compact locus within that of \red\ \highn galaxies.  It is therefore possible to cut the \N-\R plane to isolate a pure, but incomplete, sample of \red\ \highn galaxies, but any selection of \blue\ \lown objects will be significantly contaminated.  It is not yet clear whether the breadth of these distributions is determined by measurement uncertainties or intrinsic variations.  Our sample has relatively low signal-to-noise and poor spatial resolution.  With higher quality data or a lower-redshift sample, it may well be that galaxy populations will separate more cleanly in the \N-\R plane.  However, for galaxy classification \N and \R are  best considered as complementary to \sersic index or colour, rather than a potential replacement.

From the behaviour of the various colour and \sersic index selected samples in the \N-\R plane it is certain that \N and \R contain additional information.  For example, within \red\ \highn objects, \N and \R may be used to isolate galaxies which show evidence for perhaps containing a blue disk (\N~$\gg 1$), or central star-formation (\R~$\ll 1$).

\section{Interpretation}
\label{stacks}

\begin{figure}
\centering
\includegraphics[scale=0.27]{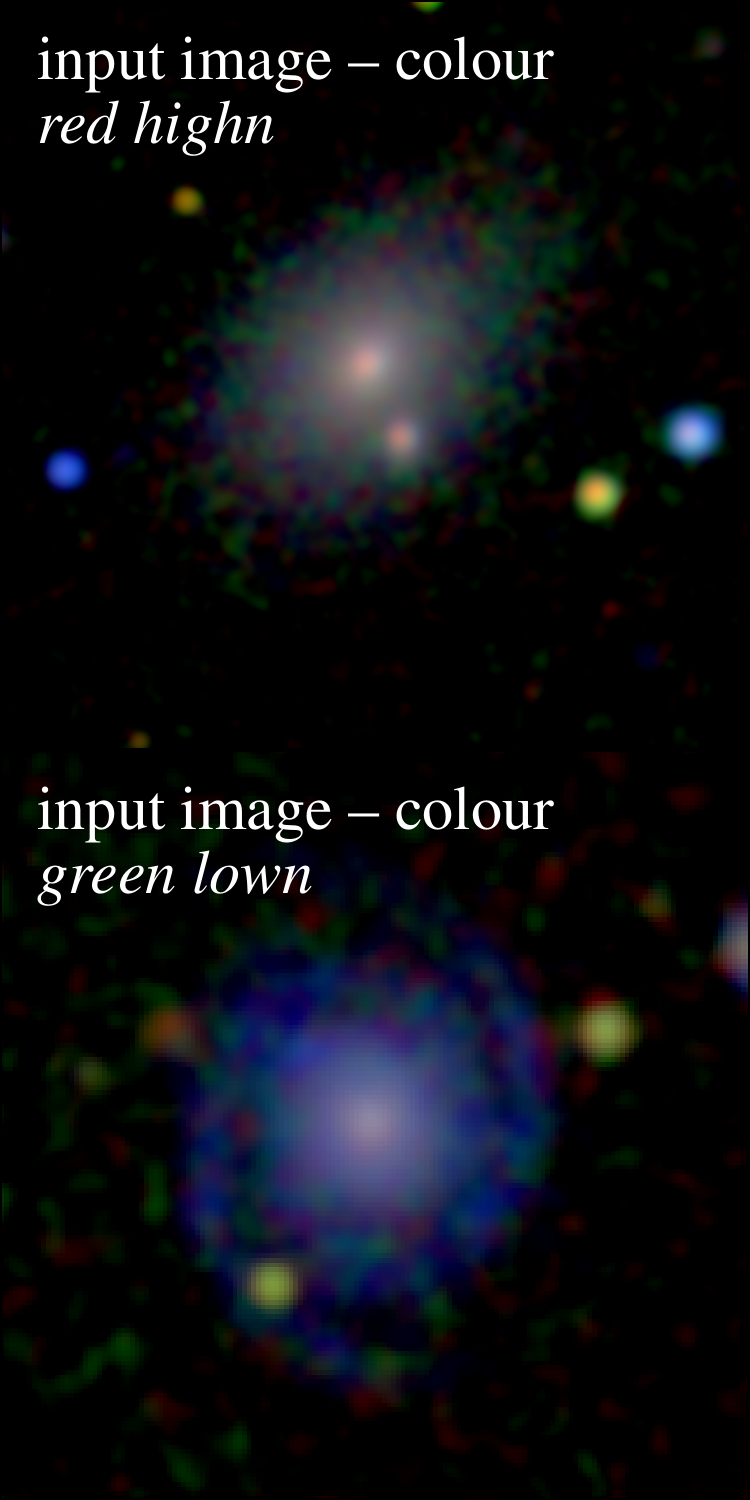}
\hfill
\includegraphics[scale=0.27]{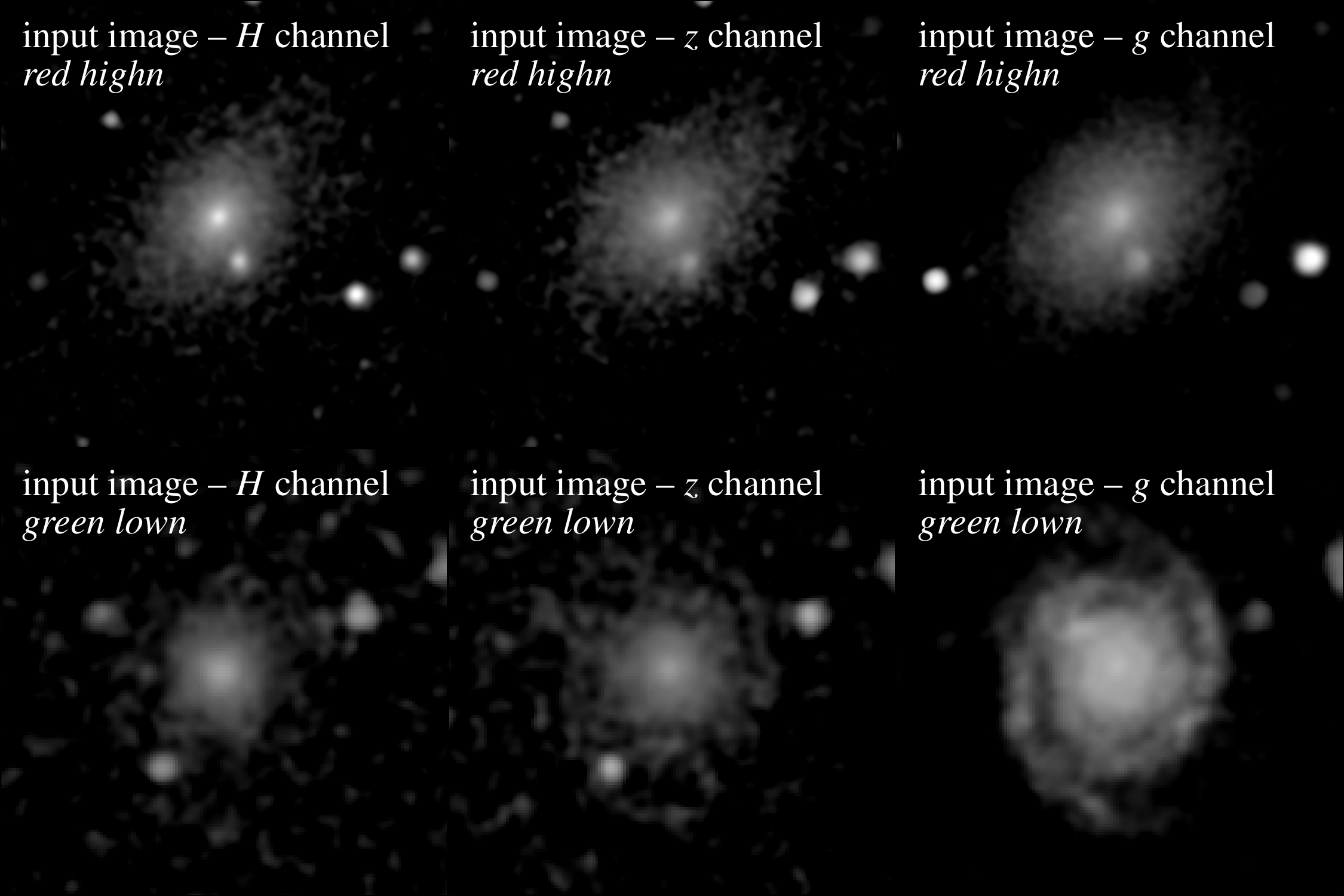}
\\[5pt]
\includegraphics[scale=0.27]{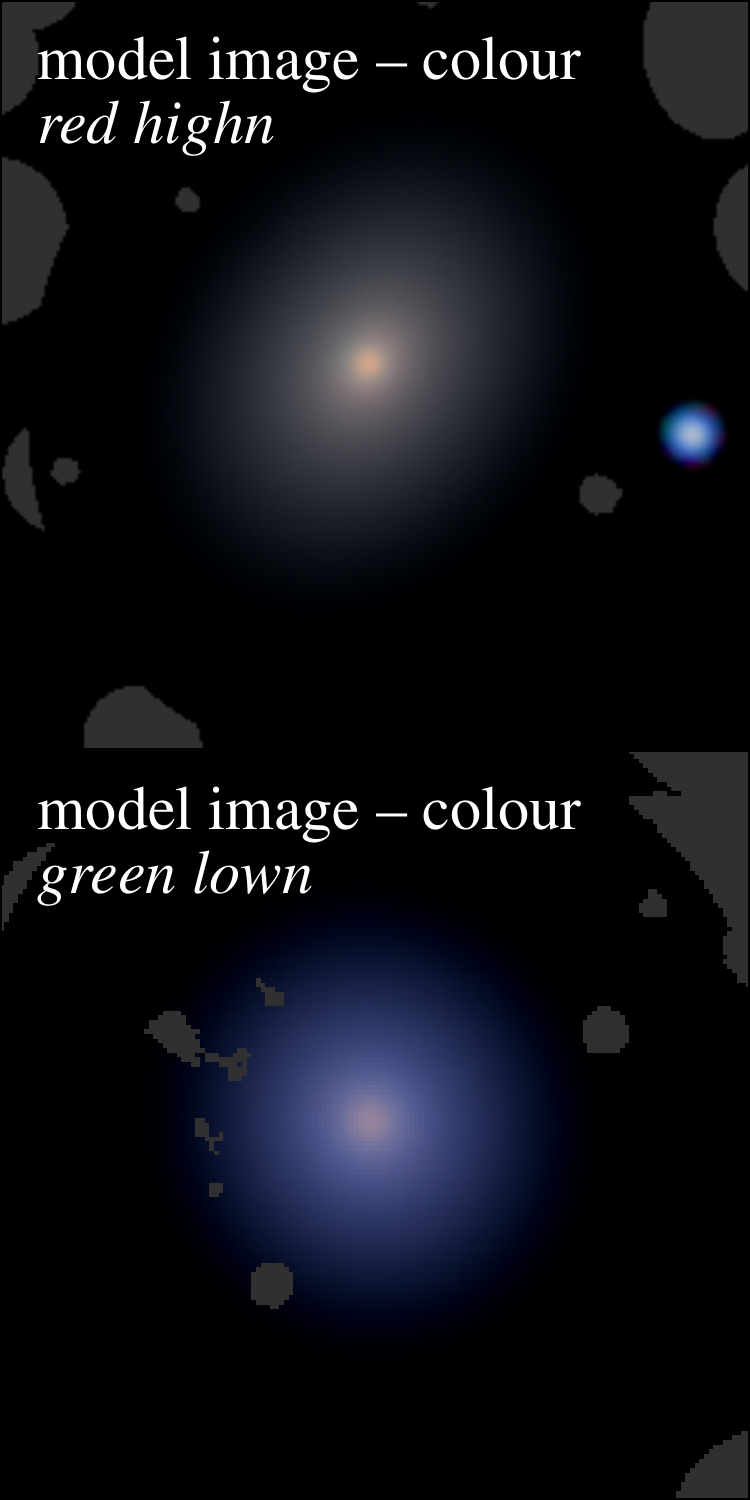}
\hfill
\includegraphics[scale=0.27]{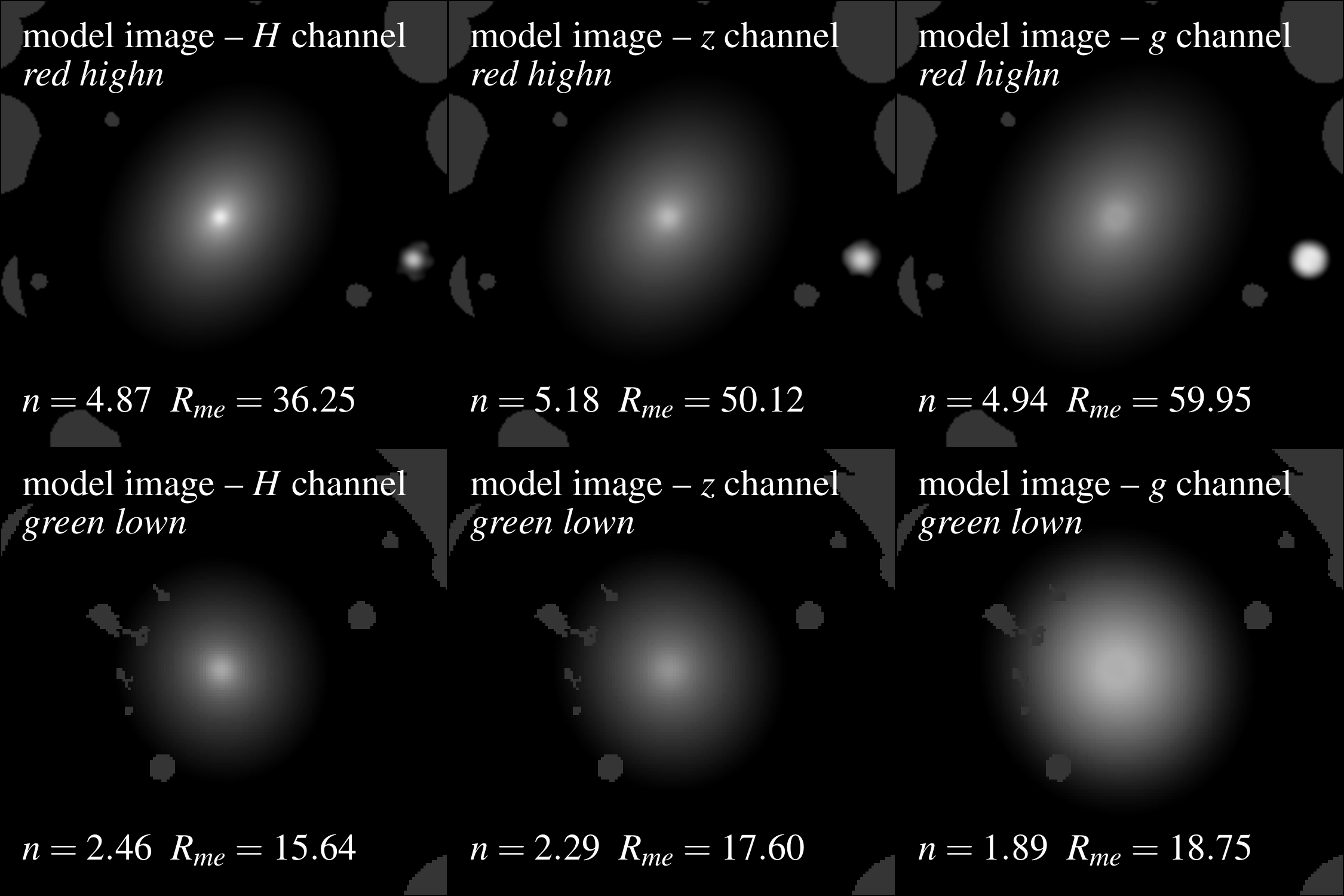}
\caption{Left: example $Hzg$ images of a \red, \highn and a \green, \lown galaxy, each with typical dependences of $n$ and $\re$ on wavelength (they lie at the peaks of their respective distributions in Fig.~\ref{nn_rr_col}).  The images are displayed using the colour-preserving method of \citet{lupton04}, with an asinh stretch to reveal the full range of surface brightness.
The upper panels show the data, while the lower panels show the \galfitm model.  Grey regions in the model images indicate where pixels were masked during the fit.
Right: individual red, green and blue channels from the colour images, which make it easier to appreciate the relative distributions of the $H$-, $z$- and $g$-band light.  For both galaxies the central region is significantly redder than the outskirts.  This is more subtle for the \highn galaxy, as $n$ is stable with wavelength, but still amounts to an increase in $\re$ by almost a factor of two between the $H$ and $g$ bands.
\label{example_images}}
\end{figure}

\begin{figure*}
\centering
\includegraphics[width=0.49\textwidth]{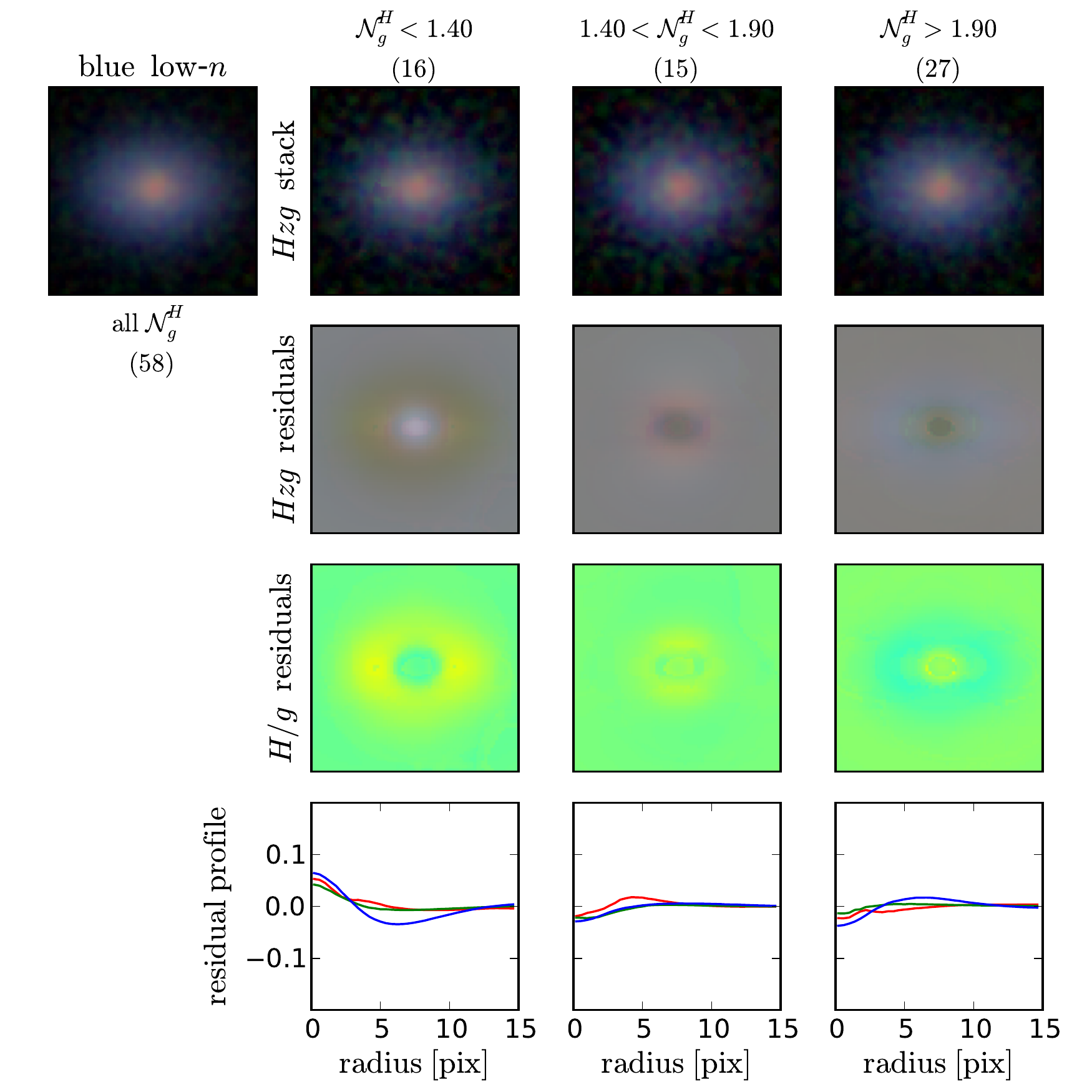}
\hfill
\includegraphics[width=0.49\textwidth]{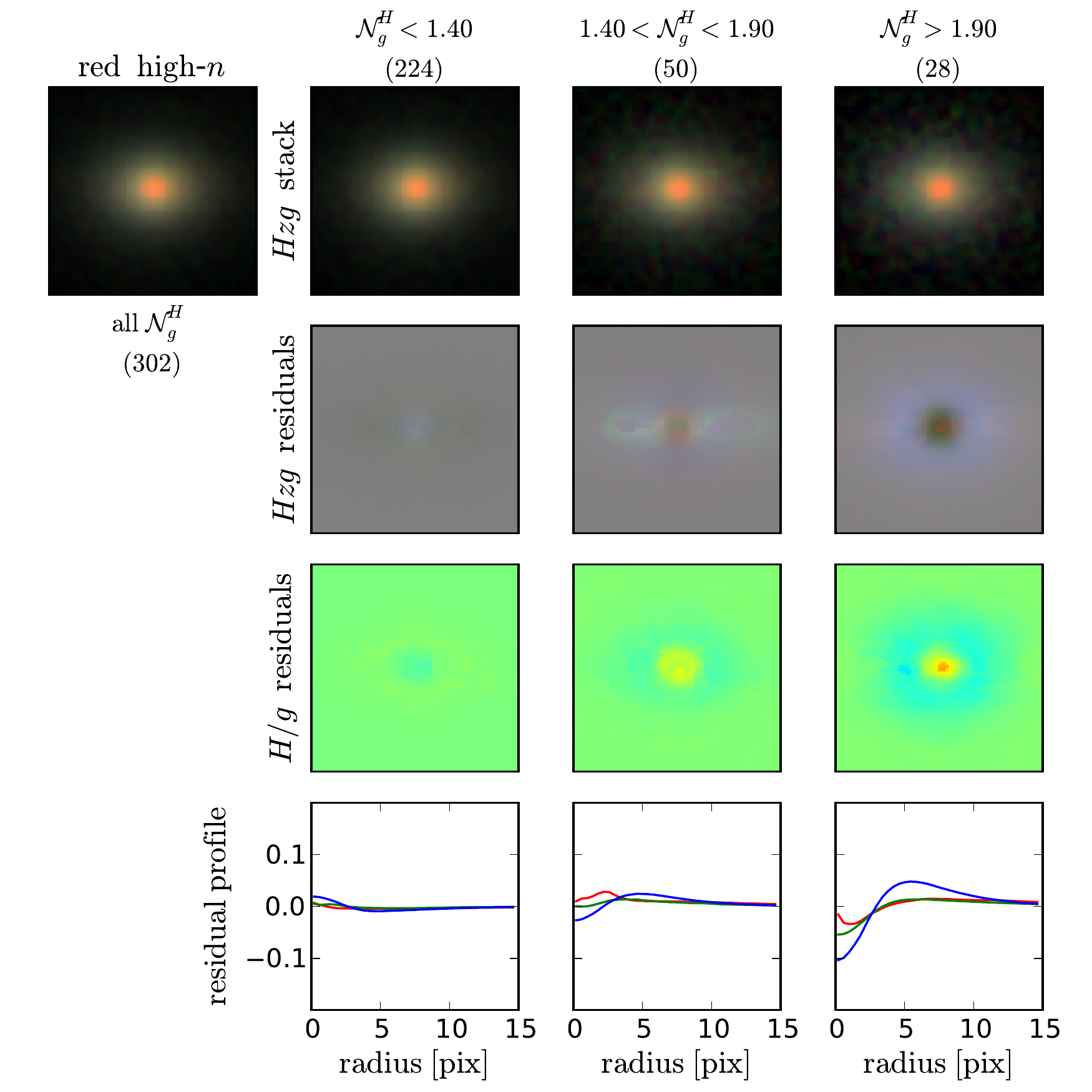}
\\[0.75cm]
\includegraphics[width=0.49\textwidth]{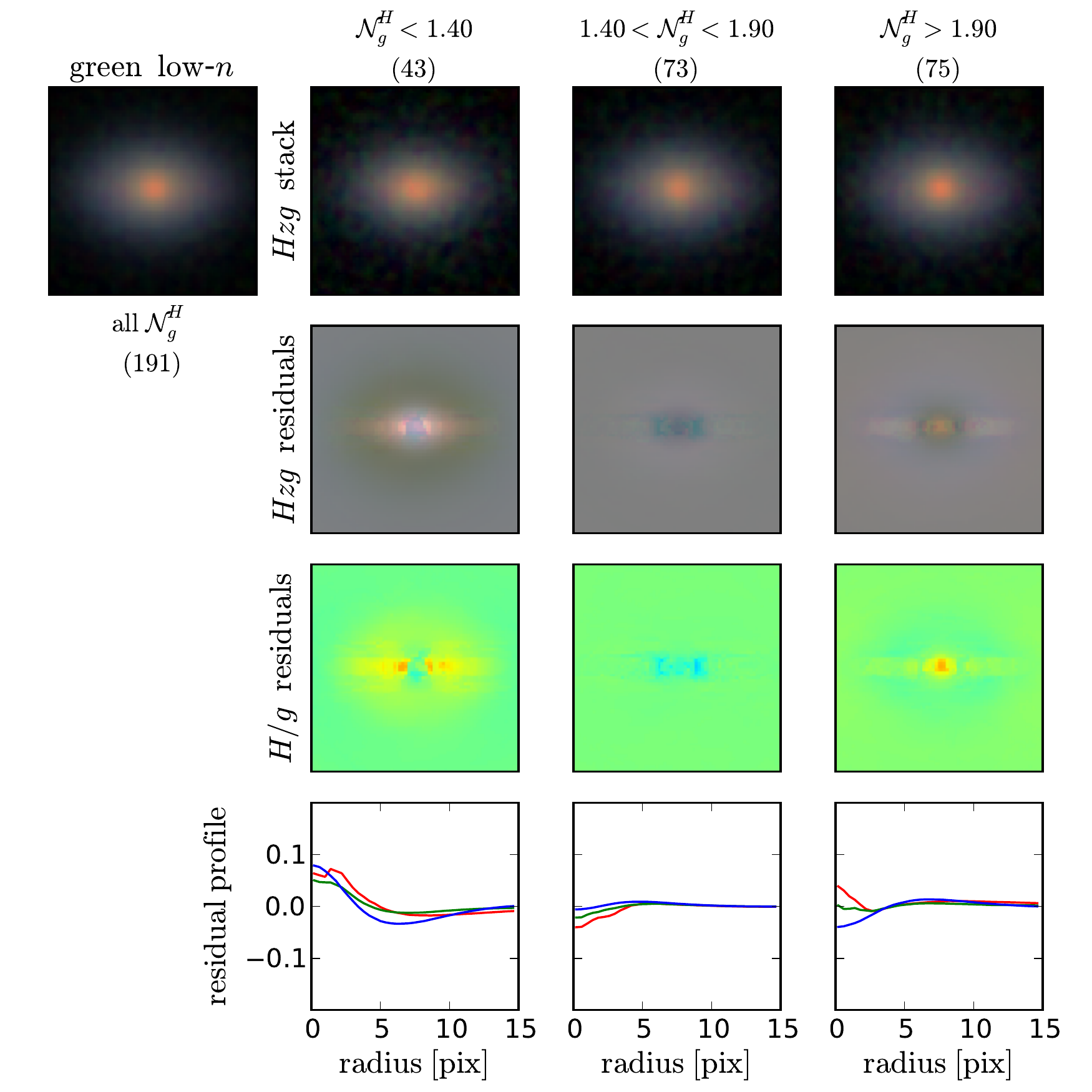}
\hfill
\includegraphics[width=0.49\textwidth]{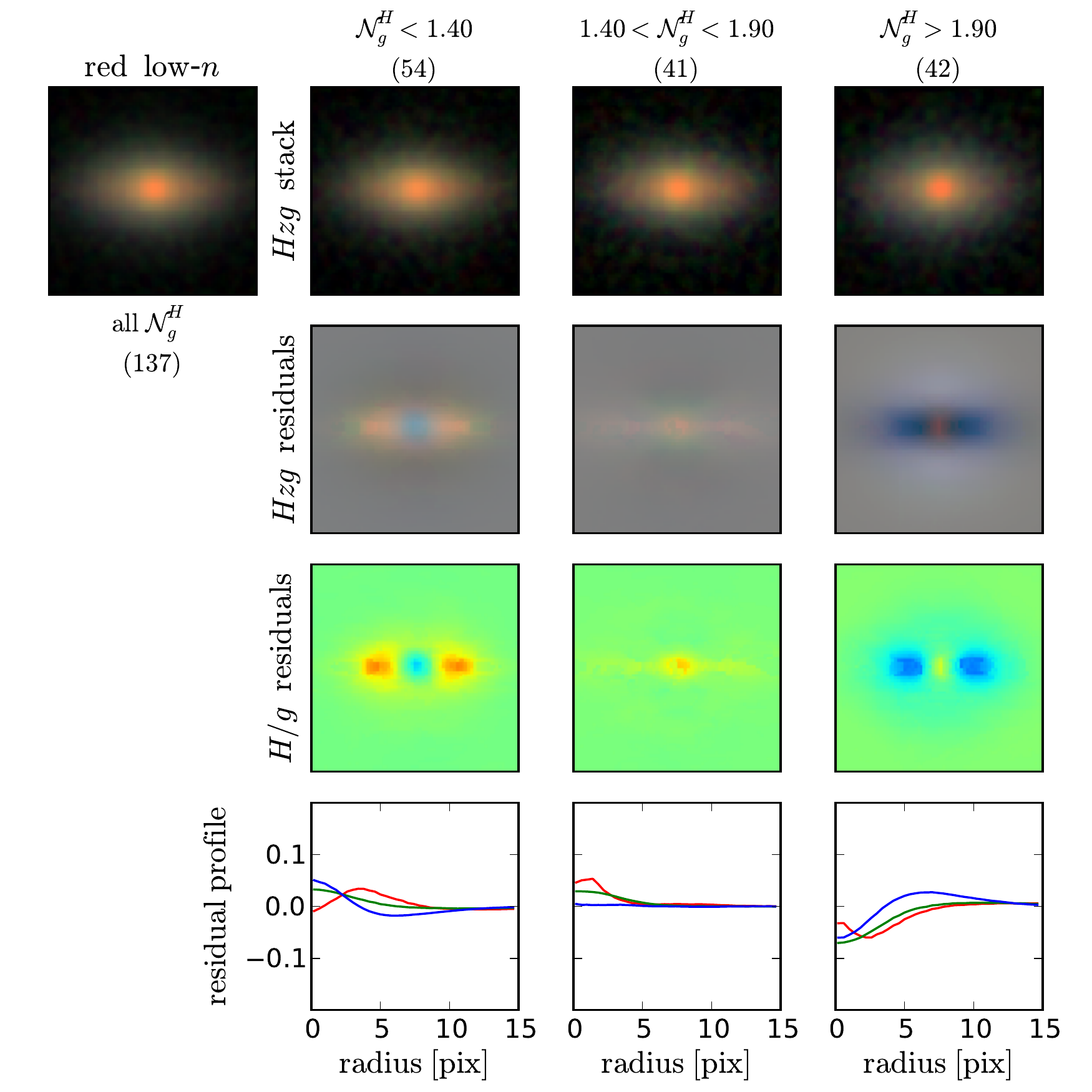}
\caption{Stacked \N[H][g] gradients, for galaxies  
with similar colour and  \sersic index.  For each set of stacks we divide the galaxy sample into three  groups, by cutting at
\N[H][g] $= 1.4$ and \N[H][g] $= 1.9$ (see text for details).
In each subplot, the top panels show the stacked images using the $Hzg$ bands, the second-to-top panels show the residuals, the second-to-bottom panels show the ratio o the $H$ and $g$ residuals and the bottom panels show the radial profiles of the $H$, $z$ and $g$ residuals. Numbers in parenthesis are the numbers of galaxies used in each stack.
{The $Hzg$ stack and residuals are displayed using the colour-preserving method of \citet{lupton04}.  The stack images have an asinh stretch, while the residual images are shown with a linear stretch. The $H/g$ residuals image has a false-colour scale, where green corresponds to equal residuals in $H$ and $g$, red indicates that the $H$ residuals are brighter than those in $g$, and blue indicates the converse. Each image is 30 pixels across.  The radial profiles show the fractional difference in flux between this image stack and that of all objects in the same colour and \sersic index subsample.
It is clear that galaxies with higher \N are fainter and redder in the centre, but bluer at intermediate radii, and vice versa for lower values of \N.  We do not show images for blue and green \highn galaxies  because there are too few to produce meaningful stacked images.} \label{stackfig_n_grad}}
\end{figure*}

\begin{figure*}
\centering
\includegraphics[width=0.49\textwidth]{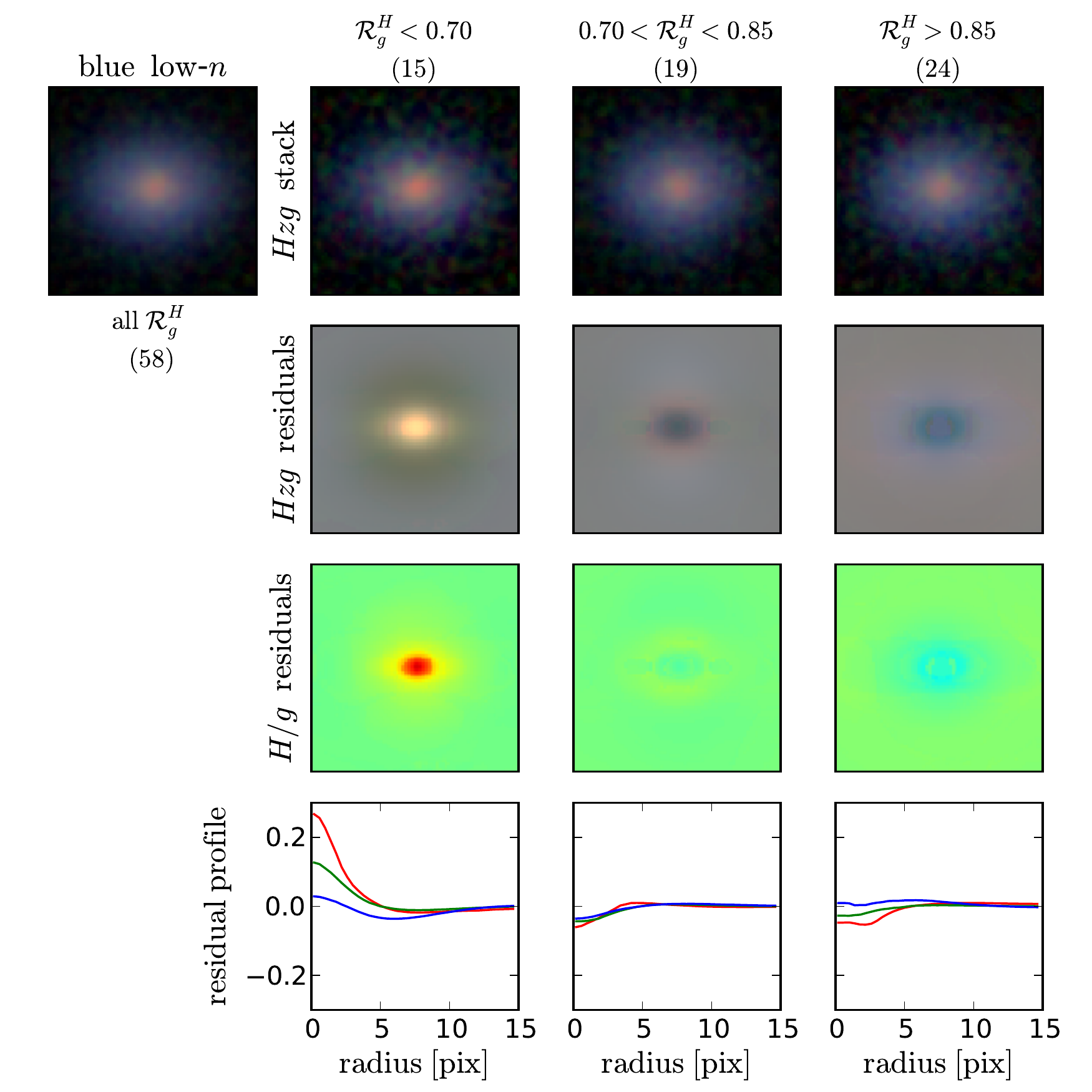}
\hfill
\includegraphics[width=0.49\textwidth]{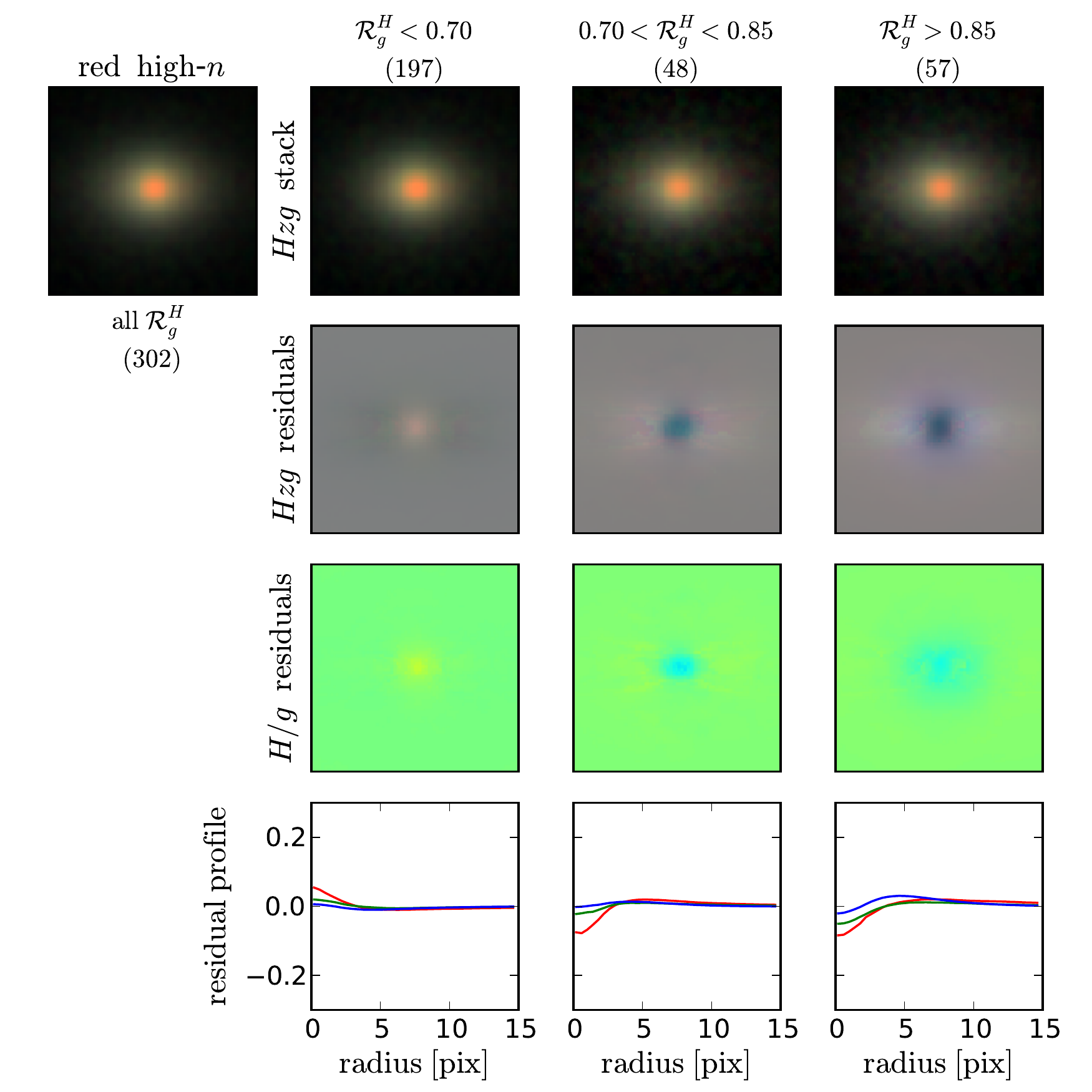}
\\[0.75cm]
\includegraphics[width=0.49\textwidth]{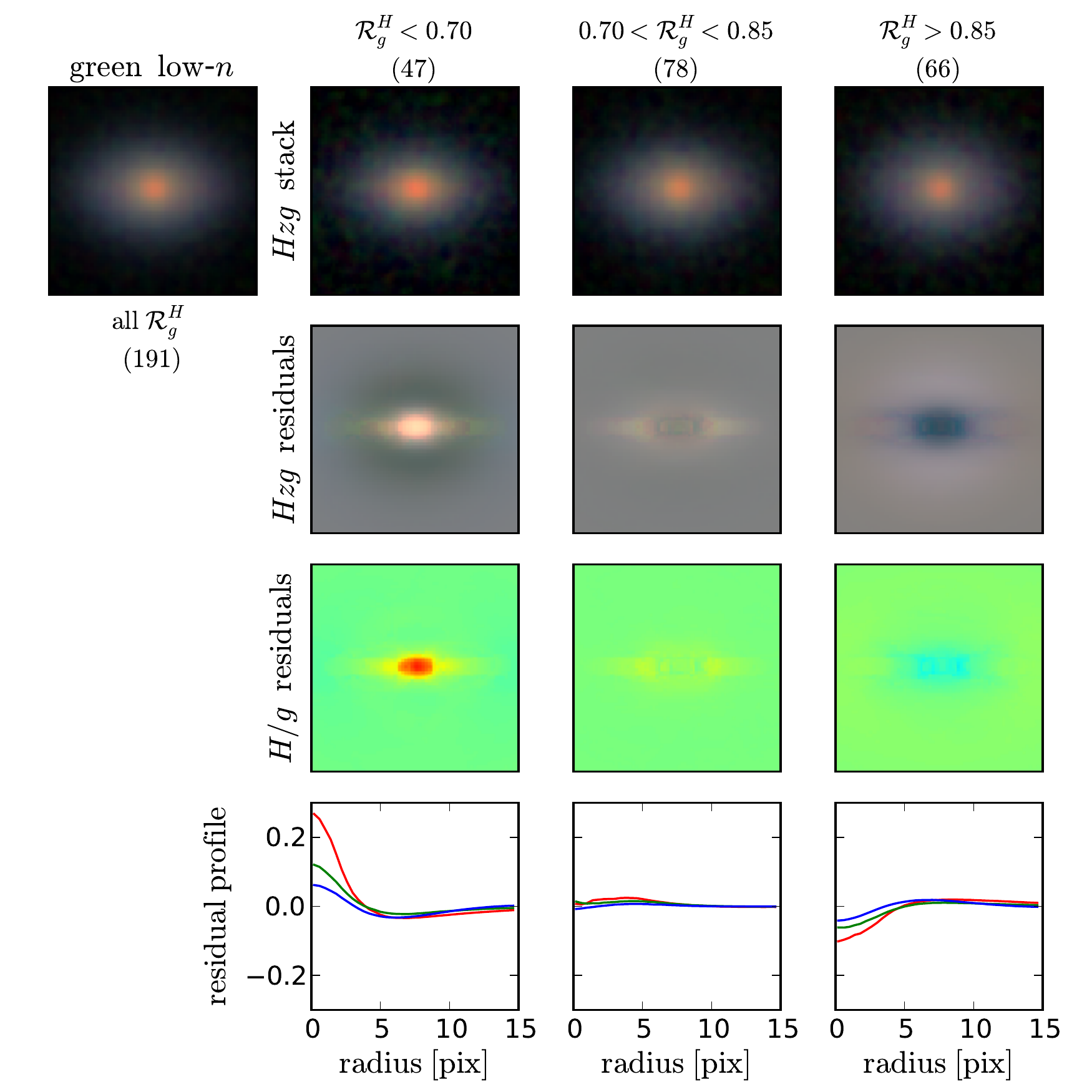}
\hfill
\includegraphics[width=0.49\textwidth]{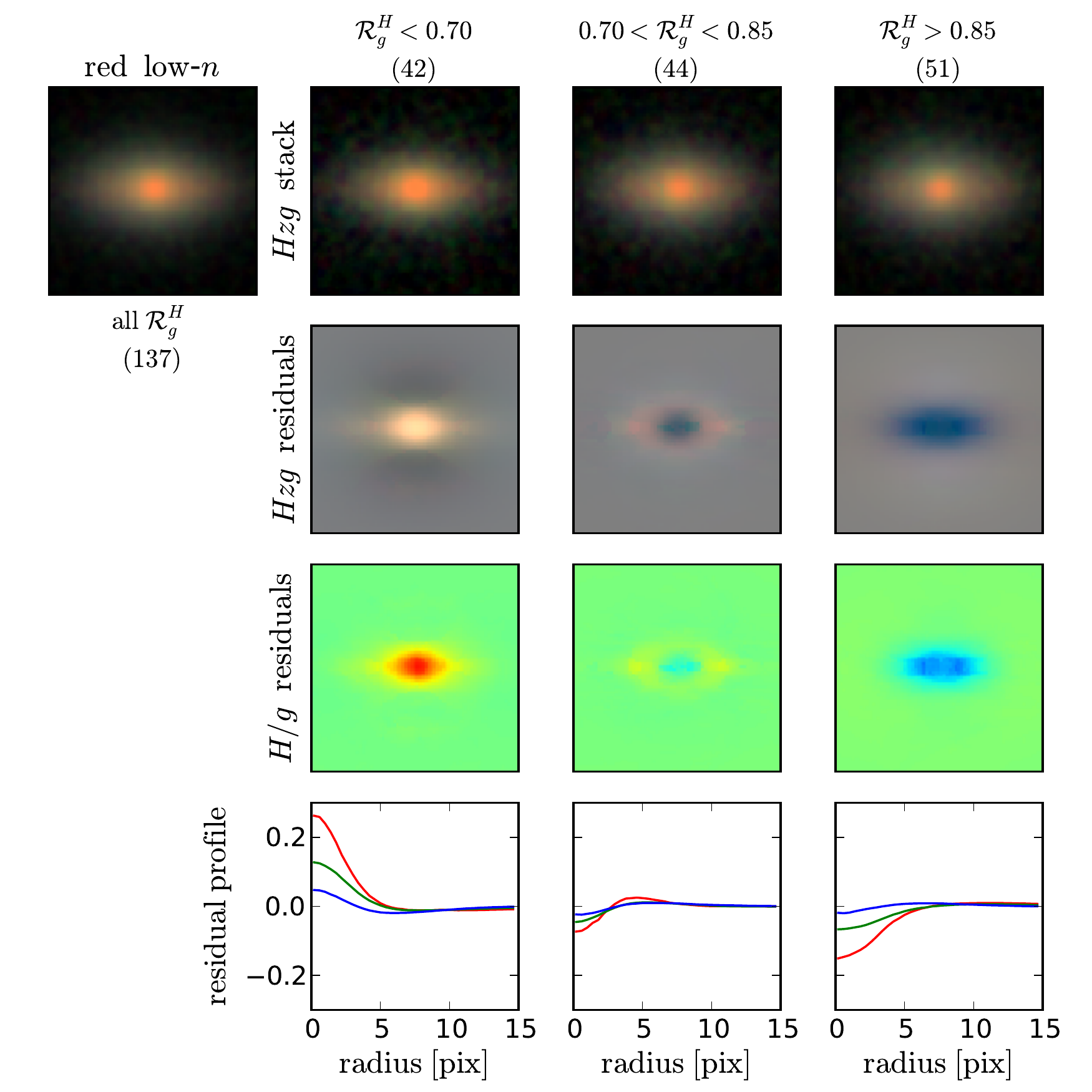}
\caption{Stacked \R[H][g] gradients, for galaxies  
with similar colour and  \sersic index.  For each set of stacks we divide the galaxy sample into three  groups, by cutting at
\R[H][g] $= 0.70$ and \R[H][g] $= 0.85$ (see text for details). Panels are as in Fig.\ref{stackfig_n_grad}.
{Each image is 30 pixels across.} Galaxies with lower \R are  redder in the centre, and vice versa for higher values of \R.  We do not show images for blue and green \highn galaxies  because they are too few to produce meaningful stacked images. \label{stackfig_re_grad}}
\end{figure*}

Our analysis has demonstrated that the wavelength dependences of \sersic index and effective radius are related to other galaxy properties.  However, it is now helpful to pause and consider what \N and \R actually mean.  What do galaxies with different values of \N and \R look like, and how do they compare with the more usual concept of colour gradients?

\N primarily indicates the degree to which the central concentration of the light profile varies with wavelength.  However, remember that for a given $\re$ a greater central concentration also implies more light in the outer regions of the profile.  A value of \N$> 1$\footnote{We remind the reader that \N and \R have been defined as the ratios of the reddest to the bluest $n$ and $\re$, respectively, hence these relations are reversed when we look at \N[u][g] and  \R[u][g].} implies that the galaxy has a peakier (higher-$n$) profile in redder bands.  All else being equal, galaxies with \N$> 1$ should therefore appear redder in their centres, bluer at intermediate radii, and then redder again in their outskirts.  In practice, however, the behaviour in the outskirts is often lost in the sky noise and the S/N-weighting scheme employed by \galfit means that it cares much less about the outer profile.\footnote{This is much like the fact that, analytically, a de Vaucouleurs bulge must always dominate over an exponential disk at sufficiently large radii, but this is rarely observed because it typically occurs at very faint surface brightness.}  \N$< 1$ implies the opposite behaviour, that galaxies are peakier in bluer bands, whereas \N$\sim 1$ indicates that the profile has a similar shape at all wavelengths.

\R quantifies the variation in size with wavelength.  For example, \R$< 1$ implies that a galaxy is smaller in redder bands.  All else being equal, galaxies with \R$< 1$ should therefore appear redder in their centres and bluer toward their outskirts.  \R$> 1$ implies the opposite behaviour, whereas \R$\sim 1$ indicates a constant color with radius.  \R thus works much like a conventional colour gradient, and in the same sense, in that a negative colour gradient would imply a bluer colour with increasing radius.

Note that (considering only the behaviour at $\lesssim \re$) variations in \N may represent similar visual effects to variations in \R, but in an opposite sense (i.e.\ in broad terms, increasing \N mimics decreasing \R).  However, although \N and \R can produce superficially the same appearance, as we have seen above they are certainly not interchangeable, providing distinct information regarding the dependence of galaxy structure on wavelength.

{Figure~\ref{example_images} presents colour images of two galaxies.  These are chosen to be bright, well-resolved objects, but are otherwise typical examples from our \red, \highn, and \green, \lown samples.  We have examined a large number of similar images, and see that the general behaviour expected from our \N and \R measurements is confirmed by visual inspection.}

When we compare individual galaxies with similar properties, but different values of \R or \N, it is possible to see differences.  However, this is somewhat obscured by other variations between individual galaxies, and the relatively low signal-to-noise of many of the objects in our sample.  In order to demonstrate that \R and \N do represent true differences in the appearance of galaxies in an incontrovertible  manner, we have stacked galaxies with similar properties.  In this way individual galaxy-to-galaxy variations are averaged out and the signal-to-noise is improved.

Recall that Figures~\ref{nn_rr} and \ref{nn_rr_col} shows that \N and \R are somewhat correlated, although the correlation is reduced when dividing by colour and \sersic index.   It is therefore difficult to completely separate out the behaviour of \N and \R in stacked images.  However, given their opposing behaviour, they will tend to cancel out each other's effect on galaxy appearance.  Thus, if anything, the stacks may under-represent the individual effects of \N and \R. 

{We combine the images using a 3-$\sigma$ clipped mean, to exclude neighbouring object and atypical galaxies from the resulting stacked image.  We further avoid any undue influence from outliers by applying the following limits to the objects in the stacks: $0.05 < z < 0.2$; $-22.5 < M_r < -21.2$; $ 6 < \re < 8$ pixels and $1.0 < (u - r) < 3.0$. We can therefore be sure that any features we see are associated with the main distribution of galaxies. So that the stack is not dominated by bright galaxies, before averaging we scale the images to the same object magnitude, and to preserve axial ratio information we rotate each image such that the galaxy's major axis is horizontal.  No scaling of the object sizes or circularisation is performed, as these operations were found to introduce artefacts due to distorting the point spread function (PSF).  However, note that only galaxies with similar apparent $\re$ are included in these stacks.}

{We stack galaxies with similar colour, \sersic index and ratio of either \sersic index or effective radius in the $H$- and $g$-bands.  In addition to our usual divisions in terms of $(u - r)$ and $n$, for each set of stacks we divide the galaxy sample into three representative groups, by cutting at
\N[H][g] $= 1.4$ and \N[H][g] $= 1.9$ and
\R[H][g] $= 0.70$ and \R[H][g] $= 0.85$, respectively.  This results in 36 stacked images.  For example, one stack contains only red, low-$n$, mid-\N galaxies, whereas another contains only blue, high-$n$, low-\R galaxies.  Each stack typically contains 10--100 galaxies. 
For comparison we also create stacked images for all galaxies of a given colour and \sersic index, irrespective of \N or \R, for example, all red, high-$n$ objects.}

The differences between the stacked images are relatively clear.  However, to further aid their comparison, for each \N- or \R-specific stack we subtract the stack of all galaxies with the same colour and \sersic index.  The results are presented as both $Hzg$ residual colour images and maps of the ratio of the $H$ and $g$ residuals.  The former retain intensity information, and therefore display variations in the total brightness or surface brightness profile between categories.  The latter remove this brightness information and just present the residual variation in $H-g$ colour with position.  As a final visualisation, we azimuthally average the residual images to compare the radial profiles for $H$, $z$ and $g$.

Figure~\ref{stackfig_n_grad} presents the behaviour of \N for the common combinations of \sersic index and colour.  As expected from the above arguments, when compared to the average, galaxies with higher \N are fainter and redder in the centre, but bluer at intermediate radii, and vice versa for lower values of \N.  In some of the stacks, particularly for \highn galaxies, the residuals are dominated by colour variations in the centre of the galaxy.  For \lown galaxies the residuals are dominated by colour differences further out and are often elongated, suggesting the presence of a disk. Low (high) values of \N appear to reflect the presence (absence) of a red disk, while the centre of the galaxy remains unchanged.

Figure~\ref{stackfig_re_grad} presents the same information for variations in \R.  Again, the behaviour is broadly as anticipated.  Compared to the average, objects with lower \R display redder centres, while higher \R implies bluer centres.  As with \N, for \lown galaxies there are indications that the value of \R partly reflects the presence of a disk.  However, unlike \N, the value of \R is also strongly related to the presence of a red central component.  Objects with lower \R possess brighter, redder centres.

These stacks visually demonstrate that the different \N and \R values we have measured represent subtle but significant differences in the appearances of the galaxies.

\section{Discussion}
\label{discussion}

In this paper we have explored how galaxy structure varies with wavelength (from rest-frame $u$ to $H$), for populations selected by \sersic index and colour, using a volume-limited sample of 14,274 bright galaxies.  We measure galaxy sizes and radial profile shapes more accurately than previous studies, by utilising a new technique which fits consistent, wavelength-dependent, two-dimensional \sersic profiles to imaging in a collection of wavebands.  Using this method, we confirm that the behaviours of individual galaxies generally follow the average trends of populations of similar objects.  We are also able to identify individual galaxies with a particular wavelength dependence of their structural parameters, and hence create samples of objects that are similar in this respect.

We find that both effective radius and \sersic index typically vary with wavelength.  The sizes of all types of galaxies decrease substantially with increasing wavelength, with high-$n$ galaxies showing the greatest variation.  The behaviour of \sersic index is more sensitive to galaxy type: for high-$n$ galaxies \sersic index is constant with wavelength, while for low-$n$ galaxies it varies dramatically.

We now discuss our findings in the context of other works: first comparing to similar studies on whose results we build, then considering the physical processes responsible for the trends we measure, and finally suggesting how subsequent studies of galaxy structure can help to further differentiate between models of galaxy formation and evolution.

\subsection{Comparison with previous results}

The SPIDER project \citep{barbera10b} has performed a careful analysis of the structural parameters of low-redshift early-type galaxies as a function of wavelength.  They fit \sersic profiles to imaging data from the same surveys as us ($griz$ from SDSS and $YJHK$ from UKIDSS LAS) using a similar tool ({\textsc 2dphot}; \citealt{LdC08}), but fitting each waveband independently.
\citeauthor{barbera10b} found that, for early-types, $\re$ decreases by $\sim 35$\% from (observed frame) $g$ to $K$. They attributed this variation to the presence of negative internal
colour gradients, such that the light profile becomes more compact as one moves from shorter to longer wavelengths. The \sersic indices of their objects span $n \sim 2$--$10$, with a median of $6$ for all wavebands. They find little variation of \sersic index with wavelength. Although their average $n$ does perhaps increase slightly ($\sim 20$\%) from $g$ to $K$, the trend is noisy and consistent with being flat.

\citet{barbera10b} select their early-type galaxies as being bright ($M_r < -20$), having a passive spectral type and being better represented by a de Vaucouleur profile than an exponential one.  This roughly corresponds to our selection of \red \highn objects, although our average $\langle n \rangle \sim 4$ is slightly lower than theirs.  For this sample we find very similar results: a $\sim 40$\% increase in $\re$ from (rest-frame) $u$ to $H$, while the \sersic index remains stable, varying by $< 5$\% over the same wavelength range.
As \citeauthor{barbera10b} point out, this behaviour is also consistent with \citet{ko05,pahre98b,pahre99} and \citet{pahre98a}; the latter of which demonstrates that the behaviour of $\re$ corresponds to an increase in the slope of the fundamental plane with wavelength, further confirmed by \citet{barbera10a} and references therein.

\citet{kelvin12} similarly perform two-dimensional single-\sersic fits to SDSS and UKIDSS LAS $ugrizYJHK$ images of galaxies in the GAMA survey, using \galfit to fit each bandpass independently.
They defined two galaxy subpopulations, based on $K$-band \sersic index and $(u - r)$ rest-frame colour, and analysed the wavelength dependence of their structural parameters.
They find that the mean effective radius of their early-types shows a smooth variation with wavelength, decreasing by $\sim 38$\% from (observed frame) $g$ through $K$.  Although their early-type selection is similar to ours, their sample is magnitude-limited ($r \la 19.4$).  Compared to the volume-limited samples of our work and \citet{barbera10b} it therefore contains additional low-luminosity, low-redshift and high-luminosity, high-redshift objects. Even so, the majority of galaxies in all these studies have similar luminosity ranges, and their findings are very similar.  \citeauthor{kelvin12} quote an increase in average \sersic index of early-types of $\sim 30$\% over $g$ to $K$.  However, as with \citeauthor{barbera10b}, the trend is rather noisy, and fairly consistent with being flat.

\citet{kelvin12} expand on the study of \citet{barbera10b} by also considering the late-type galaxy population.  These exhibit a more dramatic change in average \sersic index, increasing by $52$\% from $g$ to $K$, and a more gentle variation in effective radius, decreasing by $25$\% over the same wavelength range.  We find an identical $25$\% variation in effective radius, but an even more extreme variation in \sersic index, of $\sim 100$\% (i.e. a factor of two) from rest-frame $u$ to $H$ band.

Our results are therefore in excellent qualitative agreement with those of \citet{kelvin12}.  It may be that the differing selections play a role in the detailed differences: in addition to the different luminosity ranges, the joint $n_K$ and $(u - r)$ selection adopted by \citeauthor{kelvin12} contrasts with our orthogonal $n_r$ and $(u - r)$ selections.  For example, \citeauthor{kelvin12} includes objects with low \sersic index ($n \sim 1$) in the early-type sample provided they are sufficiently red.  We also expect our individual measurements to display significantly less scatter, due to our multi-wavelength method (H13, \citealt{vika13}), which may help us recover smoother, more accurate trends.

Note that \citeauthor{kelvin12} chose to classify galaxies using the \sersic index in the $K$-band, in order to avoid any potential misclassifications due to the effects of dust attenuation.  However, we assert that it is cleaner to use an optical \sersic index for this purpose, as the separation between early and late-type galaxies is greater at shorter wavelengths.

Since our analysis is based on the same imaging, we can directly compare our results to \cite{kelvin12}  We have tried applying their selection criteria to our measurements and find excellent agreement between the recovered trends.  When averaging over a large sample, our different approaches to measuring the wavelength dependence of \sersic index and effective radius give very similar results.

\subsection{Physical implications}

The patterns revealed in Fig.~\ref{nn_rr} provide important insights regarding the links between the stellar populations and internal structures of galaxies.  Some of the revealed trends are rather surprising and deserve deeper investigation.

\subsubsection{Early-type galaxies}

For \highn galaxies, the constancy of \sersic index with wavelength (\N~$\sim 1$), despite strong variations in effective radius (\R~$< 1$), points to very substantial differences in the stellar populations of early-types as a function of radius, but which are not associated with a change in profile shape.

Our clear result that \highn galaxies appear substantially smaller at longer wavelengths appears to largely rule out the scenario in which gas rich mergers build high-$n$ galaxies that are a combination of an old (red), large-$\re$ component together with a younger (bluer), inner stellar component formed from the gas \citep{hopkins08,hopkins09}.  If this is a significant mechanism in the formation of massive ellipticals, it must primarily occur at higher redshifts, and its signatures must be eclipsed by subsequent evolutionary processes, such as dry minor mergers (e.g., two phase galaxy formation: \citealt{oser10}).

Lower mass galaxies (both star-forming and passive) contain younger and lower metallicity stellar populations, and are therefore bluer in colour (e.g., \citealt{kauffmann03}).  At later times, accreting material will fall on to the primary galaxy at higher velocities, as both the total mass of the primary and the distance from which merging galaxies originate increases with time.  Stars accreted at later times relax into a more extended structure than the stars produced in the initial formation of the primary galaxy \citep{naab09}.  This is analogous to the preferential accretion of collisionless dark matter to the outskirts of haloes, leaving their inner profiles unchanged \citep{hiotelis03,salvadorsole05}.
Recent simulation studies have demonstrated similar behaviour for stars accreted on to galaxies during minor mergers \citep{cooper13,hilz13}.  Stars added by minor mergers end up in extended \sersic profiles resulting in a substantial increase in their effective radii for a modest increase in mass.  This `inside-out' growth scenario has gained widespread support as the mechanism responsible for the dramatic evolution of compact high-redshift ellipticals to their present, extended, forms (e.g., \citealt{hopkins10}).

Early-type galaxies are thus expected to comprise a compact red population of stars formed in-situ and a more diffuse, bluer population of stars formed in accreted systems.  It is therefore reasonable for their effective radius to decrease with wavelength.  However, 
as well as increasing the effective radius, the addition of surface brightness to the outskirts of galaxies would also be expected to increase the \sersic index.  Indeed, this effect is seen in previously mentioned simulations \citep{cooper13,hilz13}, and is consistent with observations of massive elliptical galaxies with high \sersic indices \citep{kormendy09}.  So, how can we reconcile the substantial variation in effective radius with wavelength with little or no corresponding change in \sersic index?

The reality is that an early-type galaxy will have experienced a variety of merger events.  The stars from each of these will be accreted over different ranges of radius, while in-situ stars will be somewhat redistributed, and the presence of gas may lead to the formation of stars in the galaxy centre.  Adding stars in the outskirts of a galaxy increases $n$ and increases $\re$.  Adding stars in the centre increases $n$ and decreases $\re$.  However, adding stars at the effective radius (i.e., $\sim\! 1\, \re$) decreases $n$ with no change in $\re$.  So, by adding material around $\re$ as well as further out, it is possible to increase $\re$ while maintaining $n$ roughly constant.  Also note that as the effective radius grows due to stellar accretion, subsequent accretion at the same physical radius (in terms of absolute distance and gravitational potential) will be closer to the effective radius, and hence will act to stabilise $n$.  It can be easily checked that the sum of multiple \sersic profiles spanning a range of $\re$, but each with the same $n$ and similar central surface brightness or total magnitude, is well approximated by another \sersic profile with a similar value of $n$ and intermediate $\re$ (though this does depend somewhat on the weighting given to different radii).  Measuring significant variations in $\re$ with wavelength (\R~$< 1$), but small variations in $n$ (\N~$\sim 1$), therefore appears to be a natural outcome of the two-stage formation scenario.

Cosmological models incorporating these various early-type evolutionary processes are now becoming available (e.g., \citealt{shankar13}).  Comparing the properties of simulated populations of spatially resolved galaxies to the results of this paper will help to determine if and when each of these processes are occurring.

\subsubsection{Late-type galaxies}

For \lown objects, there is a substantial increase in the average \sersic index from $\langle n \rangle \sim 1$ in the blue to $\langle n \rangle \sim 2$ in the near-infrared.  This suggests that at short wavelengths, which are sensitive to young stellar populations, the galaxy profile is dominated by an exponential disk, while at longer wavelengths, which are more representative of total stellar mass, the surface brightness profile is steepened by the presence of a bulge.

Section \ref{stacks} demonstrates that the effect of \N~$> 1$ and \R~$< 1$ is to produce a red galaxy centre and blue outskirts.  As mentioned in the introduction, such colour gradients of disk galaxies are typically attributed to radial age profiles, in that they are younger at larger radii.  Considering colour gradients in terms of the wavelength-dependence of \sersic profile parameters gives a complementary perspective, and emphasises the association between the stellar populations and structural components of these galaxies.

The fact that \lown galaxies display only a small change in effective radius with wavelength, while showing a significant variation in \sersic index, suggests that the bulge component in these systems is a spheroid with $n \ga 2$, but which is not substantially smaller than the disk.  This goes against conventional wisdom, but may be a result of our selection of very luminous galaxies, which are typically bulge-dominated.

The different trends in \sersic index versus wavelength seen for \red and \blue/\green \lown galaxies supports the interpretation of \N and \R in terms of a bulge and disk. \Red \lown galaxies show very little variation in \sersic index (\N~$\sim 1$) because their bulge and disk components have similar red colours.  \Green and \blue \lown galaxies show more substantial variation (\N~$> 1$) because they possess a bluer disk, which contrasts with a red bulge.

Although the above argument provides a simple explanation, one must not underestimate the role of dust attenuation in driving, or at least complicating, the trends we observe for \lown galaxies.
Dust attenuation can significantly modify the apparent structure of a galaxy, and introduce substantial variations with wavelength.
Obscuring the central region of a galaxy shifts the balance of total flux towards larger radii, artificially increasing the effective radius and reducing the \sersic index, particularly at shorter wavelengths.  This predicted effect has been well discussed in the literature \citep{evans94,cunow01,moll06,graham08}.  The effects on the recovered parameters of multi-wavelength \sersic-model fits have been particularly studied by \citet{Pastrav13,Pastrav13b}.  We include their predictions in our Figs.~\ref{nn_med} and \ref{rr_med}, which show that while some of the behaviour we observe may be attributable to dust, their model is not able to account for all of the measured variation in $n$ and $\re$ with wavelength.

Dust is not expected to be a dominant factor in the attenuation of light within high-$n$ (spheroid-dominated) systems, hence variations in structural parameters with wavelength cannot be ascribed to its presence. Still, 
we find that these systems exhibit a larger variation in effective radius than that found in low-$n$ (disk-dominated) systems.
Given that such substantial dependences of structure on wavelength are seemingly possible from stellar population gradients alone, it is difficult to judge the importance of dust in driving the trends in low-$n$ galaxies.  Combining measurements of structural parameters versus wavelength with constraints on the mass and distribution of dust from sub-mm observations and appropriate models (e.g., \citealt{popescu11,smith12}) will greatly help in this regard.

Finally, it should be mentioned that another possible explanation for peculiar variations in $n$ and $\re$ with wavelength is the interdependency between recovered \sersic index and effective radius. A change in the \sersic index arising due to relatively subtle unmodelled effects, e.g. a small core dust component, additional unresolved or disturbed structure in the core, the presence of an active galactic nucleus or uncertainty in the PSF, may all induce compensatory changes in the measured effective radius.  However, to be responsible for the average trends we describe, which represent the behaviour of most galaxies in each subsample, such perturbances must be the norm.  Future work fitting more complex multi-component models to specific galaxy samples will shed light on this issue, as well as addressing many of the science questions highlighted above.

\subsection{Potential applications and further studies}

In this paper we have demonstrated how the wavelength-dependence of \sersic index (\N) and effective radius (\R) provide an alternative way of considering galaxy colour gradients, and enable a deeper appreciation of the physical origins of such gradients.  For simplicity we have focussed on a sample of bright galaxies, with the added advantage that this selection demonstrates our method's ability to recover meaningful information even for poorly resolved galaxy images.  Obvious next steps are to study the variation of \N and \R as a function of luminosity (or stellar mass), to examine how the distinctions between galaxies of different colour and concentration (e.g., Fig.~\ref{nn_rr}) are modified for lower-luminosity systems.  The distribution of colour gradients, and hence \N and \R, also varies significantly with environment (e.g., \citealt{weinmann09}), as implied by the dramatic change in the relationship between overall colour and morphology with environment (e.g., \citealt{bamford09}).  The relative behaviour of colour, \sersic index, and their respective dependencies on wavelength will help us better understand the physical origins of the environmental trends observed (e.g., \citealt{vulcani12}).

In addition to considering the overall distribution of \N and \R for particular galaxy samples, in this paper we have discussed using \N and \R to classify galaxies and identify objects with specific properties.
Our approach is somewhat similar to that presented by \citet{park05}, which uses $(u-r)$ colour versus $(g-i)$ colour gradient space to perform a morphological classification of galaxies in SDSS. In this space, galaxies form well-separated early- and late-type branches. The location of galaxies along the branches reflects the degree and locality of star formation activity, and monotonically corresponds to a sequence in morphology.  An analysis of the correspondence between \N, \R and morphology is underway (Vika et al., in prep.), using the same small, but very well characterised, sample of galaxies studied in \citet{vika13}.   One major advantage of utilising colour gradient information is that blue early-type galaxies are separated from spiral galaxies, as the former tend to have either a nearly constant color profile or bluer cores, while the latter tend to have a red central bulge plus blue outer disk.  We can make similar distinctions using the different \N and \R behaviour for galaxies with contrasting internal structures.  Both \citeauthor{park05} and our method are capable of identifying a variety of unusual objects, such as central starbursts and passive disk galaxies.  We therefore foresee much potential in studying the occurrence of such galaxies as a function of stellar mass and environment.

Of course, another extension of this work would be to go beyond single-\sersic profiles, to fit multiple component models.  These are already starting to be used to great effect (e.g., \citealt{hudson10,claire13,huang13b}).  Using a multi-component, wavelength-dependent model is straightforward with \galfitm, and we are currently working with bulge-disk decompositions to help identify the correct interpretation of the trends in single-\sersic parameters.  However, while multi-component models provide more detail, they require more assumptions, not least that the chosen multi-component combination is appropriate to each galaxy in question. They are also more sensitive to practical details of galaxy data, e.g. the presence of spiral arms or small, undeblended companions, and ensuring one finds the true optimal fit is considerably more difficult.  The simplicity and robustness of single-\sersic fitting means that it will long continue to be a valuable approach, especially when pushing the signal-to-noise and resolution boundaries of any dataset.

\section{Summary}

We have studied the dependence of \sersic index ($n$) and effective radius ($\re$) on wavelength using a volume-limited sample of bright galaxies ($M_r < -21.2$, $z<0.3$), with homogeneous $ugrizYJHK$ imaging compiled by the GAMA survey \citep{hill11}.  The \sersic profile parameters were obtained by exploiting the capability of \galapagos-2 and \galfitm to fit consistent, wavelength-dependent galaxy models to a collection of multi-wavelength imaging.  These measurements have been shown to be more accurate and robust than those obtained by fitting galaxies independently in each waveband (see H13, \citealt{vika13}).

We presented the wavelength dependence of the \sersic index and effective radius distributions for the entire sample, and for subsamples split by rest-frame $(u-r)$ colour and the $r$-band \sersic index ($n_r$).  Importantly, our multi-wavelength measurements are accurate and meaningful on an individual galaxy basis.  Our analysis is thus able to go a step further than previous studies and consider the wavelength variation of $n$ and $\re$ for individual galaxies.  We describe this using the ratio of the values in two wavebands, which we denote as \N and \R, respectively.  In this paper we primarily focus on variations between $g$- and $H$-band, which is representative of the behaviour for other pairs of optical and NIR bands.  We present the distributions of \N and \R for 
the various colour and \sersic index 
combinations, and show that their unimodality implies that results based on population trends well represent the behaviour of individual galaxies in each subsample.  The dependence of median \Nang and \Rang on wavelength shows contrasting trends for different galaxy types, which reveal their different physical natures.  The \N-\R plane can be used to classify galaxies and, particularly in combination with colour and \sersic index, identify objects with interesting properties.  Finally, we have illustrated the implications of \N and \R on the appearance of galaxies, and demonstrated their relationship to colour gradients, by coadding images for various selections.

Our main results from all this analysis can be summarised as follows.  All these results are based on a volume-limited sample of bright galaxies.  The studied trends are likely to vary for the less-luminous, dwarf galaxy population, but we leave an exploration of this issue for future work.

\begin{itemize}
\item[$\bullet$]
Galaxies of opposing colour are characterised by very different \sersic index distributions, which depend differently upon the wavelength at which the \sersic index is measured.
The median \sersic index for \red galaxies does not depend on wavelength, while for our \blue and \green galaxy subsamples the median \sersic index increases significantly at longer wavelengths.
In contrast, galaxies with different colours display similar distributions of effective radius.  While galaxies in our \red subsample are typically slightly smaller than the others, the distributions of all colours shift consistently toward lower values at longer wavelengths.

\item[$\bullet$] Using \sersic index to divide the galaxy population, \highn galaxies display much less \sersic index variation with wavelength than \lown galaxies.  \Red, \highn galaxies show no significant dependence of \sersic index on wavelength at all.  On average, they possess a classic de Vaucouleur profile at all wavelengths. Bluer \highn galaxies show mild dependences on wavelength.   However, the median \sersic index for our \green \highn subsample increases toward longer wavelengths, while our more extreme \blue \highn subsample shows a significant decrease.  The \sersic indices of \lown galaxies increase dramatically toward longer wavelengths, irrespective of colour, though our \red \lown subsample shows slightly less variation.

\item[$\bullet$] With regard to effective radius, \lown samples of all colours display very similar behaviour.  Our \highn samples show stronger trends, which vary for different colour selections. In all the cases, sizes decrease toward redder wavelengths. 

\item[$\bullet$] By measuring for the first time the dependence of \sersic index on wavelength for individual galaxies (\N), we find that this quantity adds important information about galaxy structure, helping to robustly separate single-component and bulge-disk systems. \Lown galaxies show \N~$>1$, with the value strongly increasing with the wavelength range considered, suggesting a combination of a blue exponential disk and smaller and/or higher-$n$ red spheroid. \Highn galaxies show rather flat trends (\N~$\sim 1$), indicating that they are typically one-component objects. Galaxies that deviate from these behaviours are likely to be at interesting evolutionary stages. 

\item[$\bullet$] Similarly, the ratio of effective radii at different wavelengths (\R) provides insight into the structural make-up of galaxies, although it does not present such disparate behaviour for the various colour and \sersic index subsamples.  The vast majority of galaxies have \R~$<1$: at longer wavelengths galaxies appear smaller.  \Lown galaxies show a mild dependence of effective radius on wavelength, which is identical for the different colour subsamples. \Highn galaxies have a more dramatic dependence, which is stronger for bluer samples.

\item[$\bullet$] Combining \N and \R together, we show that the medians of different galaxy types (in terms of colour and \sersic index) display strikingly different variations in \Nang and \Rang with wavelength, which succinctly describe the correlations between their internal structures and stellar populations.

\item[$\bullet$] While galaxy subsamples do overlap in the \N-\R plane, the location of a galaxy in this diagram puts useful constrains on its structural properties.  This is particularly true if \sersic index and colour are considered too, allowing one to identify, e.g., central starbursts, blue ellipticals, passive disk galaxies.   The separation of galaxy populations in \N-\R may improve for higher resolution and signal-to-noise imaging data. 

\item[$\bullet$] On average, bright \highn galaxies appear consistent with possessing multiple superimposed $n \sim 4$ components, each with different effective radii, where the larger profiles are composed of bluer stellar populations.
Since the vast majority of galaxies are significantly redder in their central regions, it appears that the proposed formation of young (blue), inner stellar components in elliptical galaxies from accreted gas is negligible for typical bright galaxies at recent times.

\end{itemize}

\section*{Acknowledgements}
This work was made possible by NPRP award 08-643-1-112 from the Qatar National Research Fund (a member of The Qatar Foundation).  It was also supported by the World Premier International Research Center Initiative (WPI), MEXT, Japan.  
SPB gratefully acknowledges receipt of an STFC Advanced Fellowship.

GAMA is a joint European-Australasian project based around a spectroscopic campaign using the Anglo-Australian Telescope. The GAMA input catalogue is based on data taken from the Sloan Digital Sky Survey and the UKIRT Infrared Deep Sky Survey. Complementary imaging of the GAMA regions is being obtained by a number of independent survey programs including GALEX MIS, VST KiDS, VISTA VIKING, WISE, Herschel-ATLAS, GMRT and ASKAP providing UV to radio coverage. GAMA is funded by the STFC (UK), the ARC (Australia), the AAO, and the participating institutions. The GAMA website is http://www.gama-survey.org/.

We thank the anonymous referee for their constructive comments.

\label{lastpage}

\begin{thebibliography}{99}

\bibitem[\protect\citeauthoryear{Andrae, Melchior, 
\& Jahnke}{2011}]{andrae11} Andrae R., Melchior P., Jahnke K., 2011, MNRAS, 417, 2465 

\bibitem[\protect\citeauthoryear{Andredakis, Peletier, 
\& Balcells}{1995}]{andredakis95} Andredakis Y.~C., Peletier R.~F., Balcells M., 1995, MNRAS, 275, 874 

\bibitem[\protect\citeauthoryear{Baes 
\& van Hese}{2011}]{baes11} Baes M., van Hese E., 2011, A\&A, 534, A69 

\bibitem[\protect\citeauthoryear{Bamford et 
al.}{2012}]{bamford12} Bamford S.~P., H{\"a}u{\ss}ler B., Rojas 
A., Vika M., Cresswell J., 2012, IAUS, 284, 301 

\bibitem[\protect\citeauthoryear{Bamford et 
al.}{2009}]{bamford09} Bamford S.~P., et al., 2009, MNRAS, 393, 
1324 

\bibitem[\protect\citeauthoryear{Barden et al.}{2012}]{barden12} 
Barden M., H{\"a}u{\ss}ler B., Peng C.~Y., McIntosh D.~H., Guo Y., 2012, 
MNRAS, 422, 449 

\bibitem[\protect\citeauthoryear{Barden et al.}{2005}]{barden05} 
Barden M., et al., 2005, ApJ, 635, 959 

\bibitem[\protect\citeauthoryear{Beckman et 
al.}{1996}]{beckman96} Beckman J.~E., Peletier R.~F., Knapen 
J.~H., Corradi R.~L.~M., Gentet L.~J., 1996, ApJ, 467, 175 

\bibitem[\protect\citeauthoryear{Bertin 
\& Arnouts}{1996}]{bertin96} Bertin E., Arnouts S., 1996, A\&AS, 117, 393 


\bibitem[\protect\citeauthoryear{Bertin et al.}{2002}]{bertin02} 
Bertin E., Mellier Y., Radovich M., Missonnier G., Didelon P., Morin B., 
2002, ASPC, 281, 228 

\bibitem[\protect\citeauthoryear{Bertin}{2010}]{bertin10} Bertin 
E., 2010, ascl.soft, 10068 

\bibitem[\protect\citeauthoryear{Blanton 
\& Moustakas}{2009}]{blanton09} Blanton M.~R., Moustakas J., 2009, ARA\&A, 47, 159 

\bibitem[\protect\citeauthoryear{Brinchmann et 
al.}{2004}]{brinchmann04} Brinchmann J., Charlot S., White S.~D.~M., 
Tremonti C., Kauffmann G., Heckman T., Brinkmann J., 2004, MNRAS, 351, 1151 

\bibitem[\protect\citeauthoryear{Cameron et 
al.}{2009}]{cameron09} Cameron E., Driver S.~P., Graham A.~W., 
Liske J., 2009, ApJ, 699, 105 

\bibitem[\protect\citeauthoryear{Caon, Capaccioli, 
\& D'Onofrio}{1993}]{caon93} Caon N., Capaccioli M., D'Onofrio M., 1993, MNRAS, 265, 1013 

\bibitem[\protect\citeauthoryear{Cappellari et 
al.}{2011}]{cappellari11} Cappellari M., et al., 2011, MNRAS, 413, 
813 

\bibitem[\protect\citeauthoryear{Ciotti}{1991}]{ciotti91} Ciotti L., 1991, A\&A, 249, 99 


\bibitem[\protect\citeauthoryear{Cooper et al.}{2013}]{cooper13} 
Cooper A.~P., D'Souza R., Kauffmann G., Wang J., Boylan-Kolchin M., Guo Q., 
Frenk C.~S., White S.~D.~M., 2013, MNRAS, 434, 3348 

\bibitem[\protect\citeauthoryear{Cunow}{2001}]{cunow01} Cunow 
B., 2001, MNRAS, 323, 130 

\bibitem[\protect\citeauthoryear{de 
Jong}{1996b}]{dejong96b} de Jong R.~S., 1996b, A\&A, 313, 377 

\bibitem[\protect\citeauthoryear{de 
Jong}{1996a}]{dejong96} de Jong R.~S., 1996a, A\&A, 313, 45 

\bibitem[\protect\citeauthoryear{den Brok et 
al.}{2011}]{brok11} den Brok M., et al., 2011, MNRAS, 414, 
3052 

\bibitem[\protect\citeauthoryear{Dressler et 
al.}{1997}]{dressler97} Dressler A., et al., 1997, ApJ, 490, 577 

\bibitem[\protect\citeauthoryear{Driver et al.}{2011}]{driver11} 
Driver S.~P., et al., 2011, MNRAS, 413, 971 

\bibitem[\protect\citeauthoryear{Driver et 
al.}{2009}]{driver09} Driver S.~P., et al., 2009, A\&G, 50, 050000 

\bibitem[\protect\citeauthoryear{Driver et al.}{2006}]{driver06} 
Driver S.~P., et al., 2006, MNRAS, 368, 414 

\bibitem[\protect\citeauthoryear{Ellis, Abraham, 
\& Dickinson}{2001}]{ellis01} Ellis R.~S., Abraham R.~G., Dickinson M., 2001, ApJ, 551, 111 

\bibitem[\protect\citeauthoryear{Evans}{1994}]{evans94} Evans 
R., 1994, MNRAS, 266, 511 

\bibitem[\protect\citeauthoryear{Ferreras et 
al.}{2009}]{ferreras09} Ferreras I., Lisker T., Pasquali A., 
Kaviraj S., 2009, MNRAS, 395, 554 

\bibitem[\protect\citeauthoryear{Freeman}{1970}]{freeman70} 
Freeman K.~C., 1970, ApJ, 160, 811 

\bibitem[\protect\citeauthoryear{Gadotti 
\& Dos Anjos}{2001}]{gadotti01} Gadotti D.~A., Dos Anjos S., 2001, ASPC, 230, 237 

\bibitem[\protect\citeauthoryear{Graham}{2013}]{graham13} Graham 
A.~W., 2013, pss6.book, 91 

\bibitem[\protect\citeauthoryear{Graham 
\& Worley}{2008}]{graham08} Graham A.~W., Worley C.~C., 2008, MNRAS, 388, 1708 

\bibitem[\protect\citeauthoryear{Graham 
\& Driver}{2005}]{graham05} Graham A.~W., Driver S.~P., 2005, PASA, 22, 118 

\bibitem[Graham et al.(2005)]{graham05b} Graham, A.~W., Driver, 
S.~P., Petrosian, V., et al.\ 2005, \aj, 130, 1535

\bibitem[\protect\citeauthoryear{Graham 
\& Guzm{\'a}n}{2003}]{GrahamGuzman03} Graham A.~W., Guzm{\'a}n R., 2003, AJ, 125, 2936 

\bibitem[\protect\citeauthoryear{Guo et al.}{2011}]{guo11} 
Guo Y., et al., 2011, ApJ, 735, 18 
 
\bibitem[\protect\citeauthoryear{H{\"a}u{\ss}ler et 
al.}{2013}]{Haussler13} H{\"a}u{\ss}ler B., et al., 2013, MNRAS, 
430, 330 

\bibitem[\protect\citeauthoryear{H{\"a}u{\ss}ler et 
al.}{2007}]{Haeussler07} H{\"a}u{\ss}ler B., et al., 2007, ApJS, 172, 
615 

\bibitem[\protect\citeauthoryear{Hill et al.}{2011}]{hill11} 
Hill D.~T., et al., 2011, MNRAS, 412, 765 

\bibitem[\protect\citeauthoryear{Hilz, Naab, 
\& Ostriker}{2013}]{hilz13} Hilz M., Naab T., Ostriker J.~P., 2013, MNRAS, 429, 2924 

\bibitem[\protect\citeauthoryear{Hiotelis}{2003}]{hiotelis03} 
Hiotelis N., 2003, MNRAS, 344, 149 

\bibitem[\protect\citeauthoryear{Hopkins et 
al.}{2010}]{hopkins10} Hopkins P.~F., Bundy K., Hernquist L., 
Wuyts S., Cox T.~J., 2010, MNRAS, 401, 1099 

\bibitem[\protect\citeauthoryear{Hopkins et 
al.}{2009}]{hopkins09} Hopkins P.~F., Lauer T.~R., Cox T.~J., 
Hernquist L., Kormendy J., 2009, ApJS, 181, 486 

\bibitem[\protect\citeauthoryear{Hopkins et 
al.}{2008}]{hopkins08} Hopkins P.~F., Hernquist L., Cox T.~J., 
Dutta S.~N., Rothberg B., 2008, ApJ, 679, 156 

\bibitem[\protect\citeauthoryear{Huang et al.}{2013b}]{huang13b} 
Huang S., Ho L.~C., Peng C.~Y., Li Z.-Y., Barth A.~J., 2013b, ApJ, 768, L28 

\bibitem[\protect\citeauthoryear{Huang et al.}{2013a}]{huang13a} 
Huang S., Ho L.~C., Peng C.~Y., Li Z.-Y., Barth A.~J., 2013a, ApJ, 766, 47 

\bibitem[\protect\citeauthoryear{Hudson et al.}{2010}]{hudson10} 
Hudson M.~J., Stevenson J.~B., Smith R.~J., Wegner G.~A., Lucey J.~R., 
Simard L., 2010, MNRAS, 409, 405 

\bibitem[\protect\citeauthoryear{Jansen et al.}{2000}]{jansen00} 
Jansen R.~A., Franx M., Fabricant D., Caldwell N., 2000, ApJS, 126, 271 

\bibitem[\protect\citeauthoryear{Johnston et 
al.}{2012}]{johnston12} Johnston E.~J., Arag{\'o}n-Salamanca A., 
Merrifield M.~R., Bedregal A.~G., 2012, MNRAS, 422, 2590 

\bibitem[\protect\citeauthoryear{Kauffmann et 
al.}{2003}]{kauffmann03} Kauffmann G., et al., 2003, MNRAS, 341, 54 

\bibitem[\protect\citeauthoryear{Kelvin et al.}{2012}]{kelvin12} 
Kelvin L.~S., et al., 2012, MNRAS, 421, 1007 

\bibitem[\protect\citeauthoryear{Ko 
\& Im}{2005}]{ko05} Ko J., Im M., 2005, JKAS, 38, 149 

\bibitem[\protect\citeauthoryear{Kobayashi}{2004}]{kobayashi04} 
Kobayashi C., 2004, MNRAS, 347, 740 

\bibitem[\protect\citeauthoryear{Kormendy et 
al.}{2009}]{kormendy09} Kormendy J., Fisher D.~B., Cornell M.~E., 
Bender R., 2009, ApJS, 182, 216 

\bibitem[Kron(1980)]{kron80} Kron, R.~G.\ 1980, \apjs, 43, 305 


\bibitem[\protect\citeauthoryear{Lackner 
\& Gunn}{2013}]{claire13} Lackner C.~N., Gunn J.~E., 2013, MNRAS, 428, 2141 


\bibitem[\protect\citeauthoryear{La Barbera et 
al.}{2012}]{barbera12} La Barbera F., Ferreras I., de Carvalho 
R.~R., Bruzual G., Charlot S., Pasquali A., Merlin E., 2012, MNRAS, 426, 
2300 

\bibitem[\protect\citeauthoryear{La Barbera et 
al.}{2010a}]{barbera10a} La Barbera F., Lopes P.~A.~A., de Carvalho 
R.~R., de La Rosa I.~G., Berlind A.~A., 2010a, MNRAS, 408, 1361 

\bibitem[\protect\citeauthoryear{La Barbera et 
al.}{2010b}]{barbera10b} La Barbera F., de Carvalho R.~R., de La 
Rosa I.~G., Lopes P.~A.~A., Kohl-Moreira J.~L., Capelato H.~V., 2010b, 
MNRAS, 408, 1313 

\bibitem[\protect\citeauthoryear{La Barbera 
\& de Carvalho}{2009}]{barbera09} La Barbera F., de Carvalho R.~R., 2009, ApJ, 699, L76 

\bibitem[\protect\citeauthoryear{La Barbera et 
al.}{2008}]{LdC08} La Barbera F., de Carvalho R.~R., 
Kohl-Moreira J.~L., Gal R.~R., Soares-Santos M., Capaccioli M., Santos R., 
Sant'anna N., 2008, PASP, 120, 681 

\bibitem[\protect\citeauthoryear{La Barbera et 
al.}{2003}]{barbera03} La Barbera F., Busarello G., Massarotti M., Merluzzi P., Mercurio A., 2003, A\&A, 409, 21 

\bibitem[\protect\citeauthoryear{La Barbera et 
al.}{2002}]{barbera02} La Barbera F., Busarello G., Merluzzi P., 
Massarotti M., Capaccioli M., 2002, ApJ, 571, 790 

\bibitem[\protect\citeauthoryear{Lee et al.}{2008}]{lee08} 
Lee J.~H., Lee M.~G., Park C., Choi Y.-Y., 2008, MNRAS, 389, 1791 

\bibitem[Lupton et al.(2004)]{lupton04} Lupton, R., Blanton, 
M.~R., Fekete, G., et al.\ 2004, \pasp, 116, 133 

\bibitem[\protect\citeauthoryear{MacArthur et 
al.}{2004}]{mac04} MacArthur L.~A., Courteau S., Bell E., 
Holtzman J.~A., 2004, ApJS, 152, 175 

\bibitem[\protect\citeauthoryear{M{\"o}llenhoff, Popescu, 
\& Tuffs}{2006}]{moll06} M{\"o}llenhoff C., Popescu C.~C., Tuffs R.~J., 2006, A\&A, 456, 941 

\bibitem[\protect\citeauthoryear{Naab, Johansson, 
\& Ostriker}{2009}]{naab09} Naab T., Johansson P.~H., Ostriker J.~P., 2009, ApJ, 699, L178 

\bibitem[\protect\citeauthoryear{Oser et al.}{2010}]{oser10} 
Oser L., Ostriker J.~P., Naab T., Johansson P.~H., Burkert A., 2010, ApJ, 
725, 2312 

\bibitem[\protect\citeauthoryear{Pahre}{1999}]{pahre99} Pahre 
M.~A., 1999, ApJS, 124, 127 

\bibitem[\protect\citeauthoryear{Pahre, de Carvalho, 
\& Djorgovski}{1998}]{pahre98a} Pahre M.~A., de Carvalho R.~R., Djorgovski S.~G., 1998, AJ, 116, 1606 

\bibitem[\protect\citeauthoryear{Pahre, Djorgovski, 
\& de Carvalho}{1998}]{pahre98b} Pahre M.~A., Djorgovski S.~G., de Carvalho R.~R., 1998, AJ, 116, 1591 

\bibitem[\protect\citeauthoryear{Park 
\& Choi}{2005}]{park05} Park C., Choi Y.-Y., 2005, ApJ, 635, L29 

\bibitem[\protect\citeauthoryear{Pastrav et 
al.}{2013b}]{Pastrav13b} Pastrav B.~A., Popescu C.~C., Tuffs R.~J., Sansom A.~E., 2013b, A\&A, 557, A137 

\bibitem[\protect\citeauthoryear{Pastrav et 
al.}{2013a}]{Pastrav13} Pastrav B.~A., Popescu C.~C., Tuffs R.~J., Sansom A.~E., 2013a, A\&A, 553, A80 

\bibitem[\protect\citeauthoryear{Peletier 
\& Balcells}{1996}]{peletier96} Peletier R.~F., Balcells M., 1996, AJ, 111, 2238 

\bibitem[\protect\citeauthoryear{Peng et al.}{2002}]{peng02} 
Peng C.~Y., Ho L.~C., Impey C.~D., Rix H.-W., 2002, AJ, 124, 266 

\bibitem[\protect\citeauthoryear{Peng et al.}{2010}]{peng10} 
Peng C.~Y., Ho L.~C., Impey C.~D., Rix H.-W., 2010, AJ, 139, 2097 

\bibitem[\protect\citeauthoryear{Pompei 
\& Natali}{1997}]{pompei97} Pompei E., Natali G., 1997, A\&AS, 124, 129 

\bibitem[\protect\citeauthoryear{Popescu et 
al.}{2011}]{popescu11} Popescu C.~C., Tuffs R.~J., Dopita M.~A., Fischera J., Kylafis N.~D., Madore B.~F., 2011, A\&A, 527, A109 

\bibitem[\protect\citeauthoryear{Rawle, Smith, 
\& Lucey}{2010}]{rawle10} Rawle T.~D., Smith R.~J., Lucey J.~R., 2010, MNRAS, 401, 852 


\bibitem[Rider(1960)]{rider60}
Rider, P.~R.\ 1960, JASA, 55 (289), 148

\bibitem[\protect\citeauthoryear{Rodr{\'{\i}}guez 
\& Padilla}{2013}]{rodriguez13} Rodr{\'{\i}}guez S., Padilla N.~D., 2013, MNRAS, 434, 2153 

\bibitem[\protect\citeauthoryear{Rudnick et 
al.}{2003}]{rudnick03} Rudnick G., et al., 2003, ApJ, 599, 847 

\bibitem[\protect\citeauthoryear{Saglia et 
al.}{2000}]{saglia00} Saglia R.~P., Maraston C., Greggio L., Bender R., Ziegler B., 2000, A\&A, 360, 911 

\bibitem[\protect\citeauthoryear{Salvador-Sol{\'e}, Manrique, 
\& Solanes}{2005}]{salvadorsole05} Salvador-Sol{\'e} E., Manrique A., Solanes J.~M., 2005, MNRAS, 358, 901 

\bibitem[\protect\citeauthoryear{Sersic}{1968}]{sersic68} Sersic 
J.~L., 1968, adga.book,  

\bibitem[\protect\citeauthoryear{Shankar et 
al.}{2013}]{shankar13} Shankar F., Marulli F., Bernardi M., Mei 
S., Meert A., Vikram V., 2013, MNRAS, 428, 109 

\bibitem[\protect\citeauthoryear{Smith et al.}{2012}]{smith12} 
Smith D.~J.~B., et al., 2012, MNRAS, 427, 703 

\bibitem[\protect\citeauthoryear{Spolaor et 
al.}{2009}]{spolaor09} Spolaor M., Proctor R.~N., Forbes D.~A., 
Couch W.~J., 2009, ApJ, 691, L138 

\bibitem[\protect\citeauthoryear{Suh et al.}{2010}]{suh10} 
Suh H., Jeong H., Oh K., Yi S.~K., Ferreras I., Schawinski K., 2010, ApJS, 
187, 374 

\bibitem[\protect\citeauthoryear{Tamura 
\& Ohta}{2000}]{tamura00} Tamura N., Ohta K., 2000, AJ, 120, 533 

\bibitem[\protect\citeauthoryear{Taylor et al.}{2009}]{taylor09} 
Taylor E.~N., et al., 2009, ApJS, 183, 295 

\bibitem[\protect\citeauthoryear{Taylor et al.}{2005}]{taylor05} 
Taylor V.~A., Jansen R.~A., Windhorst R.~A., Odewahn S.~C., Hibbard J.~E., 
2005, ApJ, 630, 784 

\bibitem[\protect\citeauthoryear{Trujillo, Graham, 
\& Caon}{2001}]{trujillo01} Trujillo I., Graham A.~W., Caon N., 2001, MNRAS, 326, 869 


\bibitem[\protect\citeauthoryear{Vika et al.}{2013}]{vika13} 
Vika M., Bamford S.~P., H{\"a}u{\ss}ler B., Rojas A.~L., Borch A., Nichol 
R.~C., 2013, MNRAS, 435, 623 

\bibitem[\protect\citeauthoryear{Vulcani et 
al.}{2012}]{vulcani12} Vulcani B., et al., 2012, MNRAS, 420, 1481 

\bibitem[\protect\citeauthoryear{York et al.}{2000}]{york00} 
York D.~G., et al., 2000, AJ, 120, 1579 

\bibitem[\protect\citeauthoryear{Waller et al.}{2003}]{waller03} 
Waller W.~H., Fanelli M.~N., Marcum P.~M., Stewart S.~G., 2003, AAS, 36, 
\#146.08 

\bibitem[\protect\citeauthoryear{Weinmann et 
al.}{2009}]{weinmann09} Weinmann S.~M., Kauffmann G., van den 
Bosch F.~C., Pasquali A., McIntosh D.~H., Mo H., Yang X., Guo Y., 2009, 
MNRAS, 394, 1213 



\end{thebibliography}
\end{document}